\documentclass[11pt]{article}

\newcommand{\beq}{\begin{equation}}
\newcommand{\eeq}{\end{equation}}
\newcommand{\ul}{\underline}

\newcommand{\eps}{{\epsilon}}
\newcommand{\abs}[1]{\left| #1 \right|}

\newcommand{\hc}{\mathrm{h.c.}}
\newcommand{\Lag}{\mathcal L}

\newcommand{\Imp}{\mathrm{Im}}

\newcommand{\Lie}[1]{\mathcal L_{#1}}

\newcommand{\piecett}[4]
        {\left\{\begin{array}{cc}
                #1 & #2 \\
                #3 & #4
        \end{array} \right.}

\newcommand{\piecettt}[6]
        {\left\{\begin{array}{cc}
                #1 & #2 \\
                #3 & #4 \\
                #5 & #6
        \end{array} \right.}

\newcommand{\CD}{\mathcal{D}}
\newcommand{\BCD}{\mathcal{\bar D}}
\newcommand{\CCD}{\nabla}
\newcommand{\BCCD}{\bar\nabla}

\newcommand{\CP}{{\mathcal P}}

\newcommand{\RCP}{\Pi}

\newcommand{\fint}{\int d^2\theta\, \mathcal{E}\; }

\newcommand{\cfint}{\int d^2\theta\, \mathcal{\conf E}\;}
\newcommand{\dint}{\int d^4\theta\, E\;}
\newcommand{\cdint}{\int d^4\theta\, \conf E\;}
\newcommand{\conf}[1]{\breve{#1}}

\newcommand{\bsigma}{\bar{\sigma}}
\newcommand{\dalpha}{{\dot{\alpha}}}
\newcommand{\dbeta}{{\dot{\beta}}}
\newcommand{\dgamma}{{\dot{\gamma}}}
\newcommand{\ddelta}{{\dot{\delta}}}
\newcommand{\dmu}{{\dot{\mu}}}
\newcommand{\dnu}{{\dot{\nu}}}
\newcommand{\dphi}{{\dot{\phi}}}
\newcommand{\btheta}{{\bar\theta}}

\newcommand{\chE}{\mathcal E}
\newcommand{\chz}{\mathfrak z}

\newcommand{\W}{\mathcal W}
\newcommand{\tF}{\tilde F}
\newcommand{\tH}{\tilde H}
\newcommand{\sohn}{\zeta}

\newcommand{\eol}{\notag \\}

\newcommand{\chM}{\mathfrak{M}}
\newcommand{\chA}{\mathfrak{a}}
\newcommand{\cha}{\mathfrak{a}}
\newcommand{\chb}{\mathfrak{b}}
\newcommand{\chc}{\mathfrak{c}}
\newcommand{\chd}{\mathfrak{d}}
\newcommand{\chD}{\mathfrak{D}}
\newcommand{\chm}{\mathfrak{m}}
\newcommand{\chn}{\mathfrak{n}}
\newcommand{\chPi}{\RCP}
\newcommand{\chDelta}{\Delta}
\newcommand{\chS}{\mathfrak{S}}

\newcommand{\lsym}[1]{\stackrel{\scriptstyle#1}{\mbox{$\smile$}}}

\newcommand{\kA}{\mathbb{A}}
\newcommand{\kg}{\mathbf{k}}
\newcommand{\BX}{\bar X}

\usepackage{amsmath}
\usepackage{amsfonts}
\usepackage{wasysym}
\usepackage{graphicx}

\numberwithin{equation}{section}

\begin{document}

\thispagestyle{empty}


\hfill UCB-PTH-09/21

\hfill arXiv:0906.4399 [hep-th]

\hfill August 21, 2009

\addvspace{45pt}

\begin{center}

\Large{\textbf{$\mathcal N=1$ Conformal Superspace in Four Dimensions}}
\\[35pt]
\large{Daniel Butter}
\\[10pt]
\textit{Department of Physics, University of California, Berkeley}
\\ \textit{and}
\\ \textit{Theoretical Physics Group, Lawrence Berkeley National Laboratory}
\\ \textit{Berkeley, CA 94720, USA}
\\[10pt] 
dbutter@berkeley.edu
\end{center}

\addvspace{35pt}

\begin{abstract}
\noindent
We construct in detail an $\mathcal N=1, D=4$ superspace with the
superconformal algebra as the structure group and discuss its relation to
prior component approaches and the existing Poincar\'e superspaces.
\end{abstract}

\setcounter{tocdepth}{2}
\newpage
\setcounter{page}{1}
\tableofcontents
\newpage
\section{Introduction}
The use of conformal techniques to address supergravity has a long history.
Not all that long after Wess and Zumino discovered the superspace formulation
of supergravity \cite{WZ:sug}, Kaku, Townsend, and van Nieuwenhuizen, along with
Ferrara and Grisaru, worked out the
conformal structure of component supergravity and demonstrated
that Poincare supergravity was a gauge-fixed version of conformal supergravity \cite{PfromC}.
Howe first proposed superspace formulations of four-dimensional $\mathcal N \leq 4$
conformal supergravities by explicitly gauging $\mathrm{SL(2,\mathcal C) \times U(\mathcal N)}$
\cite{Howe:csg}. Work continued on conformal supergravity over the next few years
(an excellent review \cite{csg} on the topic was written by Fradkin and Tseytlin)
eventually culminating in the work of Kugo and Uehara, who not only popularized
the conformal compensator approach to supergravity and matter systems \cite{Kugo:1982cu}
but also made a comprehensive analysis of the component transformation rules
and spinorial derivative structure of $\mathcal N=1$ conformal
supergravity \cite{Kugo:1983mv}.

In large part, the results presented here are a superspace response to
this last work. Here we will take a complementary approach, treating superspace as
an honest supermanifold with a conformal structure. Unlike Howe, we will
seek to gauge the \emph{entire} superconformal algebra. Prior experience with
superspace hints that this approach would be a foolish one -- that the constraints
required with a larger structure group would be more numerous and their evaluation
more cumbersome. What we find is the opposite: the covariant derivatives of
conformal supergravity have a Yang-Mills structure, with the algebra
\begin{gather*}
\{\nabla_\alpha, \nabla_\beta\} = 0, \;\;\;
\{\nabla_\dalpha, \nabla_\dbeta\} = 0 \\
\{\nabla_\alpha, \nabla_\dalpha\} = -2i \nabla_{\alpha \dalpha} \\
\{\nabla_\beta, \nabla_{\alpha \dalpha}\} = -2i \eps_{\beta \alpha} \W_\dalpha, \;\;\;
\{\nabla_\dbeta, \nabla_{\alpha \dalpha}\} = -2i \eps_{\dbeta \dalpha} \W_\alpha
\end{gather*}
where $\W_\alpha$ are the ``gaugino superfields'' for the superconformal group.
The constraints of conformal superspace involve setting most of the $\W_\alpha$
to zero, and the evaluation of these constraints is no more difficult than in a
conventional Yang-Mills theory, leading the non-vanishing $\W_\alpha$ to be
expressed in terms of the single superfield $W_{\alpha \beta \gamma}$.
When the theory is ``degauged'' to a $U(1)$ Poincar\'e supergravity,
the extra gauge superfields can be reinterpreted as the familiar
superfields $R$, $G_c$, and $X_\alpha$. This is the main result of this
work.

It is well known that the various equivalent formalisms of superspace supergravity --
the minimal Poincar\'e \cite{oldMin},
the minimal K\"ahler \cite{bgg}, and even the new minimal Poincar\'e \cite{newMin} --
are all derivable from a conformal superspace under different gauge-fixing
constraints. We review one way of seeing how this occurs in our approach.

This paper is divided into two sections. In the first, we
discuss conformal representations of superfields on superspace and
construct the constraints necessary for the existence of such a space.
We also give the explicit form of all the curvatures from solving the
Bianchi identities.
In the second, we demonstrate how the auxiliary structure of $U(1)$
superspace is identical to a certain gauge-fixed version of
conformal superspace. In addition, we explicitly construct the
superspace of minimal supergravity, K\"ahler supergravity, and
new minimal supergravity.
Included in the appendix is an elementary review of the structure of global
and local spacetime symmetry groups as well as the structure of actions over
both the full manifold and submanifolds of such theories.

Throughout this paper we use the superspace notations and conventions
of Binetruy, Girardi, and Grimm \cite{bgg} (which are a slight modification of
those of Wess and Bagger \cite{wb}) -- with our own slight modification:
we choose the superspace $U(1)$ connection to be Hermitian.
That is, our connection $A_M$ here is equivalent to $-i A_M$ of \cite{bgg};
similarly, our corresponding generator $A$ is equivalent to their $iA$.
(The unfortunate coincidence of the generator and connection names will,
we hope, not overly confuse the reader.)

Although the theory discussed here ought to be properly denoted
``superconformal superspace,'' this is an awkward term that we
would like to avoid. Instead we use ``conformal''
when the subject is superspace. (Similarly, supertranslations on
superspace are simply called translations.) When the component
theory is under consideration, we restore the ``super.''

\newpage
\section{Conformal superspace}
In the ensuing section we describe the gauge structure, geometric
constraints, and curvatures of conformal superspace, which is
defined as a normal $\mathcal N=1$ superspace with the structure group
of the superconformal algebra. We discuss representations of that algebra,
invariant actions and chiral submanifold actions. As usual, constraints must be
imposed to eliminate unwanted fields; we will discuss their construction and solution.
But the first place to start is at the component level, where conformal supergravity
is well-known and its properties well-established.

Some use will be made throughout of results presented in the appendix. Specific
references will be made when especially relevant.

\subsection{Conformal supergravity at the component level}
Conformal supergravity at the component level begins with the extension
of the Poincar\'e to the super-Poincar\'e algebra by the addition of
fermionic internal symmetries $Q_\alpha$. These anticommute to give spacetime
translations:
\beq
\{Q_\alpha, \bar Q_\dalpha \} = -2i \sigma^a_{\alpha \dalpha} P_a
\eeq
The rest of the super-Poincar\'e algebra is
\begin{gather}
[M_{ab}, M_{cd}] = \eta_{bc} M_{ad} - \eta_{ac} M_{bd} - \eta_{bd} M_{ac} + \eta_{ad} M_{bc}\eol{}
[M_{ab}, P_c] = P_a \eta_{bc} - P_b \eta_{ac} \eol{}
[M_{ab}, Q_\gamma] = {(\sigma_{ab})_\gamma}^{\beta} Q_\beta
\end{gather}
The bosonic part of the algebra can be extended to include the conformal algebra.
This requires the introduction of the conformal scaling\footnote{This operation is
often called ``dilatation.''} operator $D$ and the
special conformal operator $K_a$, which loosely speaking can be understood
as a translation conjugated by inversions. A brief review of the conformal
algebra is given in Appendix \ref{conf_gp}.

These two generators can be added to the
super-Poincar\'e algebra provided one also includes two new operators, the fermionic
special conformal operator $S_\alpha$ (and its conjugate $\bar S^\dalpha$) and the chiral rotation operator
$A$. (This last generator is the $U(1)$ R-symmetry.)
It should be noted that the special conformal generators possess the
same Lorentz transformation properties as the corresponding translation and
supersymmetry generators, but have opposite weights under scalings and chiral
rotations:
\begin{gather}
[D,P_a] = P_a, \;\;\; [D, Q_\alpha] = \frac{1}{2} Q_\alpha, \;\;\; [D, \bar Q^\dalpha] = \frac{1}{2} \bar Q^\dalpha \eol{} 
[D,K_a] = -K_a, \;\;\; [D, S_\alpha] = -\frac{1}{2} S_\alpha, \;\;\; [D, \bar S^\dalpha] = -\frac{1}{2} \bar S^\dalpha \eol{}
[A, Q_\alpha] = -i Q_\alpha,\;\;\; [A, \bar Q^\dalpha] = +i \bar Q^\dalpha \eol{} 
[A, S_\alpha] = +i S_\alpha, \;\;\; [A, \bar S^\dalpha] = -i \bar S^\dalpha \eol{}
[M_{ab}, K_c] = K_a \eta_{bc} - K_b \eta_{ac} \eol{}
[M_{ab}, S_\gamma] = {(\sigma_{ab})_\gamma}^{\beta} S_\beta
\end{gather}
The special conformal generators have an algebra among each other that is
similar to the supersymmetry algebra:
\begin{align}
\{S_\alpha, \bar S_\dalpha \} = +2i \sigma^a_{\alpha \dalpha} K_a
\end{align}
Finally, the commutators of the special conformal generators with the translation
and supersymmetry generators are
\begin{gather}
[K_a, P_b] = 2 \eta_{ab} D - 2 M_{ab} \eol{}
[K_a, Q_\alpha] = i \sigma_{a \alpha \dbeta} \bar S^{\dbeta}, \;\;\;
[K_a, \bar Q^\dalpha] = i \bsigma_a^{\dalpha \beta} S_{\beta} \eol{}
[S_\alpha, P_a] = i \sigma_{a \alpha \dbeta} \bar Q^{\dbeta},\;\;\;
[\bar S^\dalpha, P_a] = i \bsigma_a^{\dalpha \beta} Q_{\beta}\eol{}
\{S_\alpha, Q_\beta\} = 2 D \epsilon_{\alpha \beta} - 2 M_{\alpha \beta} - 3 i A \epsilon_{\alpha \beta} \eol{}
\{\bar S^\dalpha, \bar Q^\dbeta\} = 2 D \epsilon^{\dalpha \dbeta} - 2 M^{\dalpha \dbeta} + 3 i A \epsilon^{\dalpha \dbeta}
\end{gather}
All other commutators vanish.

We have made use of the convenient shorthand $M_{\alpha \beta} = (\sigma^{ba} \epsilon)_{\alpha \beta} M_{ab}$
and $M^{\dalpha \dbeta} = (\bsigma^{ba} \epsilon)^{\dalpha \dbeta} M_{ab}$.
These are projections of the Lorentz generator; $M_{\alpha \beta}$ rotates
undotted spinors while $M^{\dalpha \dbeta}$ rotates dotted spinors. For example,
\begin{gather*}
[M_{\alpha \beta}, Q_{\gamma}] = -Q_{\alpha} \eps_{\beta \gamma} - Q_{\beta} \eps_{\alpha \gamma} \\
[M_{\alpha \beta}, Q_{\dgamma}] = 0 \\
[M_{\alpha \beta}, P_{(\gamma \dgamma})] = -P_{\alpha \dgamma} \eps_{\beta \gamma} - P_{\beta \dgamma} \eps_{\alpha \gamma}
\end{gather*}
where $P_{(\gamma \dgamma)} \equiv P_c \sigma^c_{\gamma \dgamma}$.
The canonical decomposition of a vector into dotted and undotted spinors is
accomplished via contraction with a sigma matrix.

Conformal supergravity in four dimensions is the gauge theory of the above
algebra. The connection forms $W_m{}^A$ can be collected with their
generators $X_A$ into the total connection form
\begin{align}
W_m = {e_m}^a P_a + \frac{1}{2} {\psi_m}^{\ul \alpha} Q_{\ul \alpha}
	+ \frac{1}{2} {\omega_m}^{ba} M_{ab} + b_m D + A_m A
	+ {f_m}^a K_a + {f_m}^{\ul \alpha} S_{\ul \alpha}
\end{align}
Here $\ul \alpha$ denotes both spinor chiralities ($\alpha$ and $\dalpha$)
and the spinor summation convention is that of \cite{wb}.
In the local theory, the generator $P_a$ is identified as the covariant
derivative when acting on a covariant field.\footnote{
A covariant field $\Phi$ transforms as $\delta_g \Phi = g^A X_A \Phi$.
This is linear in $\Phi$ and involves no derivatives of the parameter $g^A$.}
The algebra of the $P_a$ among themselves is altered by the introduction of curvatures. One finds
on a covariant field $\Phi$
\begin{align}
[P_a, P_b] \Phi \equiv [\CCD_a, \CCD_b] \Phi = -R_{ab} \Phi
\end{align}
where the curvatures are
\begin{align}
R_{nm} = {T_{nm}}^a P_a + {T_{nm}}^{\ul \alpha} Q_{\ul \alpha}
		+ \frac{1}{2} {R_{nm}}^{ba} M_{ab}
		+ H_{nm} D + F_{nm} A
		+ {R(K)_{nm}}^a K_a + {R(S)_{nm}}^{\ul \alpha} S_{\ul \alpha}
\end{align}
Here we are using ${T_{nm}}^{\ul \alpha}$ as the supersymmetry curvature
(anticipating that in superspace this will
become part of the torsion), $H$ and $F$ as the curvatures associated
with scalings and chiral rotations, and $R(K)$ and $R(S)$ as the curvatures
associated with special conformal and fermionic special conformal
transformations. (The curvatures -- with Lorentz form indices -- are also
covariant fields in the sense that a curvature transforms into another
curvature.) The construction of a local gauge theory from a generic
global theory is detailed in Appendix \ref{local_g}.

Constraints are imposed on these curvatures in such a way as to eliminate the spin
connection ${\omega_m}^{ba}$ and the special conformal connections ${f_m}^a$ and
${f_m}^{\ul \alpha}$ in terms of the other fields. This procedure is
summarized in the review literature \cite{csg} but the details do not
concern us here.

The transformation rules of the various gauge fields are straightforward
to calculate and are given in \cite{csg}. For our purposes, the only
ones which will matter are the supersymmetry transformations of the unconstrained
fields:
\begin{gather}
\delta_Q {e_m}^a = i (\xi \sigma^a \bar\psi_m - \psi_m \sigma^a \bar\xi) \\
\label{psiQ}
\delta_Q {\psi_m}^\alpha = 2 \CCD_m \xi^\alpha \\
\label{bQ}
\delta_Q b_m = 2 {f_m}^{\ul \alpha} \xi_{\ul \alpha} \\
\label{aQ}
\delta_Q A_m = -3i {f_m}^\alpha \xi_\alpha + 3i f_{m \dalpha} \bar \xi^{\dalpha}
\end{gather}
The derivative $\CCD_m$ is covariant with respect to spin, scalings, and chiral
rotations and $\xi^\alpha$ is assumed to transform with \emph{opposite} scaling and
chiral weights as $Q_\alpha$. The gravitino transformation rule is especially simple.

It is also useful to record the transformation rules of chiral matter
coupled to conformal supergravity. For the chiral multiplet
$\Phi = (\phi, \psi, F)$,
\begin{align}
\delta_Q \phi = \sqrt 2 \xi \psi, \;\;\;
\delta_Q \psi = \sqrt 2 \xi F + i \sqrt 2 \sigma^a\bar\xi \CCD_a \phi, \;\;\;
\delta_Q F = i \sqrt 2 (\bar \xi \bsigma^a \CCD_a \psi)
\end{align}
which is identical to the supersymmetry algebra except for the
replacement of the regular derivative with the covariant one.

These sets of component transformation rules must be reproduced at the
superfield level once we move to superspace; this will help us to find
the proper constraints on the curvatures in superspace.

\subsection{Conformal superspace and representations of the algebra}

$\mathcal N=1$ superspace is a manifold combining four-dimensional
Minkowski coordinates $x^m$ with four Grassmann coordinates
$\theta^\alpha, \btheta_\dalpha$ into a single supermanifold
with coordinate $z^M = (x^m,\theta^\mu,\btheta_\dmu)$.
The superconformal algebra can be represented as a set of transformations
on these coordinates. In differential form they read \cite{sohn}
\begin{gather}
p_a = \partial_a, \;\;\;
q_\alpha = \partial_\alpha - i (\btheta \bsigma^a\epsilon)_\alpha \partial_a, \;\;\;
\bar q^\dalpha = \partial^\dalpha - i (\theta \sigma^a\epsilon)^\dalpha \partial_a \eol
m_{ab} = -x_{[a} \partial_{b]} + \theta \sigma_{ab} \partial_\theta + \btheta \bsigma_{ab} \partial_\btheta \eol
d = x^m \partial_m + \frac{1}{2} \theta \partial_\theta + \frac{1}{2} \btheta \partial_\btheta, \;\;\;
a = -i \theta \partial_\theta + i \btheta \partial_\btheta \eol
s_\alpha = -2 \theta^2 \partial_\alpha + i (x_b - i \sohn_b) (\sigma_b \partial_\btheta)_\alpha
           - (x_b + i \sohn_b)(\theta \sigma_c \bsigma_b \epsilon)_\alpha \partial_c \eol
s^\dalpha = -2 \btheta^2 \partial^\dalpha + i (x_b + i \sohn_b) (\bsigma_b \partial_\theta)^\dalpha
           - (x_b - i \sohn_b)(\btheta \bsigma_c \sigma_b \epsilon)^\dalpha \partial_c \eol
k_a = 2 x_a (x \cdot \partial) - x^2 \partial_a
      - 2 \sohn_a (\sohn \cdot \partial) + \sohn^2 \partial_a
      - (x_b + i \sohn_b) (\theta \sigma_a \bsigma_b \partial_\theta)
      - (x_b - i \sohn_b) (\btheta \bsigma_a \sigma_b \partial_\btheta)
\end{gather}
where $\sohn^a \equiv \theta\sigma^a\btheta$.
These operators can be found in several ways. The most straightforward
is to write down the supersymmetry line element
$ds^2 = \left(dx^a + i \theta \sigma^a d\btheta + i \btheta \bsigma^a d\theta \right)^2$
and require that it be preserved up to a conformal factor. This yields
the coordinate representations we have given above. The elements
$p_a$, $q_\alpha$, $\bar q^\dalpha$, $m_{ab}$ and $a$ preserve the line
element exactly; the others, $d$, $k_a$, $s_\alpha$ and $\bar s^\dalpha$
preserve it only up to a conformal factor.

The field representation possesses the same algebra as the coordinate
representation but with the opposite sign. We will be most interested in
the field representation, which is the only sensible approach when the
symmetry is made a local one.

As it will be useful to collect terms in a way which makes manifest
the supersymmetry, we will denote by $P_A$ the
set of generators $P_a$, $Q_\alpha$, and $\bar Q^\dalpha$;
$P_A$ represents the super-translation generator on superspace. Similarly, the
special conformal generators may be collected into a single $K_A$.
The algebra as in Section 2.1 can then be written
\begin{gather}
[D, P_A] = \lambda(A) P_A, \;\;\; [A, P_A] = -i \omega(A) P_A \eol{}
[D, K_A] = - \lambda(A) K_A, \;\;\; [A, K_A] = +i \omega(A) K_A \eol{}
[P_A, P_B] = -{C_{A B}}^C P_C, \;\;\; [K_A, K_B] = {C_{AB}}^C K_C \eol{}
[K_A, P_B] = +2 \tilde \eta_{AB} D - 2 M_{AB} + 3 i A \eta_{AB} \omega(A)
             - \frac{1}{2} K^C C_{C B A}  - \frac{1}{2} P^C C_{C A B}
\end{gather}
The commutators and other objects are to be understood as carrying an implicit
grading, explained further in Appendix \ref{imp_g}.

The various objects defined above are
\begin{gather}
P_A = (P_a, Q_\alpha, Q^\dalpha), \;\;\;\; K_A = (K_a, S_\alpha, S^\dalpha) \eol
M_{AB} = \left(M_{ab}, M_{\alpha \beta}, M^{\dalpha \dbeta}\right) \eol
\eta_{AB} = (\eta_{ab}, -\epsilon_{\alpha \beta}, -\epsilon^{\dalpha \dbeta}), \;\;\;\;
\tilde \eta_{AB} = (\eta_{ab}, +\epsilon_{\alpha \beta}, +\epsilon^{\dalpha \dbeta})
\end{gather}
where mixed objects such as $M_{a \beta}$ and $\eta_{a \beta}$ are defined to be zero.
Note that $\tilde\eta_{AB} = (-)^A \eta_{AB}$; this will be the origin of graded
signs $(-)^A$ in subsequent formulae.

We also have the flat-space torsion tensor
\begin{gather}
{C_{A B}}^C = -{C_{BA}}^C = \piecett{- 2i{(\sigma^c \epsilon)_\alpha}^\dbeta}{\textrm{if }\; A=\alpha, B=\dbeta, C=c}{0}{\textrm{otherwise}}
\end{gather}
as well as the numerical coefficients
\begin{gather}
\lambda(A) = \piecett{1}{\textrm{if }\; A=a}{\frac{1}{2}}{\textrm{if }\; A=\alpha,\dalpha} \eol
\omega(A) = \piecettt{0}{\textrm{if }\; A=a}{+1}{\textrm{if }\; A=\alpha}
		{-1}{\textrm{if }\; A=\dalpha}
\end{gather}
The tensor $C$, like all explicitly supersymmetric objects, possesses
an implicit grading.\footnote{That is, we interpret its antisymmetry condition to mean
${C_{ab}}^C = - {C_{ba}}^C$ but ${C_{\alpha \beta}}^C = +{C_{\beta \alpha}}^C$.
The implicit grading works by appending an extra sign whenever two
fermionic objects (fields, indices, etc.) are permuted past each other.}
The matrix $\eta_{AB}$ is used to raise and lower indices; $\tilde \eta_{AB}$ is
its transpose, and is equivalent to $\eta_{AB} (-)^{ab}$, the sign coming
from the implicit grading.

The main limitation of this form is that the would-be super-rotation
generator $M_{AB}$ is highly constrained: only the boson-boson part
$M_{ab}$ is independent. The fermion-boson part $M_{\alpha b}$ vanishes,
and the fermion-fermion part $M_{\alpha \beta}$ is just a left-handed projection
of $M_{ab}$. Nevertheless, we may write its commutator with $P_A$ in
the elegant form
\beq
[M_{AB}, P_C] = P_A \eta_{B C} - P_B \eta_{AC}
\eeq
where it is to be understood that the $A$, $B$, and $C$ are all
of the same type and the implicit grading is understood.

The representation theory of fields under the conformal group is
discussed in \cite{df} as well as in Appendix \ref{conf_gp} and is rather straightforward.
The only difference from Poincar\'e representations is that we require primary
fields $\Phi$ to have constant conformal weight under $D$ and
to be annihilated by the special conformal generator $K_a$.

We may extend that discussion to the superconformal group with little
effort. Let $\Phi$ be a primary superfield. It may or may not
possess Lorentz indices, but we will suppress them for notational elegance.
The action of the superconformal group is
\begin{gather}
P_A \Phi = \CCD_A \Phi, \;\;\;
M_{ab} \Phi = \mathcal S_{ab} \Phi \eol
D \Phi = \Delta \Phi, \;\;\; A \Phi = i w \Phi \eol
K_A \Phi = 0
\end{gather}
The action of $P_A$ is the statement that the translation generator acts
as the covariant derivative. The matrix $\mathcal S_{ab}$
is appropriate for whatever representation of the Lorentz algebra $\Phi$
belongs to. $\Delta$ and $w$ represent its conformal and chiral weights.

\subsubsection{Primary chiral superfields}
The superconformal algebra by itself does not itself tell us anything
more about an arbitrary superfield than the conformal algebra tells us
in spacetime. The advantage comes when restrictions are imposed.
For example, the most important theoretical and phenomenological
superfields are chiral ones. These obey a constraint $\BCCD^\dalpha \Phi = 0$,
where again we are suppressing Lorentz indices which $\Phi$ may possess.
Requiring this to be superconformally invariant leads to
a nontrivial condition on $\Phi$:
\beq
0 = \{S^\dalpha, \BCCD^\dbeta\} \Phi = \epsilon^{\dalpha \dbeta} (2 D + 3 i A) \Phi - 2 M^{\dalpha \dbeta} \Phi
 =  \epsilon^{\dalpha \dbeta} (2 \Delta - 3 w) \Phi - 2 M^{\dalpha \dbeta} \Phi
\eeq
The first term is antisymmetric in the indices, the second is symmetric.
Therefore each must vanish separately. The first tells us $2 \Delta = 3 w$,
that is, the $U(1)$ weight and scaling dimension of the field $\Phi$ have
a fixed ratio. The second term tells us that $\Phi$, when decomposed
into irreducible spinors, contains no dotted
indices, since $M^{\dalpha \dbeta}$ acts only on these.
Thus, $\Phi_{\alpha}$, $\Phi_{\alpha \beta}$, and $\Phi_{\alpha \beta \gamma}$
are valid chiral superfields, but
$\Phi_{(\alpha \dbeta)} = {\sigma_{\alpha \dbeta}}^c \Phi_c$ is not.
(We will use the notation $(\alpha \dalpha)$ to denote a vector
index contracted with a sigma matrix.)

\subsubsection{Primary gauge vector superfields}
The next most important superfield is the gauge vector superfield $V$.
It is related to the chiral superfield $W_\alpha$ by
$W_\alpha = \CP[\CCD_\alpha V]$ where $\CP$ is
the chiral projection operator.\footnote{It is convention in literature
to call this object the ``projection'' operator even though it is not
truly a projection operator, since $\CP^2 \neq \CP$. We denote
$\RCP$ as the \emph{actual} projection operator where it matters.}
In flat supersymmetry this condition reads
$W_\alpha = -\frac{1}{4} \bar D^2 D_\alpha V$ where $D_A$ is the flat
space covariant derivative; we will
assume without (yet) a proof that this expression holds in the case of a
nontrivial geometry simply by replacing $D_A$ with $\nabla_A$.
If we demand that $W_\alpha$ be primary in addition to $V$ being primary,
we can deduce a nontrivial condition on $V$. The primary condition is actually
three: the vanishing of $K$, $S$, and $\bar S$ on $W_\alpha$. Since the anti-commutator
of $S$ and $\bar S$ yields $K$, we only need to check that $S$ and $\bar S$
vanish. Consider $S$ first:
\begin{align*}
0 = -4 S_\beta W_\alpha = S_\beta \BCCD^2 \CCD_\alpha V = \BCCD^2 S_\beta \CCD_\alpha V
	= \BCCD^2 \left(2D \eps_{\beta \alpha} - 2 M_{\beta \alpha} - 3i A \eps_{\beta\alpha} \right)V 
\end{align*}
Since $V$ is real, its chiral weight vanishes. Furthermore, it is a scalar
so $M$ annihilates it. We are left with the condition $D V = 0$, that is $V$
must have conformal dimension zero. This forces $W_\alpha$ to have
conformal dimension 3/2, which is sensible since it must
possess the gaugino as its lowest component. The check that $\bar S^{\dbeta}$
also annihilates $W_\alpha$ is straightforward; no further constraints are
required. It therefore follows that $W_\alpha$ is conformally primary precisely
when $V$ has conformal dimension zero.

\subsubsection{Primary F-term superfields}
The last special case we will discuss is where $V$ is a superfield
and we demand that its chiral projection $U = \CP[V]$ is primary. (This is
of interest since if $V$ is a D-term then $U$ is the corresponding F-term.)
Generalizing the flat space result gives $U = -\frac{1}{4} \BCCD^2 V$
(which we will show is the case later). We assume that
$V$ is primary with conformal weight $\Delta$ and chiral weight $w$. Then
the primariness of $U$ reduces to checking that $\bar S$ annihilates $U$,
since it is clear that $S$ annihilates $U$. This is a simple exercise:
\begin{align*}
-4 \bar S^\dbeta U &= -\{\bar S^\dbeta, \BCCD^\dalpha \} \BCCD_\dalpha V
				- \BCCD_{\dalpha} \{\bar S^\dbeta, \BCCD^\dalpha \}V \\
	&= - \left( 2 D \eps^{\dbeta \dalpha} - 2 M^{\dbeta \dalpha} + 3i A \eps^{\dbeta \dalpha}\right) \BCCD_\dalpha V
		 - \BCCD_\dalpha \left(2 D \eps^{\dbeta \dalpha} - 2 M^{\dbeta \dalpha} + 3i A \eps^{\dbeta \dalpha}\right) V \\
	&= (8 - 4 \Delta + 6 w) \BCCD^\dbeta V
\end{align*}
It follows that $2\Delta - 3 w = 4$ is the condition on $V$
so that $U$ is primary. If we also require that the antichiral projection
of $V$ be primary, then we would find $2\Delta + 3w = 4$, concluding
that $w=0$ and $\Delta = 2$. This latter case is most important
since we will see if $V$ is a D-term action, then $U$ is the
F-term action equivalent to $V$.

\subsection{Local superconformal symmetry}
A space of local symmetries includes a rule for parallel transport,
which allows one to compare group elements at neighboring points. One demands
that the supertranslation generators $P_A$ act as parallel transport
operators with the supervierbein ${E_M}^A$ as their corresponding gauge field.
For each of the other generators $X_A$, one also introduces a gauge field  ${W_M}^A$:
\begin{align}
{W_M}^A X_A = {E_M}^A P_A + \frac{1}{2} {\phi_M}^{ba} M_{ab} + B_M D + A_M A + {f_M}^A K_A
\end{align}
In practice, it is useful to decompose the algebra into the generators of
parallel transport and the other generators, which annihilate pure functions
(ie. scalar primary fields with vanishing chiral and scaling weights).
We denote the latter group as $\mathcal H$, its generators by $X_{\ul a}$, and
its gauge fields by ${h_M}^{\ul a}$. In this manner, the total gauge connection
is simply
\begin{align}
{W_M}^A X_A = {E_M}^A P_A + {h_M}^{\ul a} X_{\ul a}
\end{align}
The action of the generators on fields is defined by
\begin{align}
\delta_G(\xi^M {W_M}^A X_A) \Phi \equiv \Lie{\xi} \Phi.
\end{align}
(See Appendix \ref{local_g} for a deeper discussion of this.)
For fields lacking Einstein indices, this reduces to the simpler
\begin{align}
\xi^M {W_M}^A X_A \Phi = \xi^M \partial_M \Phi
\end{align}
Since the action of the non-translation generators is defined
already, this defines the action of $P_A$. One finds the
standard definition of the covariant derivative
\begin{align}
\xi^A P_A \Phi = \xi^M \CCD_M \Phi = \xi^M \left(\partial_M - {h_M}^{\ul a} X_{\ul a}\right) \Phi 
\end{align}
If $\Phi$ possesses an Einstein index, then an additional
transformation rule for that index is required. For example, on
the vierbien,
\begin{align}
\delta_P(\xi) {E_M}^A = \xi^N \CCD_N {E_M}^A + \partial_M \xi^N {E_N}^A;
\end{align}
this rule can be used to define $\delta_P$ on any other Einstein
tensor.

The algebraic structure of conformal superspace is identical to the flat case
except for the introduction of curvatures to the $[P,P]$ commutator.
We include here the results quoted in the most supersymmetric language:\footnotemark
\begin{align}
[P_A, P_B] &= - {T_{A B}}^C P_C - \frac{1}{2}{R_{AB}}^{dc} M_{cd} - H_{AB} D - F_{AB} A - {R(K)_{AB}}^C K_C\eol{}
[M_{ab}, M_{cd}] &= \eta_{bc} M_{ad} - \eta_{ac} M_{bd} - \eta_{bd} M_{ac} + \eta_{ad} M_{bc} \eol{} 
[D, P_A] &= + \lambda(A) P_A \eol{}
[A, P_A] &= -i \omega(A) P_A \eol{}
[K_A, K_B] &= + {C_{AB}}^C K_C \eol{}
[D, K_A] &= - \lambda(A) K_A \eol{}
[A, K_A] &= +i \omega(A) K_A \eol{}
[K_A, P_B] &= +2 \tilde \eta_{AB} D - 2 M_{AB} + 3 i A \eta_{AB} \omega(A)
             - \frac{1}{2} K^C C_{C B A}  - \frac{1}{2} P^C C_{C A B}
\end{align}
\footnotetext{We have adopted the notation ${R(K)_{AB}}^C$ for the special conformal
curvature. One could similarly write ${R_{AB}}^{dc}$ as ${R(M)_{AB}}^{dc}$ but
we have chosen to use the conventional name for the Lorentz curvature.}


\subsection{Invariant superconformal actions}
Superspace actions fall into two types. The first is the D-type
Lagrangian, involving an integration over the full Grassmannian
manifold. The local action reads
\begin{align}
S_D = \int d^4x \, e \, \Lag_D = \int d^4x\, d^4\theta \; E \; V
\end{align}
Here $e=\det({e_m}^a)$ is the normal four dimensional measure factor,
while $E = \det({E_M}^A)$ is the appropriate generalization for
a D-term.\footnote{This determinant becomes a
super-determinant when the implicit grading is taken into account}
(Setting $E=e=1$ retrieves the global action.) Invariance requires
$X_{\ul b} V = -(-)^A {f_{\ul b A}}^A V$, which amounts to
\begin{gather*}
D \,V = 2 V, \;\;\; A \,V = 0, \;\;\; M_{ab} V = 0, \\
K_a V = 0, \;\;\; S_\alpha V = 0, \;\;\; \bar S^\dalpha V = 0
\end{gather*}
$V$ must have scaling dimension two; its chiral weight must vanish;
it must be a Lorentz scalar; it must be superconformally primary.
One should also in general check the action of $P_a$, $Q_\alpha$,
and $\bar Q^\dalpha$ to ensure that it is translation invariant
and supersymmetric. Each of these gives a set of derivative operations;
since the entire space is integrated over, each of these vanishes.
A review of this material is presented in Appendix \ref{actions_m}.

The second Lagrangian of concern is the F-type, which involves an integration
over the chiral submanifold $\chM$ corresponding to $\btheta=0$ (or to
any other constant $\btheta$ slice):
\begin{align}
S_F = \int d^4x \;e \;\Lag_F = \int d^4x \,d^2\theta \,\chE \,W
\end{align}
(We will for brevity's sake write only the chiral part; but in physical theories
one must of course include the antichiral part.)
The chiral measure $\chE$ is to be understood as
$\det ({E_{\chm}}^\chA)$ where $\chm$ is the Einstein index over
the chiral coordinates $y$ and $\theta$ and $\chA = (a, \alpha)$.
This is a loose definition since the chiral $y$ and $\theta$ need
to be better defined. Appendix \ref{actions_sm} contains a discussion
of this.

For this action to be invariant under the non-translation part of the gauge group,
$W$ must obey
\begin{gather*}
D\, W = 3 W, \;\;\; A\, W = 2i W, \;\;\; M_{ab} W = 0 \\
K_a W = 0, \;\;\; S_\alpha W = 0, \;\;\; \bar S^\dalpha W = 0
\end{gather*}
For invariance under $P$, $Q$, and $\bar Q$, $W$ must be chiral,
$\CCD_{\dalpha} W=0$. In addition, consistency of this result
(ie. $\{\CCD_\dalpha, \CCD_\dbeta\}W = 0$) leads to
the following conditions on torsions and curvatures:
\begin{gather}
{T_{\dalpha \dbeta}}^c = {T_{\dalpha \dbeta}}^{\gamma} = 0, \;\;\;
H_{\dalpha \dbeta} + \frac{2i}{3} F_{\dalpha \dbeta} = 0
\end{gather}
These constraints are invariant under the action of $\mathcal H$, as is
expected, and should be understood as the \emph{minimal} set of constraints
for a conformal superspace.


\subsection{Constraints}
Since every superfield contains sixteen independent components, the
number of degrees of freedom represented by unconstrained gauge fields
is staggering. The vierbein ${E_M}^A$ alone consists of $64$ superfields,
each possessing $16$ independent components for a total of $1024$ component
fields. This problem can be circumvented by the imposition of certain constraints in
superspace, followed by solving the Bianchi identities subject to these constraints.
Conformal superspace must reduce to a Poincar\'e superspace upon the breaking of the
conformal symmetry, so the constraints on its geometry ought to be
more severe than those normally imposed. We will guess the constraints
necessary by identifying the properties we would like to have. If this
overconstrains the theory, so be it; the Bianchi identities will
tell us if this occurs.

Perhaps the most fundamental constraint is the existence of chiral
primary superfields, the absence of which would render any superspace theory
practically useless. The chiral requirement, $\CCD_\dalpha \Phi = 0$,
implies that the anticommutator $\{\CCD_\dalpha, \CCD_\dbeta\} \Phi$ vanishes.
(We have suppressed any Lorentz indices which $\Phi$ may possess.)
This commutator in turn gives the following constraints:
\begin{align}
{T_{\dalpha \dbeta}}^c = {T_{\dalpha \dbeta}}^\gamma = 0, \;\;\;
H_{\dalpha \dbeta} + \frac{2i}{3} F_{\dalpha \dbeta} = 0, \;\;\;
{R_{\dalpha \dbeta}}^{\gamma \delta} = 0
\end{align}
(The complex conjugates are implied for the existence of anti-chiral
superfields.) All of these conditions except the last we already
knew; the last is required if chiral superfields with undotted
spinor indices (such as $W_\alpha$ and $W_{\alpha \beta \gamma}$)
should exist.

If we consider the component level behavior, more 
constraints may be deduced. The component conformal supergravity
multiplet for a chiral matter scalar, $\phi$, possesses the same
transformation laws as in flat supersymmetry, only with the
regular derivative replaced by a covariant one: 
\beq
\delta_Q \phi = \sqrt 2 \xi \psi, \;\;\;
\delta_Q \psi = \sqrt 2 \xi F + i \sqrt 2 \sigma^a\bar\xi \CCD_a \phi, \;\;\;
\delta_Q F = i \sqrt 2 (\bar \xi \bsigma^a \CCD_a \psi)
\eeq
These equations can be interpreted as superspace formulae
with the superfields $\psi_\alpha \equiv \frac{1}{\sqrt 2} \CCD_\alpha \phi$
and $F \equiv -\frac{1}{4} \CCD^2 \phi$, and the formal definition of
$\delta_Q \equiv \xi^\alpha \CCD_\alpha + \bar \xi_\dalpha \CCD^\dalpha$. 
Requiring that this variation $\delta_Q$ act on each of the superfields
as indicated by the component transformation rules leads to a
number of further constraints on the superspace curvatures:
\beq
{T_{\alpha \beta}}^\gamma = {T_{\alpha \dbeta}}^{\ul \gamma} = {T_{\ul \alpha b}}^c = 0, \;\;\;
{T_{\alpha \dbeta}}^c = 2i \sigma^c_{\alpha \dbeta}
\eeq
Other more complicated conditions are also implied, but they end up
being satisfied automatically by the Bianchi identities, so we do not
bother listing them here in detail.

We can further restrict the superspace structure by requiring the component
transformation laws for the gravitino, $U(1)$ gauge field, and scaling
gauge field to behave as in component conformal supergravity.
For example, the gravitino ought to transform under supersymmetry into a covariant
derivative of the supersymmetry parameter, $\delta_Q \psi_m = 2\CCD_m \xi$, without
the need for any explicit auxiliary fields as in \eqref{psiQ}. Since we already know the
transformation law for the gravitino is
\beq
\delta_Q {E_m}^\alpha = \CCD_m \xi^\alpha
	+ {E_m}^c \xi^\beta {T_{\beta c}}^\alpha
	+ {E_m}^c \xi_\dbeta {{T^\dbeta}_c}^\alpha
\eeq 
we are left to conclude ${T_{\ul \beta c}}^{\alpha} = 0$.
(These are the torsion components which in the minimal multiplet give the
superfields $R$ and $G_c$ whose lowest components are the supergravity
auxiliaries $M$ and $b_m$.) A similar analysis using the $U(1)$ and
scaling gauge fields using \eqref{bQ} and \eqref{aQ}
tells us $F_{\beta c} = H_{\beta c} = 0$.

One can continue in this manner to bootstrap constraints which seem
reasonable. The ones discussed above are nearly sufficient to
completely determine a minimal superspace formulation of conformal
supergravity. It turns out only one additional constraint is needed:
${R(K)_{\alpha \beta}}^C=0$ and its conjugate.

We summarize here the constraints we take. For torsion we have
\begin{gather}
{T_{\gamma \beta}}^A = {T_{\dgamma \dbeta}}^A = 0 \eol
{T_{\gamma \dbeta}}^a = 2i \sigma_{\gamma \dbeta}^a \eol
{T_{c \beta}}^A = {T_{c \dbeta}}^A = 0 \eol
{T_{cb}}^a = 0
\end{gather}
These define all torsion except for ${T_{cb}}^\alpha$ and ${T_{cb}}^\dalpha$
which remain undetermined. For the Lorentz curvature, we have
\begin{gather}
{R_{\alpha \beta}}^{cd} = {R_{\alpha \dbeta}}^{cd} = {R_{\dalpha \dbeta}}^{cd} = 0
\end{gather}
For the chiral curvature,
\begin{gather}
F_{\alpha \beta} = F_{\alpha \beta} = F_{\dalpha \dbeta} = 0 \eol
F_{\alpha b} = F_{\dalpha b} = 0  
\end{gather}
Similarly for the scaling curvature:
\begin{gather}
H_{\alpha \beta} = H_{\alpha \beta} = H_{\dalpha \dbeta} = 0 \eol
H_{\alpha b} = H_{\dalpha b} = 0  
\end{gather}
For the special conformal curvature, we take
\begin{gather}
{R(K)_{\alpha \beta}}^C = {R(K)_{\dalpha \dbeta}}^C = {R(K)_{\alpha \dbeta}}^C = 0 
\end{gather}

This set of conditions for the curvatures is especially interesting for
one particular reason: it includes the condition $R_{\ul \alpha \ul \beta} = 0$
for all curvatures \emph{except} for torsion, where we choose the flat
result ${T_{\alpha \dbeta}}^c = 2i \sigma_{\alpha \dbeta}^c$. This is
consistent with making the following demands on the fermionic covariant derivatives:\footnotemark
\begin{gather}
\{\CCD_\alpha, \CCD_\beta\} = \{\CCD_\dalpha, \CCD_\dbeta\} = 0 \\{}
\{\CCD_\alpha, \CCD_\dbeta\} = -2i \CCD_{\alpha \dbeta}
\end{gather}
The first of these implies the existence of a gauge choice where
$\CCD_\alpha = \partial_\alpha$ and the second implies the conjugate;
the third implies that no gauge exists where \emph{both} these
conditions hold.
Moreover, in flat supersymmetry, the chiral projector $\CP$ is proportional to
$\bar D^2$. The condition that it should be $\BCCD^2$ in conformal
supergravity is equivalent to the constraints
$\{\CCD_\alpha, \CCD_\beta\} = \{\CCD_\dalpha, \CCD_\dbeta\} = 0$.

\footnotetext{These conditions alone are probably sufficient to define a conformal superspace
with dynamical spin connection and torsion as well as their superpartners;
we conjecture that the extra constraints are to eliminate the spin connection and its associated
multiplet but as yet are unaware of any direct evidence for this.}

These constraints may at first glance seem exceedingly restrictive,
certainly more so than those assumed in deriving Poincar\'e supergravity.
It helps to recall that each of these objects, the torsion and the other
curvatures, are internally more complicated than their non-conformal
cousins due to the presence of the extra gauge fields.
We will find that it is these fields, in particular those
associated with the special conformal generators, which 
allow us to reconstruct normal Poincar\'e supergravity with its relaxed
constraints after gauge fixing.

\subsection{Jacobi and Bianchi identities}\label{curv_solns}
The discussion of the Jacobi and Bianchi identities in an arbitrary
theory is given in \ref{jac_bi} and merely needs to be specialized here.
The Jacobi identity serves to define the gauge transformation
properties of the curvatures:
\begin{align}
D\; {T_{CB}}^A &= \left(\Delta(C) + \Delta(B) - \Delta(A)\right) {T_{CB}}^A \eol
D\; {R(K)_{CB}}^A &= \left(\Delta(C) + \Delta(B) + \Delta(A)\right) {R(K)_{CB}}^A \eol
D\; {R_{DC}}^{BA} &= \left(\Delta(D) + \Delta(C)\right) {R_{DC}}^{BA} \eol
D\; F_{BA} &= \left(\Delta(B) + \Delta(A)\right) F_{BA} \eol
D\; H_{BA} &= \left(\Delta(B) + \Delta(A)\right) H_{BA}
\end{align}
(With the exception of the $K$-curvature, these are entirely straightforward.)
The $U(1)$ transformations are similarly simple:
\begin{align}
A\; {T_{CB}}^A &= -i \left(w(C) + w(B) - w(A)\right) {T_{CB}}^A \eol
A\; {R(K)_{CB}}^A &= -i \left(w(C) + w(B) + w(A)\right) {R(K)_{CB}}^A \eol
A\; {R_{DC}}^{BA} &= -i\left(w(D) + w(C)\right) {R_{DC}}^{BA} \eol
A\; F_{BA} &= -i\left(w(B) + w(A)\right) F_{BA} \eol
A\; H_{BA} &= -i\left(w(B) + w(A)\right) H_{BA}
\end{align}
The transformations under $K_A$ are, however, less than obvious:\footnotemark
\begin{align}
K_D\;\; {T_{CB}}^A &= \frac{1}{2} \Delta {T_{CB}}^F {C^A}_{D F}
				+ \frac{1}{2} {C^F}_{D [C} \Delta {T_{B] F}}^A \eol
K_D\;\; H_{CB} &= -(-)^D 2 \Delta T_{CBD} + \frac{1}{2} {C^F}_{D[C} H_{B] F} \eol
K_D\;\; F_{CB} &= -3i w(D) \Delta T_{CBD} + \frac{1}{2} {C^F}_{D[C} F_{B] F} \eol
K_D\;\; {R(K)_{CB}}^A &= {R(K)_{CB}}^F {C_{FD}}^A + \frac{1}{2} {C^F}_{D[C} {R(K)_{B]F}}^A
	- \frac{1}{2} \Delta {T_{CB}}^F {{C_F}^A}_D \eol
	& - \lambda(D) H_{CB} {\delta_D}^A
	+ i w(D) F_{CB} {\delta_D}^A
	+ {R_{CBD}}^A \eol
\frac{1}{2} \left(K_D\;\; {R_{CB}}^{a' a}\right) M_{a a'} &= 2 \Delta {T_{CB}}^A M_{AD} 
	- \frac{1}{4} {C^F}_{D[C} {R_{B]F}}^{a' a} M_{a a'}
\end{align}
The notation $[C B]$ in the above denotes graded antisymmetrization of the respective
indices. The rule for the Lorentz curvature has been left in a form with the explicit
Lorentz generators since they are not independent of each other.
\footnotetext{Note that gradings arising from the order of the indices
have been left off for simplicity of notation.
To replace them, note the order of the indices on the left
side of the equation and add appropriate gradings to arrive at the same order.
Also, contracted indices must be placed next to each other with the raised
index on the left.
For example, in the first line, the order of indices on the left is $DCBA$.
If we replace the gradings, we would have
$K_D\; {T_{CB}}^A = \frac{1}{2} \Delta {T_{CB}}^F {C^A}_{D F} (-)^{AF + D(A+F+B+C)}
				+ \frac{1}{2} {C^F}_{D C} \Delta {T_{B F}}^A (-)^{F(D+C+B)}$.
}
Since $K_A$ is in a sense the inverse of $P_A$, these rules
are like inverted Bianchi identities, and they
provide a nontrivial check of consistency once the curvatures
are specified.

We do not list explicitly the Lorentz transformation rules for
the curvatures since each transforms as its indices indicate.

Invariance under parallel transports is equivalent to checking the Bianchi
identities. These are significantly more complicated:
\begin{gather}
0 = \sum_{[DCB]} \CCD_D F_{CB} + {T_{DC}}^F F_{FB} -3i R(K)_{DCB} w(B) \eol
0 = \sum_{[DCB]} \CCD_D H_{CB} + {T_{DC}}^F H_{FB} -2 R(K)_{DCB} (-)^B \eol
0 = \sum_{[DCB]} \CCD_D {T_{CB}}^A + {T_{DC}}^F {T_{FB}}^A
	- {R_{DCB}}^A + \lambda(A) H_{DC} {\delta_B}^A 
	+ i w(A) F_{DC} {\delta_B}^A
	- \frac{1}{2} {R(K)_{DC}}^F {{C_F}^A}_B \eol
0 = \sum_{[DCB]} \CCD_D R(K)_{CBA} + {T_{DC}}^F {R(K)_{FBA}}
	-\frac{1}{2} {R(K)_{DC}}^F C_{BAF} \eol
0 = \sum_{[FDC]} \CCD_F R_{DC\beta\alpha} + {T_{FD}}^H R_{HC\beta\alpha}
	-\frac{1}{2} {R(K)_{FD}}_{\{\beta \dphi} {(\bsigma_c\eps)^{\dphi}}_{\alpha\}}
	+ 2 R(K)_{FD\{\beta} \,\eps_{\alpha\} \delta}
\end{gather}
The sum over $[DCB]$ denotes a sum over graded cyclic permutations of those
indices. Also, the notation $\{ \}$ on indices denotes symmetrization;
for example, $X_{\{\alpha} Y_{\beta \}} \equiv X_\alpha Y_\beta + Y_\beta X_\alpha$.
(The last identity involving the Lorentz curvature has been projected
to the left-handed part of the Lorentz group. The right-handed part
is found by complex conjugation.)

As in \cite{wb} the constraints we have chosen restrict our gauge potentials;
we must ensure that the Bianchi identities are satisfied in order for these
constraints to be valid. Though our constraints are stronger than in
$\cite{wb}$, our curvatures and Bianchi identities are more numerous.
We avoid recounting the derivation in detail here (see Appendix
\ref{app_Bianchi} for that) and merely quote the result: every curvature either
vanishes or is expressed in terms of
a single chiral superfield $W_{\alpha \beta \gamma}$. It obeys
\beq
D \, W_{\alpha \beta \gamma} = \frac{3}{2} W_{\alpha \beta \gamma}, \;\;\;
A \, W_{\alpha \beta \gamma} = i W_{\alpha \beta \gamma}, \;\;\;
K_A W_{\alpha \beta \gamma} = 0
\eeq
That is, $W_{\alpha \beta \gamma}$ possesses the same scaling and
$U(1)$ weights as it does in Poincar\'e supergravity and is
conformally primary. Furthermore, it is constrained
by its own Bianchi identity
\beq
{\CCD^\gamma}_{\dbeta} \CCD^\phi W_{\phi \gamma \beta}
	= -{\CCD_\beta}^{\dgamma} \CCD^\dphi W_{\dphi \dgamma \dbeta}
\eeq

The results for the curvatures follow below.

\subsubsection{Torsion}
The conformal torsion two-form is defined in terms of the gauge connections:
\beq
T^A = dE^A + \lambda(A) E^A B - i w(A) E^A A + E^B {\phi_B}^A - \frac{1}{2} C^{ACB} E_B f_C
\eeq
We group the various components in terms of their scaling dimension.
\begin{itemize}
\item {Dimension $0$ }
\begin{gather}
{T_{\gamma \beta}}^a =0, \;\;\; {T^{\dgamma \dbeta a}} = 0 \\
{T_{\gamma}}^{\dbeta a} =-2i {(\sigma^a \eps)_{\gamma}}^{\dbeta}
\end{gather}
\item{Dimension $1/2$ }
\begin{gather}
{T_{\ul {\gamma \beta}}}^{\ul \alpha} = 0, \;\;\;
{T_{\ul {\gamma} b}}^{a} = 0
\end{gather}
\item{Dimension $1$}
\begin{gather}
{T_{\gamma b}}^{\alpha} = 0, \;\;\;
{T^{\dgamma}}_{b \dalpha} = 0 \\
T_{\gamma b \dalpha} = 0, \;\;\;
{{T^{\dgamma}}_b}^{\alpha} = 0 \\
{T_{cb}}^a = 0
\end{gather}
\item{Dimension $3/2$}
\begin{gather}
{T_{cb}}^\alpha \ \  \leadsto \ \
	T_{(\gamma \dgamma) (\beta \dbeta) \alpha} = + 2 \epsilon_{\dgamma \dbeta} W_{\gamma \beta \alpha} \\
{T_{cb}}_\dalpha \ \  \leadsto \ \
	T_{(\gamma \dgamma) (\beta \dbeta) \dalpha} = - 2 \epsilon_{\gamma \beta} \bar W_{\dgamma \dbeta \dalpha} 
\end{gather}
\end{itemize}

\subsubsection{Lorentz curvature}
The conformal Lorentz curvature two-form is
\beq
R^{ba} = d\phi^{ba} + \phi^{bc} {\phi_c}^a - 2 E^{[b} f^{a]}
	- 4 E^{\beta} f^{\alpha} (\sigma^{ba} \eps)_{\alpha\beta}
	- 4 E_{\dbeta} f_{\dalpha} (\bsigma^{ba} \eps)^{\dalpha\dbeta}
\eeq
The notation $[b ... a]$ denotes antisymmetrization of those indices;
for example, $X_{[b} Y_{a]} \equiv X_b Y_a - X_a Y_b$.

Because the form is valued in the Lorentz group, it may be canonically
decomposed:
\beq
{R_{DC}}^{ba} \ \ \leadsto \ \ R_{DC (\beta\dbeta) (\alpha \dalpha)} =
2 \eps_{\dbeta \dalpha} R_{DC \beta \alpha}
- 2 \eps_{\beta \alpha} R_{DC \dbeta \dalpha} 
\eeq
It is simplest to express the curvature results in terms of these components.
\begin{itemize}
\item {Dimension 1}
\begin{gather}
R_{\delta \gamma \, \beta \alpha} = 0, \;\;\;
R_{\delta \gamma \, \dbeta \dalpha} = 0 \\
R_{\dot \delta \dgamma \, \beta \alpha} = 0, \;\;\;
R_{\dot \delta \dgamma \, \dbeta \dalpha} = 0 \\
R_{\delta \dgamma \, \beta \alpha} = 0, \;\;\;
R_{\delta \dgamma \, \dbeta \dalpha} = 0
\end{gather}
\item {Dimension 3/2}
\begin{gather}
{R_{\delta (\gamma \dgamma)}}_{\beta \alpha} = 0, \;\;\;
{R_{\delta (\gamma \dgamma)}}_{\dbeta \dalpha} = +4i \eps_{\delta \gamma} \bar W_{\dgamma \dbeta \dalpha} \\
{R_{\dot \delta (\gamma \dgamma)}}_{\dbeta \dalpha} = 0, \;\;\;
{R_{\dot \delta (\gamma \dgamma)}}_{\beta \alpha} = -4i \eps_{\dot \delta \dgamma} W_{\gamma \beta \alpha}
\end{gather}
\item{Dimension 2}
\begin{align}
R_{(\delta \dot\delta)(\gamma \dgamma)\beta \alpha} &=
	+ 2 \eps_{\ddelta \dgamma} \chi_{\lsym{\delta\gamma\beta\alpha}}
	- \frac{1}{4} \eps_{\ddelta \dgamma}
		\sum_{(\delta \gamma)} \sum_{(\beta \alpha)} \eps_{\delta \beta} \CCD^\phi W_{\phi \gamma \alpha} \eol
	&= + \eps_{\ddelta \dgamma} \nabla_{\{\beta} W_{\alpha\} \delta \gamma}  \\
R_{(\delta \dot\delta)(\gamma \dgamma)\dbeta \dalpha} &=
	- 2 \eps_{\delta \gamma} \chi_{\lsym{\ddelta\dgamma\dbeta\dalpha}}
	+ \frac{1}{4} \eps_{\delta \gamma} 
		\sum_{(\ddelta \dgamma)} \sum_{(\dbeta \dalpha)} \eps_{\ddelta \dbeta} \CCD^\dphi \bar W_{\dphi \dgamma \dalpha} \eol
	&= -\eps_{\delta \gamma} \nabla_{\{\dbeta} W_{\dalpha\} \ddelta \dgamma}
\end{align}
The totally symmetric symbol $\chi$ is itself the spinorial curl of the superfield $W$:
\begin{align}
{\chi}_{\lsym{\delta \gamma \beta \alpha}} &=
	\frac{1}{4} (\CCD_\delta W_{\gamma\beta\alpha} + \CCD_\gamma W_{\delta\beta\alpha} +
				\CCD_\beta W_{\gamma\delta\alpha} + \CCD_\alpha W_{\gamma\beta\delta})\\
{\chi}_{\lsym{\ddelta \dgamma \dbeta \dalpha}} &=
	\frac{1}{4} (\CCD_\ddelta \bar W_{\dgamma\dbeta\dalpha} + \CCD_\dgamma \bar W_{\ddelta\dbeta\dalpha} +
				\CCD_\dbeta \bar W_{\dgamma\ddelta\dalpha} + \CCD_\dalpha \bar W_{\dgamma\dbeta\ddelta})
\end{align}
\end{itemize}

\subsubsection{Scaling and $U(1)$ curvatures}
The conformal field strengths for scalings and chiral rotations are
\begin{align}
H = dB + 2 E^A F_A (-)^a \\
F = dA + 3i E^A F_A w(A)
\end{align}
\begin{itemize}
\item {Dimension 1}
\begin{gather}
H_{\delta \gamma} = F_{\delta\gamma} = 0 \\
H_{\delta \dgamma} = F_{\delta \dgamma} = 0 \\
H_{\dot \delta \dgamma} = F_{\dot\delta \dgamma} = 0
\end{gather}
\item {Dimension 3/2}
\begin{gather}
H_{\mu (\gamma \dgamma)} = F_{\mu (\gamma \dgamma)} = 0 \\
H_{\dmu (\gamma \dgamma)} = F_{\dmu (\gamma \dgamma)} = 0
\end{gather}
\item{Dimension 2}
\begin{align}
H_{cb} \ \ \leadsto \ \ H_{(\gamma \dgamma)(\beta \dbeta)} &=
    2 \eps_{\dgamma \dbeta} {\tH}_{\gamma \beta} - 2 \eps_{\gamma \beta} \tH_{\dgamma \dbeta} \\
F_{cb} \ \ \leadsto \ \ F_{(\gamma \dgamma)(\beta \dbeta)} &=
    2 \eps_{\dgamma \dbeta} {\tF}_{\gamma \beta} - 2 \eps_{\gamma \beta} \tF_{\dgamma \dbeta}
\end{align}
The components $\tH$ and $\tF$ are themselves related to the spinorial divergence of
the superfield $W$:
\begin{align}
\CCD^\gamma W_{\gamma \beta \alpha} &= \frac{4i}{3} \tF_{\beta \alpha} = +2 \tH_{\beta \alpha} \\
\CCD^\dgamma W_{\dgamma \dbeta \dalpha} &= \frac{4i}{3} \tF_{\dbeta \dalpha} = -2 \tH_{\dbeta \dalpha}
\end{align}
\end{itemize}

\subsubsection{Special conformal curvature}
The special conformal curvatures are
\[
R(K)^a = d f^A - \lambda(A) f^A B + i w(A) f^A A + f^B {\phi_B}^A
	+ \frac{1}{2} C^{ACB} f_C E_B
	+ \frac{1}{2} f^B f^C {C_{CB}}^A
\]

We will group them by their form indices.
\begin{itemize}
\item {Fermion/fermion}
\begin{gather}
R(K)_{\gamma \beta \alpha} = 0, \;\;\;
R(K)_{\gamma \dbeta \alpha} = 0, \;\;\;
R(K)_{\dgamma \dbeta \alpha} = 0 \\
R(K)_{\gamma \beta \dalpha} = 0, \;\;\;
R(K)_{\gamma \dbeta \dalpha} = 0, \;\;\;
R(K)_{\dgamma \dbeta \dalpha} = 0 \\
R(K)_{\gamma \beta a} = 0, \;\;\;
R(K)_{\gamma \dbeta a} = 0, \;\;\;
R(K)_{\dgamma \dbeta a} = 0
\end{gather}
\item {Fermion/boson}
\begin{gather}
R(K)_{\alpha (\beta \dbeta) \gamma} = 0, \;\;\;
R(K)_{\dalpha (\beta \dbeta) \dgamma} = 0 \\
R(K)_{\alpha (\beta \dbeta) \dgamma} = +i \eps_{\alpha \beta} \CCD^{\dot \phi} W_{\dot \phi \dbeta \dgamma},\;\;\;
R(K)_{\dalpha (\beta \dbeta) \gamma} = +i \eps_{\dalpha \dbeta} \CCD^{\phi} W_{\phi \beta \gamma} \\
R(K)_{\alpha (\beta \dbeta) (\gamma \dgamma)} =
    -2i \eps_{\alpha \beta} \CCD_{\gamma \dot\phi} {W^{\dot\phi}}_{\dbeta \dgamma}, \;\;\;
R(K)_{\dalpha (\beta \dbeta) (\gamma \dgamma)} =
    -2i \eps_{\dalpha \dbeta} \CCD_{\phi \dgamma} {W^{\phi}}_{\beta \gamma}
\end{gather}
\item{Boson/boson}
\begin{gather}
R(K)_{cb \mu} = -\frac{i}{3} \CCD_\mu F_{cb}, \;\;\;
R(K)_{cb \dmu} = +\frac{i}{3} \CCD_\dmu F_{cb} \\
R(K)_{(\gamma \dgamma)(\beta \dbeta) (\alpha\dalpha)} =
	-\eps_{\gamma \beta} \BCCD_\dgamma {\CCD_{\alpha}}^\dphi W_{\dphi \dbeta \dalpha}
	-\eps_{\dgamma \dbeta} \CCD_\gamma {\CCD^\phi}_{\dalpha} W_{\phi \beta \alpha}
\end{gather}
where the chiral curvature $F_{cb}$ has been used for notational simplicity.
\end{itemize}

%
\subsection{Chiral projectors and component actions}
One can use the details of Appendix \ref{actions_sm}, specifically equation \eqref{cp}
to construct an explicit form for the chiral projector in conformal superspace:
\beq
\CP[V] = \int d^2 \btheta \, \bar\Sigma \, V
\eeq
where $\bar\Sigma$ is the superdeterminant constructed out
of ${E^{\dmu}}_\dalpha$ in the gauge where $E_{m \dalpha}$
and $E_{\mu \dalpha}$ vanish. Let us explicitly construct
the vierbein (and other connections) in this gauge.

Recall that the variation of the connections $W^{\dmu A}$ is
\beq
\delta_G W^{\dmu A} = \partial^\dmu g^A + W^{\dmu B} g^C {f_{CB}}^A.
\eeq
The gauge parameter $g^A$ is a superfield and so has a larger
parameter space than what survives at the component level.
In principle, every $\theta$ and $\btheta$-dependent part
of $g^A$ can be exhausted to put the connections in a
desirable form without affecting the component Lagrangian.
We will here use the $\btheta$-dependence of
$g^A$ to fix $W^{\dmu A}$ to a specific form. (This will correspond
to a chiral version of Wess-Zumino gauge. Later on we shall fix
the $\theta$-dependence.)

Let $g^A = \btheta_{\dmu} g^{\dmu A} + \frac{1}{2} \btheta^2 g_2^A$
where the functions $g^{\dmu A}$ and $g_2^A$ depend on $x$ and $\theta$
but not $\btheta$. It is immediately clear by inspection of the gauge
transformation law that $g^{\dmu A}$ can be chosen to fix the gauge
$W^{\dmu A}\vert_{\btheta=0}=\delta^{\dmu A}$, meaning the vierbein
is gauged to $\delta^{\dmu A}$ at lowest component and all other
gauge fields set to zero. The gauge connection $\btheta$-expansion
then becomes
\beq
W^{\dmu A} = \delta^{\dmu A} + \btheta_{\dnu} W^{\dnu \dmu A}
	+ \frac{1}{2} \btheta^2 W_2^{\dmu A}
\eeq
for fields $W^{\dnu \dmu A}$ and $W_2^{\dmu A}$ which depend on
only $x$ and $\theta$. The remaining gauge parameter $g_2^A$
can be used to eliminate the antisymmetric part of 
$W^{\dnu \dmu A}$, leaving $W^{\dnu \dmu A} = W^{\dmu \dnu A}$.
This exhausts our $\btheta$-dependent gauge freedom.
The curvatures then uniquely determine the remaining
bits of the connection. By taking the
definition of the curvature $R$ and projecting to
$\btheta=0$, one finds $W^{\dnu \dmu A} = \frac{1}{2} R^{\dnu \dmu A}\vert_{\btheta=0}$.
The remaining component of $W$ is determined by taking the derivative of the
curvature formula and projecting to $\btheta=0$. One finds
$
W_2^{\dmu A} = -\frac{1}{3} \CCD_{\dalpha} R^{\dalpha \dmu A} \vert_{\btheta=0}
	- \frac{1}{6} {R_{\dalpha}}^{\dmu \ul b} {f_{\ul b}}^{\dalpha A}\vert_{\btheta=0}.
$
This gives the formula
\begin{align}
W^{\dmu A} = \delta^{\dmu A} + \frac{1}{2} \btheta_{\dalpha} R^{\dalpha \dmu A}\vert_{\btheta=0}
	- \frac{1}{6} \btheta^2 \CCD_{\dalpha} R^{\dalpha \dmu A} \vert_{\btheta=0}
	- \frac{1}{12} \btheta^2{R_{\dalpha}}^{\dmu \ul b} {f_{\ul b}}^{\dalpha A}\vert_{\btheta=0}.
\end{align}
Within conformal superspace, all of the $\btheta$-dependent terms vanish, giving
\begin{align}
E^{\dmu A} = \delta^{\dmu A}, \;\;\; h^{\dmu \ul a} = 0
\end{align}
Therefore the chiral projector is simply defined as
\beq
\CP[V] = \int d^2\btheta V = -\frac{1}{4} \partial_\dmu \partial^\dmu V
	= -\frac{1}{4} \BCCD^2 V
\eeq
where the last equality follows due to the simplicity of the connections
in this gauge. Since the left and right sides of this equation transform
the same way under gauge transformations, their equality in this
special gauge implies their equality in any.

Since the result is suspiciously simple, we should check that this
approach works for minimal supergravity where the chiral projector
is known to be not so simple. There the vierbein should take the
general form
\begin{align}
E^{\dmu A} = \delta^{\dmu A}
	- \frac{1}{12} \btheta^2{R_{\dalpha}}^{\dmu \ul b} {f_{\ul b}}^{\dalpha A}\vert_{\btheta=0}
\end{align}
since the relevant torsion components vanish. The only curvature in
Poincar\'e superspace is the Lorentz curvature, and it is straightforward
to evaluate the term appearing here. One finds
\begin{align}
E^{\dmu A} = \delta^{\dmu A} - \delta^{\dmu A} \btheta^2 R
\end{align}
for the vierbein (as well as a non-vanishing spin connection
which we will ignore since it turns out not to matter).
The chiral projection formula becomes
\beq
\CP[V] = \int d^2\btheta (1 + 2\btheta^2 R) V
	= 2 R V - \frac{1}{4} \partial_\dmu \partial^\dmu V
	= -\frac{1}{4} (\BCCD^2 - 8 R) V
\eeq
Here the spin connection is not zero but it contributes nothing
when $\BCCD^2$ acts on a field without dotted indices, and so $\BCCD^2$ in this gauge is
as simple in Poincar\'e superspace as it is in conformal superspace.

In either formalism, the conversion from a D to an F-term proceeds
straightforwardly. Using \eqref{DtoF}, we find
\begin{align}
\int d^4x d^4\theta E \,V = \int d^4x d^2\theta \mathcal E \,\CP[V].
\end{align}
where the second integration is understood to occur at
$\btheta=0$. Although the operations above were performed in a specific
$\btheta$ gauge, the final results have been written in a
gauge-invariant manner. In fact, since the gauge-fixing procedure
undertaken had no effect on the fields at $\btheta=0$, the
right hand side of the above equation \emph{must} be
independent of our gauge choices.

\subsubsection{F-term integrations}
We have shown that any $D$-term can be written as an $F$-term.
It is still necessary to evaluate the component Lagrangian
corresponding to an $F$-term. A chiral integral has the form
\begin{align}
\int d^4x \, d^2\theta\, \mathcal E \,W,
\end{align}
an integral over the superspace slice where $\btheta=0$.
$W$ is a chiral superfield transforming under the gauge
group in order to leave the full action invariant.

We can evaluate this integral by the method of gauge-fixing,
much like how we derived the $D$ to $F$ integral conversion
formula. The first step is to use the $\theta$-dependent part
of the gauge transformations to fix the connections.\footnote{
It is useful to note that whether or not we gauge-fixed
the $\btheta$-dependent part of the connections is irrelevant
for evaluating an $F$-term as its integral occurs at
$\btheta=0$.} In a way entirely analogous to what we did in the previous section, we
may choose\footnotemark
\begin{align}
{W_{\mu}}^{A} = {\delta_\mu}^{A} + \frac{1}{2} \theta^{\alpha} {R_{\alpha \mu}}^A\vert_{\theta=0}
	- \frac{1}{6} \theta^2 \CCD^{\alpha} {R_{\alpha \mu}}^A \vert_{\theta=0}
	+ \frac{1}{12} \theta^2{R_{\mu}}^{\alpha \ul b} {f_{\ul b \alpha}}^{A}\vert_{\theta=0}.
\end{align}
by exhausting the remaining $\theta$-dependence of $g^A$.
Here the projection to $\btheta=0$ has also already been done, so we will
avoid indicating it explicitly.\footnotetext{This last gauge-fixing has
an interesting effect on $\theta$. Their Einstein index is now effectively
a Lorentz index, since every Lorentz rotation which would
alter the vierbein must be countered by a $P$-gauge (or coordinate)
transformation. The $\theta$'s are therefore the same as the
$\Theta$ variables of \cite{wb}. Their $F$-terms are written
$\int d^2\Theta \mathcal E$ where $\Theta$ is equivalent to $\theta$
and their $\mathcal E$ is half of ours when we go to this gauge.}

In conformal superspace, this expression is extremely simple. It gives
\begin{align}
{E_{\mu}}^{A} = {\delta_{\mu}}^{A}, \;\;\; {h_{\mu}}^{\ul a} = 0
\end{align}
The $F$-term integration then becomes
\begin{align}
\Lag_F = \int d^4x\, d^2\theta \,e\, W = 
	-\frac{1}{4} e \partial^\mu \partial_\mu W
	- \frac{1}{2} \partial^\mu e \partial_\mu W
	- \frac{1}{4} (\partial^\mu \partial_\mu e) W
\end{align}
The first term is rather simple. In our gauge choice, it is easy
to see that $\CCD^\alpha \CCD_\alpha W = \partial^\alpha \partial_\alpha W$
when $\theta=\btheta=0$. The other
terms are usually constructed in the literature from supersymmetric
completion of this term; here we will evaluate them directly in this
gauge. For example,
\beq
\partial_\mu e = e (\partial_\mu {E_m}^a) {e_a}^m
	= e (\partial_m {E_\mu}^a + {T_{\mu m}}^a) {e_a}^m
	= 0 + e {T_{\mu \dbeta}}^a {E_m}^\dbeta {e_a}^m
	= i e (\sigma^a \bar\psi_a)_\mu
\eeq
where we have used ${E_\mu}^a\vert=0$ as well as the torsion
constraint ${T_{\gamma \beta}}^a = {T_{\gamma b}}^a = 0$.
This allows us to evaluate the second term of $\Lag_F$;
we find $ie (\bar \psi_a \bsigma_a)^\alpha \CCD_\alpha W/2$
(since $\partial_\alpha W = \CCD_\alpha W$ at $\theta=\btheta=0$
in this gauge.)

The remaining third term is slightly more complicated. One begins
with
\beq\label{eq:ddE}
\partial^\mu \partial_\mu e = \partial^\mu (e {T_{\mu \dalpha}}^a {E_m}^\dalpha {e_a}^m)
\eeq
The outer spinorial derivative acts on each term in parentheses except the torsion
(which is constant). From differentiating $e$, we find the term
$e (\bar\psi_a \bsigma^a \sigma^b \bar\psi_b)$. From the inverse vierbein,
one gets $-e (\bar\psi_a \bsigma^b \sigma^a \bar\psi_b)$. From the gravitino
one finds no additional terms. This leads to
\beq
\partial^\mu \partial_\mu e 
	= 4 e (\bar\psi_a \sigma^{ab} \bar\psi_b)
\eeq
which gives the chiral Lagrangian
\beq
\Lag_F = \int d^2\theta\, \mathcal E \,W = e \left(
	-\frac{1}{4} \CCD^\alpha \CCD_\alpha W
	+ \frac{i}{2} (\bar \psi_a \bsigma_a)^\alpha \CCD_\alpha W
	- (\bar\psi_a \sigma^{ab} \bar\psi_b) W\right)
\eeq
where the projection to $\theta=\btheta=0$ is implicit.

Again, we may repeat this process for Poincar\'e superspace.
One finds
\begin{align}
{E_\mu}^{A} = {\delta_\mu}^{A} - {\delta_\mu}^{A} \theta^2 \bar R
\end{align}
and for the F-term
\begin{align}
\Lag_F = \int d^2\theta \,e\, (1+2\theta^2 \bar R) W
	= -\frac{1}{4} e \partial^\mu \partial_\mu W
	- \frac{1}{2} \partial^\mu e \partial_\mu W
	- \frac{1}{4} (\partial^\mu \partial_\mu e) W + 2 \bar R W
\end{align}
The first and second terms are evaluated as before. The
third gains an extra contribution of $-16 e \bar R$ 
from \eqref{eq:ddE} when the spinorial derivative hits the gravitino.
This gives the chiral Lagrangian
\beq
\Lag_F = \int d^2\theta\, \mathcal E \,W = \,e \left(
	-\frac{1}{4} \CCD^\alpha \CCD_\alpha W
	+ \frac{i}{2} (\bar \psi_a \bsigma_a)^\alpha \CCD_\alpha W
	- (\bar\psi_a \sigma^{ab} \bar\psi_b) W
	+ 6 \bar R W \right)
\eeq
where the projection to $\theta=\btheta=0$ is implicit.

\subsubsection{D-term integrations}
Within conformal superspace, the F-term component Lagrangian is
\begin{align}
\Lag_F = \int d^2\theta\, \mathcal E \,W =
	e \left(F
	+ \frac{i\sqrt 2}{2} (\bar \psi_a \bsigma_a \rho)
	- (\bar\psi_a \bsigma^{ab} \bar\psi_b) W \right)
\end{align}
where
\begin{gather}
F \equiv -\frac{1}{4} \CCD^2 W \vert \;\;\; \textrm{and} \;\;\;
\rho_\alpha \equiv \frac{1}{\sqrt 2} \CCD_\alpha W\vert
\end{gather}
A D-term can be divided into two terms, one evaluated
via a chiral integration and the other via an antichiral
integration in order to give a manifestly Hermitean action:
\begin{align}
\int d^4\theta \,E\, V = \frac{1}{2} \int d^2\theta \,\chE \, U
		+ \frac{1}{2} \int d^2\btheta \, \bar\chE \, \bar U
\end{align}
where $U \equiv -\frac{1}{4} \BCCD^2 V$ and $\bar U \equiv -\frac{1}{4} \CCD^2 V$
are the chiral and antichiral projections of $V$. These two
F-terms can then be evaluated using the F-term formula giving
the general D-term formula
\begin{align}
\Lag_D = \int d^4\theta \, E \, V = e \left(
	 \frac{1}{2} (F + \bar F)
	+ i \frac{\sqrt 2}{4} \left(\bar\psi_a \bsigma^a \rho + \psi_a \sigma^a \bar \rho \right)
	- \frac{1}{2} (\bar\psi_a \bsigma_{ab} \bar\psi_b) U
	- \frac{1}{2} (\psi_a \sigma_{ab} \psi_b) \bar U
	\right)
\end{align}
where
\begin{gather}
U \equiv -\frac{1}{4} \BCCD^2 V\vert, \;\;\;
F \equiv \frac{1}{16} \CCD^2 \BCCD^2 V \vert, \;\;\; \textrm{and} \;\;\;
\rho_\alpha \equiv -\frac{1}{4\sqrt 2} \CCD_\alpha \BCCD^2 V \vert
\end{gather}
The fields $F$ are actually not quite independent fields. In terms of
the D-term of $V$, they are
\begin{align}
F &= D + \frac{1}{2} \CCD_c \CCD^c V + \frac{i}{2} \CCD_c V^c \\
\bar F &= D + \frac{1}{2} \CCD_c \CCD^c V - \frac{i}{2} \CCD_c V^c
\end{align}
where\footnotemark
\begin{gather}
D \equiv \frac{1}{16} \CCD^\alpha  \BCCD^2 \CCD_\alpha V = \frac{1}{16} \BCCD_\dalpha  \CCD^2 \BCCD^\dalpha V \\
V_c \equiv -\frac{1}{2} \bsigma_c^{\dalpha \alpha} [\CCD_\alpha, \BCCD_\dalpha] V
\end{gather}
The imaginary part of the fields $F$ and $\bar F$ is
the divergence of the vector component of $V$. When evaluating a
$D$-term integral, it is occasionally useful to use the
fields $D$ rather than $F$.

\footnotetext{As in normal superspace, one must be careful to
note that $\CCD_c$ is covariant even with respect to supersymmetry.
That is,
\[
\CCD_c = {e_c}^m \left(\CCD_m - \frac{1}{2} {\psi_m}^{\ul \alpha} \CCD_{\ul \alpha}\right) 
	= {e_c}^m \left(\partial_m - \frac{1}{2} {\psi_m}^{\ul \alpha} \CCD_{\ul \alpha}
		- {h_m}^{\ul a} X_{\ul a}\right)
\]
where $\ul \alpha$ denotes both spinor indices.
In fact, were we to treat supersymmetry as a gauge theory in normal
space with internal symmetry operators $Q$ which happened to include
translations in their algebra, we would denote $\frac{1}{2} {\psi_m}^{\ul\alpha}$
as the gauge field associated with the generator $Q_{\ul \alpha}$.
Then the above formula is simply the covariant derivative.
There is a further mild complication in conformal superspace: $\CCD_c$ will include
the gauge action of $S_{\ul \alpha}$; therefore, a superconformal
covariant derivative includes not only terms higher in the
multiplet (due to $Q$), but also terms lower in the multiplet
(due to $S$).}

\subsection{K\"ahler structure of conformal superspace of chiral superfields}
It turns out that the conformal superspace of an arbitrary set of scalar
chiral superfields possesses a simple K\"ahler structure due to its relation
to the K\"ahler manifold $\mathbb{C}P^n$.

Suppose we are furnished with a set of chiral primary superfields $\Phi_I$
where $I=0,1,\ldots,n$. Our action consists in general of a D-term and an F-term which
respectively take the forms
\begin{gather}
\Lag_D = -3 \int d^4\theta \,E \,Z(\Phi_I, \bar\Phi_I),\;\;\;\;
\Lag_F = \int d^2\theta \,\chE \,P(\Phi_I)
\end{gather}
where $Z$ is some real non-negative function of the fields with $\Delta(Z)=2$
and $P$ is some chiral function with $\Delta(P)=3$ and $w(P)=2$. (The assumption of
non-negativity of $Z$ is ultimately for stability of the underlying
Einstein-Hilbert term. The factor of 3 is for convenience.) We can take the
$\Phi_i$ as parametrizing some complex manifold. In order for $Z$ to have
a nonzero scaling weight, at least one of the $\Phi_i$ must have $\Delta_i\neq 0$.
We will assume without loss of generality that this is $\Phi_0$
(by renaming the fields if necessary) and that $\Delta_0 = 1$
(by redefining $\Phi_0 \rightarrow (\Phi_0)^{1/\Delta_0}$ if necessary).

It is then possible to trade the fields $\Phi_j$ with $j\geq1$ for projective fields
$\xi_j$ which have zero weight. (The simplest way of doing this is by defining
$\xi_j \equiv \Phi_j / \Phi_0^{\Delta_j}$.) Since the fields $\xi_j$ have vanishing
scaling weight, the fields $Z$ and $P$ in this parametrization are
restricted in their form to
\begin{align}
Z = \Phi_0 \bar\Phi_0 \exp\left(-K/3\right), \;\;\;\;
P = \Phi_0^3 W
\end{align}
where $K = K (\xi_j, \bar\xi_j)$ is some real function
of the projective fields and $W = W(\xi_j)$ is some
chiral function.\footnote{Since $\Phi_0$ has scaling weight
$1$ and chiral weight $2/3$ (their ratio is fixed at $3/2$ for
any primary chiral superfield) $P$ has the correct scaling and
chiral weights for an F-term.} (The choice of this definition
for real $K$ is possible only if $Z$ is assumed to be
non-negative.) It is obvious that both $Z$ and $P$,
viewed as functions of the complex manifold spanned by the $\Phi_i$,
are independent of the projective representation chosen. A different
representation is induced on the projective coordinates by the mapping
\beq
\Phi_0 \rightarrow \Phi_0 \exp(F/3), \;\;\;
K \rightarrow K + F + \bar F, \;\;\;
W \rightarrow e^{-F} W
\eeq
where $F = F(\xi_j)$ is a holomorphic function of the projective
parameters.
(For example, trading $\Phi_0$ for $\Phi_1$ as the field to project
with is accomplished by choosing $F = 3 \log (\Phi_1 / \Phi_0) = 3 \log(\xi_1)$.)
The above transformation law is simply a K\"ahler transformation, and
the manifold under discussion is the complex projective space
$\mathbb{C}P^n$, a simple example of a K\"ahler manifold.

The two actions then take the form
\begin{gather}
\Lag_D = -3 \dint \bar \Phi_0 e^{-K/3} \Phi_0, \;\;\;\;
\Lag_F = \fint \Phi_0^3 \, W
\end{gather}
where $W$ is chiral and $K$ is real. The factor of $e^{-K/3}$ is
reminiscent of $e^V$ for a theory with an internal $U(1)_K$ symmetry;
this $U(1)_K$ is gauged not by an independent gauge multiplet but by
the other chiral fields. We may make the $U(1)_K$ more manifest in the
following manner. Define a new complex superfield $\Psi_0$ by
\begin{align}\label{temp2}
\Psi_0 \equiv e^{-K/6} \Phi_0, \;\;\;
\bar \Psi_0 \equiv e^{-K/6} \bar \Phi_0
\end{align}
under which the actions become
\begin{gather*}
-3 \dint \bar \Psi_0 \Psi_0, \;\;\;\;
\fint \Psi_0^3 \, e^{K/2} W
\end{gather*}
The new field $\Psi_0$ and effective superpotential $e^{K/2} W$ are the only
objects (besides $K$) which transform under K\"ahler transformations:
\begin{gather}
\Psi_0 \rightarrow \exp\left(+\frac{i}{3} \Imp F \right)\, \Psi_0, \;\;\;\;
\bar \Psi_0 \rightarrow \exp\left(-\frac{i}{3} \Imp F \right)\,\bar \Psi_0  \\
e^{K/2} W \rightarrow \exp\left(-i \Imp F \right)\, e^{K/2} W, \;\;\;\;
e^{K/2} \bar W \rightarrow \exp\left(+i \Imp F \right)\, e^{K/2} \bar W
\end{gather}
We normalize the generator of K\"ahler transformations, $\kg$, by
requiring the above K\"ahler transformation to correspond to
$\exp\left(-\Imp F \,\kg /2\right)$. In this way the K\"ahler weights
of $\Psi_0$ and $e^{K/2} W$ are set to be -2/3 and 2, respectively:
\[
\kg \, \Psi_0 = -i \frac{2}{3}\Psi_0, \;\;\;
\kg \, e^{K/2} W = +2i e^{K/2} W
\]
(Note that $e^{K/2} W$ is chiral from the point of view of the K\"ahler
covariant derivative, which carries a K\"ahler connection.)
This normalization is purely a matter of convention; it is chosen so that
$e^{K/2} W$ possesses the same K\"ahler and $U(1)$ weights.

The K\"ahler covariant derivative then takes the form
\begin{equation}
\nabla^K \equiv \CCD - \kA \kg
\end{equation}
where $\kg$ is the generator of the K\"ahler transformations.
The K\"ahler connection $\kA$ is defined in terms of the K\"ahler potential $K$:
\begin{gather}
\kA_\alpha = +\frac{i}{4} \CCD_\alpha K, \;\;\;
\kA_\dalpha = -\frac{i}{4} \CCD_\dalpha K \eol
\kA_{\alpha \dalpha} = \frac{i}{2} (\CCD_\alpha \kA_\dalpha + \CCD_\dalpha \kA_\alpha)
	= \frac{1}{8} [\CCD_\alpha, \CCD_\dalpha] K
\end{gather}
(In these formulae, the function $K$ is a primary scalar superfield and
is therefore invariant under all the generators of the superconformal
algebra.) The definition of $\kA_{\alpha \dalpha}$ is conventional;
it is chosen so that $\{\nabla^K_\alpha, \nabla^K_\dalpha\} = -2i \nabla^K_{\alpha \dalpha}$.

\newpage
\section{Degauging to Poincar\'e}
Poincar\'e superspace lacks the explicit scaling and conformal
symmetries enjoyed by conformal superspace. It may also,
depending on the flavor of supergravity chosen, lack the
$U(1)$ R-symmetry. Converting conformal supergravity to
one of the flavors of Poincar\'e supergravity must then involve
some measure of gauge-fixing. We will demonstrate how this
is accomplished by first reducing the conformal multiplet
to a theory with an explicit $U(1)$ symmetry and a nonlinearly
realized conformal symmetry. To guide our path, we first review
in broad strokes how it works without supersymmetry;
the details of this can be found in \cite{csg}.

\subsection{Review: Conformal gravity and the Einstein-Hilbert Lagrangian}
Conformal gravity consists of the following gauge connections:
\begin{align}
W_m = {e_m}^a P_a + \frac{1}{2} {\omega_m}^{ba} M_{ab} + b_m D + {f_m}^a K_a
\end{align}
We will denote by $\conf R$ the curvatures of the conformal theory and
by $R$ the Poincar\'e curvatures. One usually takes the constraint of vanishing
conformal torsion (which is equivalent to vanishing Poincar\'e torsion)
to determine the spin connection ${\omega_m}^{ba}$
in terms of the vierbein and the scaling gauge field $b_m$. One also would
like to express the special conformal gauge field ${f_m}^a$ in terms of
other fields; this can be done by taking the conformal Ricci tensor to
vanish, ${\conf R_{mn}}\,^{ba} {e_b}^n = 0$. Having done so, one finds
\begin{align}
{f_m}^a = -\frac{1}{4} \left({\mathcal R_{m}}^a - \frac{1}{6}{e_m}^a \mathcal R\right)
\end{align}
where ${\mathcal R_m}^a = {R_{mn}}^{ba} {e_b}^n$ is the Poincar\'e Ricci tensor
and $\mathcal R = {\mathcal R_m}^a {e_a}^m$ the Poincar\'e Ricci scalar. One further, for
simplicity, usually adopts the $K$-gauge choice $b_m=0$. (This is possible
since $\delta_K(\epsilon) b_m = -2 {e_m}^a \epsilon_a$ allows one to
gauge $b_m$ away.)

Having made these constraints and gauge choices, one then examines the
simplest conformal action for a scalar field $\phi$ with $\Delta=1$:
\begin{align}
e^{-1} \Lag = \frac{1}{2} \phi \nabla^a \nabla_a \phi
	= -\frac{1}{2} \nabla^a \phi \nabla_a \phi
		- \frac{1}{2} {T_{ba}}^a \phi \nabla^b \phi - {f_a}^a \phi^2
\end{align}
(We have integrated the covariant d'Alembertian by parts.) The torsion
term vanishes by assumption. The term involving $\nabla_a \phi$ also
vanishes if we fix the remaining $D$-gauge by gauging $\phi$ to the
constant $\phi_0$:
\[
\nabla_a \phi_0 = {e_a}^m \partial_m \phi_0 = 0
\]
(There is no scaling connection in the above expression since $b_m = 0$
has been chosen as our $K$-gauge.)
This leaves for the Lagrangian
\begin{align}
e^{-1} \Lag = \frac{1}{2} \phi_0 \nabla^a \nabla_a \phi_0
	=  -{f_a}^a \phi_0^2 = +\frac{1}{12} \mathcal R \phi_0^2
\end{align}
This is almost the Einstein-Hilbert term $-\mathcal R/2$
(in units where the reduced Planck mass is unity). We need
only start with the wrong sign for the kinetic term and then
choose $\phi_0^2 = 6$.

If we had started with a complex gauge field $\phi$, the
Lagrangian would have been
\begin{align}
e^{-1} \Lag = \phi^* \nabla^a \nabla_a \phi
	=  -\nabla^a \phi^* \nabla_a \phi - 2 {f_a}^a \abs{\phi}^2
\end{align}
We may gauge $\abs{\phi}=\phi_0$ but not the phase of $\phi$,
which we shall denote $\omega$. This gives
\begin{align}
e^{-1} \Lag = \phi^* \nabla^a \nabla_a \phi
	=  - \phi_0^2 \partial^m \omega \partial_m \omega
		+ \frac{1}{6} \mathcal R \phi_0^2
\end{align}
Gauging $\phi_0^2=3$ and choosing to flip the sign of the
Lagrangian gives back the Einstein-Hilbert term; unfortunately
this also leaves an unstable kinetic term for $\omega$.
A model with an additional gauged $U(1)$ symmetry would be able to
dispense with this phase. The superconformal algebra has such a symmetry,
and we will find it is the supersymmetric version of this
model with a complex $\phi$ which reproduces the minimal version of
Poincar\'e supergravity.

\subsection{$U(1)$ superspace}
In conformal gravity, the scaling gauge field $b_m$ was the only
field that transformed under the special conformal symmetry; moreover,
this symmetry was precisely enough to allow the choice $b_m=0$.
The latter property is also enjoyed in the superconformal case,
even though not every other field is $K$-inert. It is here that
we begin our gauge fixing procedure.

Recall that under the action of $K_A$ with parameter $\epsilon^A$,
$
\delta_K B_M = -2 \epsilon^A E_{MA}  (-)^a.
$ 
If we choose $\epsilon^A = \eta^M {E_M}^A (-)^a$, then we find
$\delta_K B_M = -2\eta_M$ and we can freely choose the gauge
$B=0$. The generator $D$ then
drops out of the covariant derivative. We also have chosen
a gauge for $K_A$ and so we ought not to consider $K_A$ a
symmetry any longer. We denote this by the breakdown of the
conformally covariant derivative $\CCD$ to the Poincar\'e
derivative $\CD$.

Since $K_A$ is no longer considered a symmetry, the fields ${f_M}^A$ are
now auxiliary. In order to analyze the various properties of these
objects, we must make use of the conformal curvatures.
Most of these (torsion, Lorentz, and $U(1)$) can be viewed as
the Poincar\'e versions with additional terms arising from the
conformal gauge fields. The remaining ones (special conformal and scaling)
have no Poincar\'e versions and so give pure constraints among the
various fields ${f_M}^A$. After examining all the conformal constraints
we will show that they reduce to the Poincar\'e constraints with
precisely the auxiliary fields of $U(1)$ superspace.

For reference, we give here the relations among the various objects
in the gauge where $B=0$. For the conformal/Poincar\'e curvatures,
\begin{align}
\conf F_{BA} &= F_{BA} + 3 i f_{BA} w(A) - 3i f_{AB} w(B) \\
{{\conf T}_{CB}}\,^A &= {T_{CB}}^A + \frac{1}{2} {f_{[C}}^D {{C_D}^A}_{B]} \\
{{\conf R}_{DC}}\,^{\beta \alpha} &= {R_{DC}}^{\beta \alpha}
	+ 2 {\delta_{[D}}^b {f_{C]}}^a (\eps \sigma_{ab})^{\beta \alpha}
	+ 2 {\delta_{[D}}^{\{\beta} {f_{C]}}^{\alpha \}} (-)^C
	\label{cRtoR}
\end{align}
For the purely conformal curvatures,
\begin{align}
{\conf H}_{BA} &= 2 f_{BA} (-)^a - 2 f_{AB} (-)^b \\
\conf {R(K)_{CB}}^A &= \CD_{[C} {f_{B]}}^A + {T_{CB}}^D {f_{D}}^A
	+ \frac{1}{2} f_{[C D} {C_{B]}}^{AD}
	- \frac{1}{2} {f_{[C}}^D {f_{B]}}^F {C_{FD}}^A
\end{align}
The covariant derivative appearing in $R(K)$ is Poincar\'e.
${f_M}^A$ is understood to transform as a Lorentz vector on
the index $A$ and to possess a scaling weight of $\lambda(A)$
and a $U(1)$ weight of $-w(A)$. (These latter two weights
mean ${f_M}^A$ transforms oppositely under scalings and the $U(1)$
as ${E_M}^A$.) In the above and subsequent formulae,
we will use the combination ${f_B}^A = {E_B}^M {f_M}^A$, which
possesses scaling and $U(1)$ weights of $\lambda(A)+\lambda(B)$ and
$-(w(A)+w(B))$, respectively.

\subsubsection{Constraint analysis}
We shall start with the scaling curvature:
\begin{align*}
{\conf H}_{BA} = (dB)_{BA} + 2 f_{BA} (-)^a - 2 f_{AB} (-)^b
\end{align*}
Since $B$ has been gauged away, the constraints on the $H_{BA}$
give constraints on the fields ${f_M}^A$.
These are:
\begin{align}
\conf H_{\beta \alpha} = 0 &\implies f_{\beta \alpha} = -f_{\alpha \beta} = - \epsilon_{\beta \alpha} \bar R \\
\conf H_{\dbeta \dalpha} = 0 &\implies f_{\dbeta \dalpha} = -f_{\dalpha \dbeta} = + \epsilon_{\dbeta \dalpha} R \\
\conf H_{\beta \dalpha} = 0 &\implies f_{\beta \dalpha} = -f_{\dalpha \beta} = -\frac{1}{2} G_{\beta \dalpha} \\ {\conf H}_{\ul \beta a} = 0 &\implies f_{\ul \beta a} = - f_{a \ul \beta}
\end{align}
The above serve as definitions of the fields $R$ and $G_c$. The
complex conjugation properties of the above identities tell us
$\bar R = R^\dag$ and $G_c = (G_c)^\dag$. The scaling weights of these
objects are $\Delta(R)=\Delta(\bar R) = 2$ and $\Delta (G_c) = 2$;
the $U(1)$ weights are $w(R) = -w(\bar R) =2$ and $w(G_c)=0$.

The next set of constraints to analyze are those of the $U(1)$
curvature. Recall
\begin{align*}
\conf F_{BA} = F_{BA} + 3 i f_{BA} w(A) - 3i f_{AB} w(B)
\end{align*}
which leads to
\begin{align}
\conf F_{\beta \alpha} = 0 &\implies F_{\beta \alpha} = 0 \\
\conf F_{\dbeta \dalpha} = 0 &\implies F_{\dbeta \dalpha} = 0 \\
\conf F_{\beta \dalpha} = 0 &\implies F_{\beta \dalpha} = 6i f_{\beta \dalpha} = -3 i G_{\beta \dalpha} \\
\conf F_{\beta a} = 0 &\implies F_{\beta a} = -3i f_{\beta a} \\
\conf F_{\dbeta a} = 0 &\implies F_{\dbeta a} = +3i f_{\dbeta a}
\end{align}

Now consider the torsion. Noting that
\beq
{{\conf T}_{CB}}\,^A = {T_{CB}}^A + \frac{1}{2} {F_{[C}}^D {{C_D}^A}_{B]}
\eeq
one can see the only torsions which differ between the conformal and Poincar\'e
cases are those with $A$ fermionic and either $C$ or $B$ (or both) bosonic.
Thus the constraints on the conformal torsions pass unchanged for the
fermion/fermion form indices:
\begin{align}
{{\conf T}_{\gamma \beta}}\,^A = 0 &\implies {T_{\gamma \beta}}^A = 0 \\
{{\conf T}_{\dgamma \dbeta}}\,^A = 0 &\implies {T_{\dgamma \dbeta}}^A = 0 \\
{{\conf T}_{\gamma \dbeta}}\,^{\ul\alpha} = 0 &\implies {T_{\gamma \dbeta}}^{\ul\alpha} = 0 \\
{{\conf T}_{\gamma \dbeta}}\,^{a} = 2 i \sigma_{\gamma\dbeta}^a 
	&\implies {T_{\gamma \dbeta}}^{\ul\alpha} = 2 i \sigma_{\gamma\dbeta}^a
\end{align}
For the fermion/boson form indices, it is only slightly more complicated:
\begin{align}
{{\conf T}_{\gamma b}}\,^\alpha = 0 &\implies
{T}_{\gamma (\beta \dbeta) \alpha} = +i \epsilon_{\beta \alpha} G_{\gamma \dbeta} \\
{{\conf T}_{\dgamma b}}\,^\alpha = 0 &\implies
{T}_{\dgamma (\beta \dbeta) \alpha} = -2i \eps_{\dgamma \dbeta} \eps_{\beta \alpha} R \\
{{\conf T}_{\gamma b}}\,^a = 0 &\implies {T_{\gamma b}}^a = 0 \\
{{\conf T}_{\dgamma b}}\,^a = 0 &\implies {T_{\dgamma b}}^a = 0
\end{align}
The only other torsion constraint was ${{\conf T}_{cb}}\,^a=0$, which
gives the same constraint on the Poincar\'e torsion
\begin{align}
{{\conf T}_{c b}}\,^a = 0 &\implies {T_{c b}}^a = 0.
\end{align}

The Lorentz curvature is quite simple to analyze:
\begin{align}
{\conf R}_{\delta \gamma \beta \alpha} = 0 &\implies
	R_{\delta \gamma \beta \alpha} = 4 (\eps_{\delta \beta} \eps_{\gamma \alpha}
		+ \eps_{\delta \alpha} \eps_{\gamma \beta}) \bar R \\
{\conf R}_{\delta \gamma \dbeta \dalpha} = 0 &\implies R_{\delta \gamma \dbeta \dalpha} = 0 \\
{\conf R}_{\ddelta \dgamma \beta \alpha} = 0 &\implies R_{\ddelta \dgamma \beta \alpha} = 0 \\
{\conf R}_{\ddelta \dgamma \dbeta \dalpha} = 0 &\implies
	R_{\ddelta \dgamma \dbeta \dalpha} = 4 (\eps_{\ddelta \dbeta} \eps_{\dgamma \dalpha}
		+ \eps_{\ddelta \dalpha} \eps_{\dgamma \dbeta}) R \\
{\conf R}_{\delta \dgamma \beta \alpha} = 0 &\implies
	R_{\delta \dgamma \beta \alpha} = - \eps_{\delta \beta} G_{\alpha \dgamma}
								- \eps_{\delta \alpha} G_{\beta \dgamma} \\
{\conf R}_{\delta \dgamma \dbeta \dalpha} = 0 &\implies
	R_{\delta \dgamma \dbeta \dalpha} = - \eps_{\dgamma \dbeta} G_{\delta \dalpha}
								- \eps_{\dgamma \dalpha} G_{\delta \dbeta}
\end{align}

The remaining curvatures are:
\begin{align}
\conf R(K)_{\gamma \beta \alpha} = 0 &\implies \CD_\alpha \bar R = 0 \\
\conf R(K)_{\gamma \beta \dalpha} = 0 &\implies
	f_{\gamma (\beta \dalpha)} + f_{\beta (\gamma \dalpha)} = -\frac{i}{2} \CD_{\{\gamma} G_{\beta\} \dalpha} \\
\conf R(K)_{\gamma \beta a} = 0 &\implies
	\CD_{\{\gamma} f_{\beta \} (\alpha \dalpha)} = +2i G_{\{\gamma \dalpha} \eps_{\beta\} \alpha} \bar R \\
\conf R(K)_{\gamma \dbeta \alpha} = 0 &\implies
	f_{\gamma (\alpha \dbeta)} - 2 f_{\alpha (\gamma \dbeta)} =
		\frac{i}{2} \CD_\gamma G_{\alpha \dbeta} - i \eps_{\gamma \alpha} \CD_\dbeta \bar R \\
\conf R(K)_{\gamma \dbeta a} = 0 &\implies
	f_{(\beta \dbeta)(\alpha \dalpha)} = \frac{i}{2} \CD_{\{\beta} f_{\dbeta\} (\alpha \dalpha)}
		+ 2 \eps_{\beta \alpha} \eps_{\dbeta \dalpha} R \bar R
		+ \frac{1}{2} G_{\alpha \dbeta} G_{\beta \dalpha}
\end{align}
(We have used the spinor notation $f_{\gamma (\beta \dalpha)} \equiv f_{\gamma c} \,\sigma^c_{\beta\dalpha}$
as well as
$f_{(\beta \dbeta) (\alpha \dalpha)} \equiv f_{ba} \,\sigma^a_{\alpha\dalpha}\,\sigma^b_{\beta\dbeta}$.)
The first condition indicates that $\bar R$ is an antichiral superfield; its complex
conjugate tells that $R$ is chiral. The second and fourth equations can be combined to yield
\beq
3 i f_{\beta (\alpha \dalpha)} =
	+\frac{1}{2} \CD_\beta G_{\alpha \dalpha} + \CD_\alpha G_{\beta \dalpha}
	+ \eps_{\beta \alpha} \CD_\dalpha \bar R
\eeq
as well as its conjugate
\beq
3 i f_{\dbeta (\alpha \dalpha)} =
	-\frac{1}{2} \CD_\dbeta G_{\alpha \dalpha} - \CD_\dalpha G_{\alpha \dbeta}
	- \eps_{\dbeta \dalpha} \CD_\alpha R.
\eeq
This result can be substituted into the third equation, yielding
\begin{align}\label{gr_eq}
\CD^2 G_c = 4i \CD_c \bar R, \;\;\;
\BCD^2 G_c = -4i \CD_c R
\end{align}
The result given for $f_{\beta a}$ allows the determination of $F_{\beta a}$:
\begin{align}
F_{\beta (\alpha \dalpha)} &= -3 i f_{\beta (\alpha \dalpha)}
	= -\frac{3}{2} \CD_\beta G_{\alpha \dalpha} - \eps_{\beta \alpha} \bar X_\dalpha \\
F_{\dbeta (\alpha \dalpha)} &= +3 i f_{\dbeta (\alpha \dalpha)}
	= -\frac{3}{2} \CD_\dbeta G_{\alpha \dalpha} - \eps_{\dbeta \dalpha} X_\alpha
\end{align}
where
\begin{equation}
X_\beta \equiv \CD_\beta R - \CD^\dbeta G_{\beta \dbeta}, \;\;\;
\bar X_\dbeta \equiv \CD_\dbeta \bar R - \CD^\beta G_{\beta \dbeta}
\end{equation}
just as in $U(1)$ superspace.
Furthermore, (\ref{gr_eq}) implies (after some algebra) the
chirality of $X_\alpha$:
\begin{equation}
\CD_\dalpha X_\alpha = 0, \;\;\; \CD_\alpha X_\dalpha = 0
\end{equation}

Finally the fourth $R(K)$ constraint gives
\begin{align}
f_{(\beta \dbeta)(\alpha \dalpha)} =& \frac{i}{2} \CD_{\{\beta} f_{\dbeta\} (\alpha \dalpha)}
		+ 2 \eps_{\beta \alpha} \eps_{\dbeta \dalpha} R \bar R
		+ \frac{1}{2} G_{\alpha \dbeta} G_{\beta \dalpha} \eol
	=& -\frac{1}{12} [\CD_\beta, \CD_\dbeta] G_{\alpha \dalpha}
		- \frac{1}{6} \CD_\beta \CD_\dalpha G_{\alpha \dbeta}
		+ \frac{1}{6} \CD_\dbeta \CD_\alpha G_{\beta \dalpha} \eol
		&- \frac{1}{12} \eps_{\dbeta \dalpha} \eps_{\beta \alpha} (\CD^2 R + \BCD^2 \bar R)
		+ 2 \eps_{\dbeta \dalpha} \eps_{\beta \alpha} R \bar R
		+ \frac{1}{2} G_{\alpha \dbeta} G_{\beta \dalpha}
\end{align}
The special conformal gauge field ${f_B}^A$ is now entirely specified
in terms of superfields $R$ and $G_c$.

It is worth pausing a moment to take stock of our position. We have
now checked that every constraint taken in conformal superspace
reproduces (in the $B=0$ gauge) a known result in $U(1)$ superspace;
in particular, we have reproduced among our relations the constraint
structure of $U(1)$ superspace. Since the $U(1)$ constraints
uniquely specify $U(1)$ superspace, the gauge $B=0$ of our
constrained conformal superspace \emph{must} correspond to
the standard $U(1)$ superspace. All further checks are merely
tests of consistency.

\subsubsection{Some consistency checks}
\begin{itemize}
\item Torsion \\
The only torsion components we have not yet discussed are those which
we did not constrain: ${T_{cb}}^{\ul \alpha}$. These also differ between
conformal and Poincar\'e theories. Using
\[
{{\conf T}_{cb}}\,^\alpha = {T_{cb}}^\alpha + i f_{[c \ddelta} \bsigma_{b]}^{\ddelta \alpha}
\]
one finds
\begin{align}
T_{(\gamma \dgamma)(\beta \dbeta)\alpha}
	= + 2 \epsilon_{\dgamma \dbeta} W_{\gamma \beta \alpha}
		+ \eps_{\alpha \beta} \left(\CD_{\dbeta} G_{\gamma \dgamma} +
			\frac{2}{3} \eps_{\dbeta \dgamma} X_\gamma\right)
		- \eps_{\alpha \gamma} \left(\CD_{\dgamma} G_{\beta \dbeta} +
			\frac{2}{3} \eps_{\dgamma \dbeta} X_\beta\right) \\
T_{(\gamma \dgamma)(\beta \dbeta)\dalpha}
	= - 2 \epsilon_{\gamma \beta} W_{\dgamma \dbeta \dalpha}
		+ \eps_{\dalpha \dbeta} \left(\CD_{\beta} G_{\gamma \dgamma} +
			\frac{2}{3} \eps_{\beta \gamma} \bar X_\dgamma\right)
		- \eps_{\dalpha \dgamma} \left(\CD_{\gamma} G_{\beta \dbeta} +
			\frac{2}{3} \eps_{\gamma \beta} \bar X_\dbeta\right)
\end{align}
This is equivalent to the corresponding formulae in equations (B-2.12) through
(B-2.18) of \cite{bgg}; therefore, the torsion of $U(1)$ supergravity is
equivalent to the $B=0$ gauge of conformal superspace.

\item Lorentz curvatures \\
The Lorentz curvatures in their canonically decomposed form are
\begin{align}
{{\conf R}_{DC}}\,^{\beta \alpha} &= {R_{DC}}^{\beta \alpha}
	+ 2 {\delta_{[D}}^b {f_{C]}}^a (\eps \sigma_{ab})^{\beta \alpha}
	+ 2 {\delta_{[D}}^{\{\beta} {f_{C]}}^{\alpha \}} (-)^C
\end{align}
The case of purely fermionic form indices has already been dealt
with. Turn next to the fermion/boson case:
\begin{align}
{{\conf R}_{\delta (\gamma \dgamma)\beta\alpha}} &= {R_{\delta (\gamma \dgamma)\beta \alpha}}
	+ \sum_{\beta \alpha} \left(-\eps_{\gamma \alpha} f_{\delta (\beta \dgamma)}
	+ 2 \eps_{\delta \beta} f_{\alpha (\gamma\dgamma)} \right)
\end{align}
Noting that ${\conf R}_{\delta (\gamma \dgamma) \beta\alpha} = 0$ and
inserting the explicit expression for $f_{\beta (\alpha \dalpha)}$, one
finds
\begin{align}
R_{\delta (\gamma \dgamma) \beta \alpha} = 
	+ i \sum_{\beta \alpha} \left(
		\frac{1}{2} \eps_{\delta \gamma} \CD_\beta G_{\alpha \dgamma}
		+ \frac{1}{2} \eps_{\delta \beta} \CD_\gamma G_{\alpha \dgamma}
		- \eps_{\delta \beta} \eps_{\gamma \alpha} \BCD_\dgamma \bar R
	\right)
\end{align}
as in $U(1)$ superspace \cite{bgg}. The other Lorentz curvature term we
need to calculate is
\begin{align*}
R_{\ddelta (\gamma \dgamma) \beta \alpha}
	&= \conf R_{\ddelta (\gamma \dgamma) \beta \alpha}
		+ \sum_{\beta \alpha} f_{\ddelta (\beta \dgamma)} \eps_{\gamma \alpha} \\
	&= -4i \eps_{\ddelta \dgamma} W_{\gamma \beta \alpha}
		+ \sum_{\beta \alpha} \eps_{\gamma \alpha} \left(
			\frac{i}{6} \CD_{\ddelta} G_{\beta \dgamma}
			+ \frac{i}{3} \CD_{\dgamma} G_{\beta \ddelta}
			+ \frac{i}{3} \eps_{\ddelta \dgamma} \CD_\beta R
		\right) \\
	&= -4i \eps_{\ddelta \dgamma} W_{\gamma \beta \alpha}
		+ i \sum_{\beta \alpha} \eps_{\gamma \alpha} \left(
			\frac{1}{2} \CD_{\ddelta} G_{\beta \dgamma} + \frac{1}{3} \eps_{\ddelta \dgamma} X_\beta\right)
\end{align*}
which is also as in $U(1)$ superspace \cite{bgg}.

At the dimension 2 level, results are a bit more interesting.
Using \eqref{cRtoR}, one finds
\begin{align}\label{PRb}
R_{(\delta \ddelta) (\gamma \dgamma) \beta \alpha} = 
	{\conf R}_{(\delta \ddelta) (\gamma \dgamma) \beta \alpha}
	+ \sum_{\beta \alpha} \left(
		f_{(\delta \ddelta)(\beta \dgamma)} \eps_{\gamma \alpha}
		- f_{(\gamma \dgamma)(\beta \ddelta)} \eps_{\delta \alpha}
	\right)
\end{align}
Recall that
\begin{equation}\label{CRb}
{\conf R}_{(\delta \dot\delta)(\gamma \dgamma)\beta \alpha} =
	+ 2 \eps_{\ddelta \dgamma} \chi_{\lsym{\delta\gamma\beta\alpha}}
	- \frac{1}{4} \eps_{\ddelta \dgamma}
		\sum_{(\delta \gamma)} \sum_{(\beta \alpha)} \eps_{\delta \beta} \CD^\phi W_{\phi \gamma \alpha}
\end{equation}
where
\[
{\chi}_{\lsym{\delta \gamma \beta \alpha}} \equiv
	\frac{1}{4} (\CD_\delta W_{\gamma\beta\alpha} + \CD_\gamma W_{\delta\beta\alpha} +
				\CD_\beta W_{\gamma\delta\alpha} + \CD_\alpha W_{\gamma\beta\delta}).
\]
We would like to show that \eqref{PRb} reduces to
\begin{align}
R_{(\delta \ddelta) (\gamma \dgamma) \beta \alpha} = 
	+ 2 \eps_{\ddelta \dgamma}
		\chi_{\delta\gamma \beta \alpha}
	- 2 \eps_{\delta \gamma} 
		\eps_{\dbeta \dalpha} \psi_{\ddelta \dgamma \beta \alpha}
\end{align}
where
\begin{gather}
\chi_{\delta \gamma \beta \alpha} = {\chi}_{\lsym{\delta \gamma \beta \alpha}}
	+ (\eps_{\delta \beta} \eps_{\gamma \alpha} + \eps_{\delta \alpha} \eps_{\gamma \beta}) \chi \\
\psi_{\delta \gamma \dbeta \dalpha} =  \frac{1}{8} \sum_{\delta \gamma} \sum_{\dbeta \dalpha}
		\left( G_{\delta \dbeta} G_{\gamma \dalpha} -
			\frac{1}{2} [\CD_\delta, \CD_\dbeta] G_{\gamma \dalpha} \right) \\
\chi = -\frac{1}{12} (\CD^2 R + \BCD^2 \bar R) + \frac{1}{48} [\CD^\alpha, \CD^\dalpha] G_{\alpha \dalpha}
		- \frac{1}{8} G^{\alpha \dalpha} G_{\alpha \dalpha} + 2 R \bar R
\end{gather}
This is a straightforward (albeit tiresome) check. Some intermediate
results help:
\begin{gather}\label{temp1}
\sum_{\beta \alpha} {f_{(\beta \dphi) (\alpha}}^{\dphi)} = -\CD^\phi W_{\phi \beta \alpha} \\
\sum_{\ddelta \dgamma} \sum_{\beta \alpha} f_{(\beta \ddelta) (\alpha \dgamma)} =
	4 \psi_{\ddelta \dgamma \beta \alpha} \\
{f_{(\phi \dphi)}}^{(\phi \dphi)} = 4 \chi
\end{gather}
which allow the complete expression of the $f$ terms from \eqref{PRb}
in terms of the relevant Poincar\'e quantities. For example,
\eqref{temp1} allows for the cancellation of the similar
$\CD^\phi W_{\phi \beta \alpha}$ terms in \eqref{CRb}; the
remaining terms involving $\psi$ and $\chi$ combine with
${\chi}_{\lsym{\delta \gamma \beta \alpha}}$ to give the
Poincar\'e Lorentz curvature.

\item Scaling and $U(1)$ curvatures \\
The only $U(1)$ curvature we haven't discussed yet is $F_{ba}$,
but this is the same in both conformal and Poincar\'e theories.
We have
\[
F_{(\gamma \dgamma)(\beta \dbeta)} =
    2 \eps_{\dgamma \dbeta} {\tF}_{\gamma \beta} - 2 \eps_{\gamma \beta} \tF_{\dgamma \dbeta}
\]
where $\CD^\phi W_{\phi \beta \alpha} = \frac{4i}{3} \tF_{\beta \alpha}$.
This is exactly as in \cite{bgg} (aside from the extra factor of $i$).

For the scaling curvature,
\[
H_{(\gamma \dgamma)(\beta \dbeta)} =
    2 \eps_{\dgamma \dbeta} {\tH}_{\gamma \beta} - 2 \eps_{\gamma \beta} \tH_{\dgamma \dbeta}
\]
where $\CD^\phi W_{\phi \beta \alpha} = +2 \tH_{\beta \alpha}$. This is easily
checked explicitly. Since
$
H_{(\gamma \dgamma)(\beta \dbeta)} =
	2 f_{(\beta \dbeta)(\alpha \dalpha)} - 2 f_{(\alpha \dalpha)(\beta \dbeta)},
$
it follows that
\[
\tH_{\beta \alpha} = -\frac{1}{2} \sum_{\beta \alpha} {f_{(\beta \dphi) (\alpha}}^{\dphi)}
	= +\frac{1}{2} \CD^\phi W_{\phi \beta \alpha},
\]
as needed.

\item Special conformal curvatures \\
These are by far the most complicated expressions remaining to check.
The ones remaining for us to examine are $R(K)_{\gamma b A}$ and
$R(K)_{c b A}$, which amount to five extra checks to perform. These
give no extra insight or relations beyond those we already have, so
we will avoid evaluating them explicitly here.

\end{itemize}

\subsubsection{Conformal symmetry of $U(1)$ superspace}
If $U(1)$ superspace is indeed a gauge-fixed version of a fully conformal
superspace, then it must permit some form of scale transformation.
This must be more than that of Howe and Tucker \cite{howe} since those authors
were restricted to a chiral parameter in order to preserve the minimal
torsion constraints. In fact, an unrestricted transformation \emph{does} exist.
Binetruy, Girardi, and Grimm \cite{bgg}
showed that the minimal matter coupling $e^{-K/3}$ could be absorbed into the
frame of the vierbein provided the minimal superspace structure was enlarged
to include a $U(1)$ superconnection. This can be understood as an unconstrained
scale transformation.\footnote{Enlarging the structure group is not the only
way to do this. Instead, one may choose fewer torsion constraints in Poincar\'e
supergravity, which allow the superfield $T_\alpha$ in addition to $R$,
$W_{\alpha \beta \gamma}$ and $G_c$. One can show (see for example \cite{Muller:1985vg})
that this $T_\alpha$ is essentially $A_\alpha$.}

They postulated a transformation for the vierbein
\beq
{E'_M}^A = {E_M}^B {X_B}^A
\eeq
with a parameter ${X_B}^A$ of the form
\beq
{X_{B}}^{A} \,=\,\left(
\begin{array}{ccc}
{\delta_{b}}^{a} X {\BX} & {X_{b}}^{\alpha} & X_{b \, \dalpha}\\ 0 &
{\delta_{\beta}}^{\alpha}X & 0\\ 0 & 0 & {{\delta}^{\dbeta}}_{\dalpha}\ {\BX}
\end{array}
\right)
\eeq
where
\begin{align}
{X_{b}}^{\alpha} \equiv\frac{i}{2} {{(\epsilon \sigma_{b})}^{\alpha}}_{\dalpha}
{\BX}^{-1} {\BCD}^{\dalpha}(X {\BX}),\;\;\;
X_{b \dalpha} \equiv\frac{i}{2} {{(\epsilon \bar \sigma_{b})}_{\dalpha}}^{\alpha}
{X}^{-1} {\CD}_{\alpha}(X {\BX})
\end{align}
is required to preserve torsion constraints. Otherwise,
the factors $X$ and $\BX$ are totally unconstrained.
By investigating the constraints of $U(1)$ superspace,
they found the required transformation
rules of the superfields
\begin{align}
R' &= (\bar X)^{-2} \left(R - \frac{1}{8} (X\BX)^{-2} \BCD^2 (X\BX)^2\right) \\
G'_{\alpha \dalpha} &= (X \bar X)^{-1} \left(G_{\alpha \dalpha} - \frac{1}{2} [\CD_\alpha, \BCD_\dalpha] \log(X \bar X)
	+ Y_\alpha \bar Y_\dalpha \right) \\
W'_{\alpha \beta \gamma} &= (X \bar X)^{-1} (\bar X)^{-1} W_{\alpha \beta \gamma}
\end{align}
where $Y_A \equiv \CD_A \log(X \bar X)$.
Although they restricted to the case where the $U(1)$ connection
was initially zero, it is simple to extend to the case of
a non-vanishing initial connection:
\begin{align}
A'_M &= A_M - i\frac{1}{2} Z_M
	- \frac{3i}{2} {E_M}^\alpha Y_\alpha
	+ \frac{3i}{2} {E_M}_\dalpha \bar Y^\dalpha
	+ \frac{3}{4} {E_M}^{\alpha \dalpha} Y_\alpha \bar Y_\dalpha
\end{align}
where $Z_M \equiv \partial_M \log(X/\bar X)$.
Without loss of generality, the superfields $X$ and $\bar X$ can be written
\beq
X = \exp(-\Lambda/2 + i \Omega), \;\;\; \bar X = \exp(-\Lambda/2 -i \Omega),
\eeq
for real superfields $\Omega$ and $\Lambda$. The infinitesimal transformation
rules are
\begin{align}\label{U1trans1}
\delta {E_m}^a &= - \Lambda {E_m}^a \\
\delta {E_m}^\alpha &= \left(- \frac{1}{2}\Lambda + i \Omega \right) {E_m}^\alpha
	- \frac{i}{2} {{(\epsilon \sigma_{m})}^{\alpha}}_{\dalpha} \BCD^\dalpha \Lambda \\
\delta R &= \left(\Lambda + 2i\Omega\right) R + \frac{1}{4} \BCD^2 \Lambda \\
\delta G_{\alpha \dalpha} &= \Lambda G_{\alpha \dalpha}
	+ \frac{1}{2} [\CD_\alpha, \BCD_\dalpha] \Lambda \\ \label{U1trans2}
\delta W_{\alpha \beta \gamma} &= \left(\frac{3}{2} \Lambda + i \Omega \right) W_{\alpha \beta \gamma} \\
\delta A_M &= \partial_M \Omega  + \frac{3i}{2} {E_M}^\alpha \CD_\alpha \Lambda
	- \frac{3i}{2} {E_{M \dalpha}} \CD^\dalpha \Lambda
\end{align}
(Of the fields in the supervierbein, we have listed only those corresponding
to the graviton and the gravitino. The other components of the supervierbein
also transform, but they are unphysical so we'll ignore them for simplicity.)
The above set of transformation rules is quite interesting.
For the most part, they have the form of scale ($\Lambda$) and chiral ($\Omega$)
transformations, with $A$ as the gauge field for the chiral transformations;
however, for every term other than ${E_m}^a$, $W_{\alpha \beta \gamma}$,
and $A_{\alpha \dalpha}$, there are modifications which depend on the
derivative of the scale parameter $\Lambda$.

These extra modifications can be viewed as requirements forced by
the torsion and curvature constraints of $U(1)$ superspace, but they can
also be viewed as having a deeper geometrical origin. Our claim was that
$U(1)$ superspace is a gauge-fixed version of conformal superspace.
This is straightforward to see. The variation of the field $B_M$ under $D$
and $K$ transformations is
\begin{align*}
\delta B_M = \partial_M \Lambda - 2 {E_M}^A \epsilon_A (-)^a
\end{align*}
where $\epsilon^A$ is the parameter for $K$ transformations
and $\Lambda$ that of $D$. If we demand that $B_M=0$ remain fixed,
then every scale transformation must be accompanied by a
$K$-transformation with $\epsilon^A = (-)^a \frac{1}{2} D_A \Lambda$.
It is this corresponding $K$-transformation which generates the
additional derivatives of $\Lambda$. 

Consider first the vierbein. Under a $K$-transformation,
$\delta_K {E_M}^A = \frac{1}{2} {E_M}^C \epsilon^B {C^A}_{BC}$, which
corresponds to
\begin{gather*}
\delta_K {E_m}^a = 0 \\
\delta_K {E_m}^\alpha = -i \epsilon_\dbeta \bsigma_m^{\dbeta \alpha}
	= \frac{i}{2} \CD_\dbeta \Lambda \bsigma_m^{\dbeta \dalpha}
\end{gather*}
for the graviton and gravitino, reproducing the additional terms exactly.
Take the $U(1)$ connection next. Under a $K$-transformation,
$\delta_K A_M = -3i w(A) {E_M}^A \epsilon_A$. Plugging in for $\epsilon$
we find
\begin{align*}
\delta_K A_M = \frac{3i}{2} w(A) {E_M}^A \CD_A \Lambda
\end{align*}
as expected.

The fields $R$ and $G_{\alpha \dbeta}$ are a bit more complicated.
Recall that they are themselves related to the $K$-gauge fields by
$f_{\dalpha \dbeta} = \eps_{\dalpha\dbeta} R$ and
$f_{\alpha \dbeta} = - G_{\alpha \dbeta}/2$.
The rule for the transformation of $f_{M \dbeta}$ is
$\delta_K f_{M \dbeta} = \CD_M \eps_\dbeta - i {E_M}^\beta \eps_{\beta \dbeta}$
which corresponds to
\begin{align*}
\delta_K G_{\alpha \dbeta} = \CD_\alpha \CD_\dbeta \Lambda + i \CD_{\alpha\dbeta} \Lambda
	= \frac{1}{2} [\CD_\alpha, \CD_\dbeta] \Lambda.
\end{align*}
For $R$, using $\delta_K f_{\dalpha \dbeta} = \eps_{\dalpha \dbeta} \BCD^2 \Lambda / 4$
gives
\begin{align*}
\delta_K R = \BCD^2 \Lambda / 4
\end{align*}
These are precisely the extra terms enforced by the torsion constraints.

Finally note that $W_{\alpha \beta \gamma}$ is a chiral primary superfield;
thus it is inert under $K$ and so has no extra terms.

\subsection{Old minimal supergravity}
We break the explicit scale invariance of the superspace theory
by following as closely as possible the non-supersymmetric case.
There a compensating matter field $\Phi$ was introduced with
unit scaling weight. The $D$-gauge was then used to scale $\Phi$
to a constant, explicitly breaking the scale invariance and collapsing
the kinetic Lagrangian into the Einstein-Hilbert term.

An analogous procedure can be undertaken in superspace. We
must make use of a compensating superfield, and the simplest
one is a chiral field. We denote it $\Phi_0$, assume it to
have a scaling weight of $\Delta(\Phi_0) = 1$ (and therefore a chiral
weight of $\omega(\Phi_0) = 2/3$). The kinetic multiplet for
$\Phi_0$ is just the superspace D-term
\beq
-3 \int \conf E \ \bar \Phi_0 \Phi_0
\eeq
(Here and in the following we use $\conf{}$ over the measure to denote
when we are in the conformal frame where the gauge is unfixed.)
We would like to gauge $\Phi_0=1$. That converts the kinetic action into the
supervolume, which reproduces the supersymmetrized Einstein-Hilbert term.

First let us note some things. After gauge-fixing $\Phi_0$ to a constant,
we are left with an issue of consistency, the equation of chirality for $\Phi_0$:
\beq
0 = \CCD_\dalpha \Phi_0 = \left(D_\dalpha - B_\dalpha - \frac{2i}{3} A_\dalpha\right) \Phi_0
\eeq
We have explicitly used all of the $K$-gauge to fix
$B=0$. When $\Phi_0$ is gauged to a constant, $A_\dalpha=0$ follows.
A corresponding analysis with $\bar \Phi_0$
leads us to conclude $A_\alpha$ vanishes as well. Using
$F_{\alpha \dalpha} = (dA)_{\alpha \dalpha} = -3 i G_{\alpha \dalpha}$,
one can immediately deduce $A_{\alpha \dalpha} = -\frac{3}{2} G_{\alpha \dalpha}$.
The $U(1)$ symmetry is broken; the bosonic component of
$A$ has become the auxiliary field $G_c$.

The superfield $R$ also ultimately has an origin in the
unfixed gauge. Recall that the
F-term of the field $\Phi_0$ was defined using the conformal
superspace derivatives. We must convert these to Poincar\'e
derivatives, giving, after gauge-fixing $\Phi_0$ to a constant,
\beq
F = -\frac{1}{4} \CCD^2 \Phi_0 = -\frac{1}{4} \left(\CD^2 - 8 \bar R\right) \Phi_0
	= 2 \bar R \Phi_0
\eeq
The anti-chiral superfield $\bar R$ is itself
nothing more than the F-term of the chiral compensator, which
is a well-known result. \footnote{
It can be shown (see for example \cite{bgg}) that the theory above,
with a remnant $U(1)$ field, can be converted to the theory
of Wess and Bagger, where the $U(1)$ connection is entirely
absent, by a simple modification of the torsion components.}

\subsubsection{The chiral compensator and super-Weyl transformations}
The normal approach to conformal supergravity \cite{Kugo:1982mr}
makes use of a chiral field $\Phi_0$, introduced
as a book-keeping device, whose bosonic component is used to
fix the normalization of the Einstein-Hilbert term while the
rest of the components are set to zero. This is completely
analogous to the theory discussed above, except in those
formulations the compensator is fixed at the component level.
This theory also possesses a residual ``super-Weyl'' symmetry.

Begin with a model where the only field
with scaling or chiral weight is the compensator $\Phi_0$. It must
therefore be employed to make the conformal D and F-terms invariant.
These take the form
\begin{gather}
\Lag_D = \int d^4\theta \; \conf E \; \Phi_0 \bar \Phi_0 \ V, \;\;\;
\Lag_F = \int d^2\theta \; \conf {\mathcal E} \; \Phi_0^3 \ W
\end{gather}
Although $V$ and $W$ are generic real and chiral superfields of vanishing
scaling and chiral weights, they possess a residual symmetry:
\begin{align}
\Phi_0 \rightarrow \Phi_0 e^{2 \Sigma}, \;\;\;
V \rightarrow e^{-2\Sigma - 2\bar\Sigma} V, \;\;\;
W \rightarrow e^{-6\Sigma} W
\end{align}
where $\Sigma$ is a chiral field of vanishing scaling and chiral weights.
If we work in the gauge where $\Phi_0=1$, the above redefinition of
the chiral compensator must be compensated by an honest conformal
transformation with a rescaling $\Lambda = -\Sigma - \bar \Sigma$
and a $U(1)$ rotation $\Omega = \frac{2i}{3} (\Sigma - \bar \Sigma)$.
This combined redefinition and conformal transformation is the
super-Weyl transformation of Howe and Tucker \cite{howe}
which preserves the form of the minimal Poincar\'e torsion
constraints. $V$ transforms as a real super-Weyl field with weight 2,
$W$ as a chiral super-Weyl field of weight 3,
and the superdeterminant of the vierbein, $E$,
as a real super-Weyl field with weight -2. (The transformation
rules on the superfields $R$, $G_c$, the graviton, and
gravitino can be derived from (\ref{U1trans1}-\ref{U1trans2}).)

The conformal transformations discussed in this article must be
contrasted with these super-Weyl transformations. The former
are unconstrained in superspace; the latter
are \emph{highly} constrained in superspace (the $\Sigma$ must be chiral)
but correspond to unconstrained superconformal
transformations at the \emph{component} level.

\subsubsection{Integral relations between various formulations}
We have several types of integrals -- D and F, gauge fixed
and unfixed -- that describe the same physics, and we should
demonstrate how they are related to each other. 

The F-term action in conformal superspace can be rewritten
\begin{align}
\int d^2 \theta \; \conf {\mathcal E} \; \Phi_0^3 \ W
= -\frac{1}{4} \int d^2 \theta \; \conf {\mathcal E} \; \BCCD^2\left( \frac{\bar \Phi_0 \Phi_0^3 W}{\bar F}\right)
\end{align}
where $\bar F \equiv -\frac{1}{4} {\BCCD}^2 \bar \Phi_0$.
(The equivalency follows since the only non-chiral term
in the parentheses is $\bar \Phi_0$, whose derivatives are cancelled
by the denominator.) This is equivalent to a D-term:
\beq
-\frac{1}{4} \int d^2 \theta \; \conf {\mathcal E} \; \BCCD^2\left( \frac{\bar \Phi_0 \Phi_0^3 W}{\bar F}\right)
= \int d^4 \theta \; \conf E \; \frac{\bar \Phi_0 \Phi_0^3 W}{\bar F}
\eeq
Now we gauge $\Phi_0$ to one. This leaves the inverse of the
F-component of $\bar \Phi_0$, but this is nothing more than the
chiral field $R$. Thus we find the following set of equalities:
\beq
\int d^2 \theta \; \conf {\mathcal E} \; \Phi_0^3 \ W
= \int d^2 \theta \; {\mathcal E} \; \ W 
=  \frac{1}{2} \int d^4 \theta \; \frac{E}{R} \;  W
\eeq
The term on the left is the expression for the chiral
F-term in the presence of a conformal multiplet.
The term in the middle is the chiral F-term after conformal
gauge-fixing. The term on the right is the form of the
chiral F-term used in \cite{bgg}. Since the difference between the
first and third terms is just a gauge-fixing, it should
make no difference \emph{when} we fix the gauge. Therefore
if we were to evaluate the first term completely within
conformal superspace and then gauge-fix, we would necessarily
arrive at the same answer as the term on the right.\footnote{
One may also note that the rather curious form of $1/2R$ as the
term converting from an $F$ to a $D$-term can be understood
as a delta function. In particular, using the result of
Apppendix \ref{actions_sm}, the chiral delta function is of
a general form $\chDelta = X / \CP[X]$. For the case of $X=1$,
this gives $\chDelta = -1 / \frac{1}{4} (\BCD^2 - 8R)(1) = 1 / 2R$.}

To address the D-term, first note that in conformal superspace
one can easily convert a D to an F-term:
\begin{align}
\int d^4\theta \; \conf E \; \bar\Phi_0 \Phi_0 \ V
&= \int d^2\theta \, \conf{\mathcal E} \Phi_0\left(\bar F V
	- \frac{1}{2} \CCD_\dalpha \bar \Phi_0 \CCD^\dalpha V
	- \bar\Phi_0 \frac{1}{4} \CCD^2 V\right) \eol
&= \int d^2\theta \, \conf{\mathcal E}\left(2R \bar\Phi_0 \Phi_0 V
	- \frac{\Phi_0}{2} \CCD_\dalpha \bar\Phi_0 \CCD^\dalpha V
	- \bar\Phi_0 \Phi_0 \frac{1}{4} \CCD^2 V\right)
\end{align}
(Here $V$ has zero scaling weight.)
Now, let us gauge fix $\Phi_0$ to unity and equate the first
and final steps. We find
\beq
\int d^4\theta \, E \, V
= -\frac{1}{4}\int d^2\theta \, \chE (\BCD^2 - 8 R) V
\eeq
This tells us that the proper way in Poincar\'e superspace
to convert a D to an F-term is through the use of the
chiral Poincar\'e projector. This is actually quite intuitive
if we use our other F to D-term conversion formula:
\beq
\int d^4\theta \, E \, V
= -\frac{1}{4}\int d^2\theta \, \chE\,
	(\BCD^2 - 8 R) V
= -\frac{1}{8}\int d^4\theta \, \frac{E}{R} \,
	(\BCD^2 - 8 R) V
\eeq
The equality of the first and third terms follows
by integration by parts in Poincar\'e
superspace.\footnote{Note the significance of these steps. Within conformal
superspace as in flat supersymmetry, one can convert from
a D to an F-term, but the reverse is not an easily defined
operation. Upon gauge-fixing to minimal Poincar\'e superspace,
we gain the field $R$ which allows us to do so.}

\subsection{K\"ahler supergravity}
A general set of chiral fields coupled to conformal supergravity
generically has D and F-terms
\beq
\Lag_D = -3\cdint \bar \Phi_0 e^{-K/3} \Phi_0, \;\;\;\;
\Lag_F = \cfint \Phi_0^3 \, W
\eeq
for chiral primary superfield $\Phi_0$ with $\Delta = 1$ and $\omega=2/3$.
$K$ is real and $W$ is chiral, both with vanishing scale and chiral weights.
The actions are invariant under a K\"ahler transformation
\begin{gather}
K \rightarrow K + F + \bar F \\
\Phi_0 \rightarrow \Phi_0 e^{+F/3}, \;\; \bar \Phi_0 \rightarrow \bar \Phi_0 e^{+\bar F/3}\\
W \rightarrow e^{-F} W, \;\; \bar W\rightarrow e^{-\bar F} W
\end{gather}
Here the superfields $F$ and $\bar F$ are chiral/antichiral respectively.
$K$ is a real function of K\"ahler chiral matter
fields $\xi^i$ and $\bar \xi^i$ with vanishing conformal weight,
and $W$ is a function of only the
chiral ones $\xi^i$. In the language of complex manifolds, $W$ is a holomorphic
function and $K$ a real function. The transformation fields $F$ and $\bar F$ are,
respectively, holomorphic and antiholomorphic functions of the chiral and
anti-chiral K\"ahler matter fields. Note that the K\"ahler transformation has no
effect a priori on the supergravity sector.

There are two straightforward ways to accomplish a conformal gauge fixing.
The first is to gauge $\Phi_0$ to one. As the K\"ahler transformations
alter $\Phi_0$, a corresponding conformal transformation must compensate
every K\"ahler transforation. This is the well-known Howe-Tucker transformation\cite{howe},
which when combined with the given K\"ahler transformations
of $K$ and $W$ render the D and F-terms invariant. Unfortunately,
the D-term action then yields a non-canonical Einstein-Hilbert term.
There are two traditional methods for dealing with this.
One may rescale fields at the component level in a quite complicated
fashion; this is the path taken in \cite{wb}. One may also
leave $\Phi_0$ unscaled until the very end of the calculation; this
is the chiral compensator approach popularized by Kugo and Uehara
\cite{Kugo:1982mr}.

A newer method is that of Binetruy, Girardi, and Grimm \cite{bgg}.
They demonstrated that enlarging to $U(1)$ superspace from a minimal Poincar\'e superspace
allowed an arbitrary super-Weyl transformation to absorb the factor
$e^{-K/3}$ into $E$. From our point of view, their approach
has a very simple interpretation. Rather than scale
$\Phi_0=1$, choose the gauge $\Phi_0 = e^{K/6}$. The equation
of chirality then reads
$
0 = \BCD_\dalpha \Phi_0 = D_\dalpha \Phi_0 - \frac{2i}{3} A_\dalpha \Phi_0
$
which implies
$
A_\dalpha = -\frac{i}{4} D_\dalpha K.
$
The antichirality of $\Phi_0$ similarly implies $A_\alpha = \frac{i}{4} D_\alpha K$.
The Poincar\'e constraint $F_{\alpha \dalpha} = -3i G_{\alpha \dalpha}$ then
gives $A_{\alpha \dalpha}$. The entire connection is given in terms of $K$
and $G_{\alpha \dalpha}$:
\begin{gather}
A_\alpha = +\frac{i}{4} D_\alpha K, \;\;\;\;
A_\dalpha = -\frac{i}{4} D_\dalpha K \eol
A_{\alpha \dalpha} = - \frac{3}{2} G_{\alpha \dalpha} + \frac{1}{8} [\CD_\alpha, \CD_\dalpha] K \label{temp3}
\end{gather}
The imaginary part of the K\"ahler transformation now plays the role
of the $U(1)$ R-symmetry; the real part is equivalent to a
super-Weyl transformation and corresponds to a rescaling of $\Phi_0$.

Alternatively, one may absorb the K\"ahler potential into the fields
$\Phi_0$ to define K\"ahler-covariant fields $\Psi_0$ as in \eqref{temp2}.
Then the gauge choice $\Psi_0=1$ gives
\beq
0 = \nabla^K_\dalpha \Psi_0 = - \frac{2i}{3} A_\dalpha + \frac{2i}{3} \kA_\dalpha
	\implies A_\dalpha = \kA_\dalpha
\eeq
where $A_\dalpha$ is the $U(1)$ connection and $\kA_\dalpha = -\frac{i}{4} D_\dalpha K
$ is the K\"ahler connection. We arrive at the same result as \eqref{temp3}.
The gauge $\Psi_0=1$ breaks one combination of
the $U(1)$ and K\"ahler symmetries, leaving the combination where
the $U(1)$ and K\"ahler transform together. Therefore an effective transformation
on the matter fields (the K\"ahler transformation) has been
extended to the entire frame of superspace (by merging it with the
$U(1)$ R-symmetry).

\subsection{New minimal supergravity}
In both of the prior cases, we have used the simplest superfield,
a chiral one with eight components, to gauge fix to Poincar\'e
supergravity. Needless to say this is not the only choice. Another
minimal choice would be a linear multiplet, which also contains eight
components. We begin with a real linear superfield $L$, obeying
\beq
\CCD^2 L = \BCCD^2 L = 0
\eeq
From the superconformal algebra, we know that $L$ must possess
a scaling weight of $\Delta(L) = 2$ and, by reality, a
vanishing $U(1)$ weight. This latter feature will leave
the $U(1)$ gauge symmetry unaffected by the gauge-fixing
procedure.

Before fixing the gauge $L=1$, one important feature of the linear multiplet
must be discussed. Due to the linearity constraint, $[\CCD^2, \BCCD^2] L = 0$,
which implies $\CCD^{\dalpha \alpha} [\CCD_\alpha, \BCCD_\dalpha] L = 0$ --
the divergence of the vector component of $L$ vanishes. In global
supersymmetry, this implies the vector component is the dual of a three-form,
but in supergravity this statement is modified by terms involving the gravitino.
The simplest way to derive this behavior is to consider the two-form potential $B_{MN}$,
whose three-form field strength $H = d B$ obeys a Bianchi identity, $dH = 0$.
Following \cite{Muller:1985vg} and \cite{bgg}, one chooses $H$ to obey the
constraints
\begin{align}
0 = H_{\ul \gamma \ul \beta \ul \alpha} = H_{\gamma \beta a} = H_{\dgamma \dbeta a}
\end{align}
Then as a consequence of the Bianchi identities, one can show that
\begin{gather}
H_{\gamma \dbeta}{}^a = 2i \sigma^a_{\gamma \dbeta} L \\
H_{\gamma b a} = 2 (\sigma_{ba})_\gamma{}^\phi \CCD_\phi L, \;\;\;
H^\dgamma{}_{b a} = 2 (\bsigma_{ba})^\dgamma{}_\dphi \BCCD^\dphi L \\
H_{cba} = \eps_{cba}{}^d \Delta_d L
\end{gather}
where $L$ is a linear superfield and where we have defined
\begin{align}
\Delta_{\alpha \dalpha} L \equiv -\frac{1}{2} [\CCD_\alpha, \BCCD_\dalpha] L.
\end{align}
It follows that the dual of the three form is
\begin{align}
\frac{1}{3!} \eps^p{}^{nml} H_{nml} &= e_a{}^p \Delta^a L - \frac{i}{2} \eps^{pnm\ell} (\psi_n \sigma_m \bar\psi_\ell) L
	+ i (\psi_n \sigma^{np})^\phi \nabla_\phi L - i (\bar\psi_n \bsigma^{np})_\dphi \nabla^\dphi L \eol
	&= \frac{1}{2} \eps^{pnm\ell} \partial_n B_{m\ell}
\end{align}

Let us now gauge fix $L=1$. The equations of linearity become,
in Poincar\'e superspace,
\beq
(\CD^2 - 8 \bar R)L = (\BCD^2 - 8 R)L = 0 
\eeq
Since $L$ is a constant, the only way this can be satisfied
is if $R = \bar R = 0$. From the relations relating $R$ to $G_c$,
this necessarily implies $\CD_c G^c = 0$. Noting that
\beq
-2 \Delta_{\alpha \dalpha} L = [\CCD_\alpha, \BCCD_\dalpha] L = [\CD_\alpha, \CD_\dalpha] L - 4 G_{\alpha \dalpha} L
\eeq
and that both $\CD_\alpha L$ and $\CD_\dalpha L$ vanish (we have
gauged $B$ to zero, and the $U(1)$ connection appears
in neither expression since $L$ has no chiral weight), we derive that
\begin{align}
\Delta_a L = 2 G_a
\end{align}
in the gauge where $L=1$. It follows that
\begin{align}
e_a{}^p G^a\vert &= \frac{1}{4} \eps^{pnm\ell} \partial_n b_{m\ell} 
	+ \frac{i}{4} \eps^{pnm\ell} (\psi_n \sigma_m \bar\psi_\ell)
\end{align}
where $b_{m \ell}$ denotes the bosonic lowest component $B_{m \ell}\vert$.

The bosonic two-form $b_{m \ell}$ corresponds to three real bosonic
components (after accounting for its gauge invariance).
The superfield $R$ vanishes so no component field $M$ is generated.
However, the $U(1)$ symmetry has not been broken, and so we will have
in our off-shell spectrum the bosonic field $A_m$ which is the
gauge field of the chiral gauge symmetry, giving three
bosonic components. As in the (old) minimal model, we
have introduced six extra bosonic degrees of freedom
to close the supergravity algebra off-shell.

The immediate candidate for the simplest D-term action is
\beq
\int d^4\theta \; \conf E \; L
\eeq
However, using the D to F conversion in conformal superspace,
this becomes
\beq
\int d^4\theta \; \conf E \; L = -\frac{1}{4} \int d^2\theta \; \mathcal E \; \BCCD^2 L = 0.
\eeq
The linearity condition tells us that this simple integral vanishes.
This immediately implies (after gauging $L$ to one) that in the new
minimal Poincar\'e superspace the integral of the supervolume vanishes:
$\int d^4\theta \; E = 0$. This is a well-known property of the new
minimal model, and nothing more meaningful than the fact that
$R=0$ \cite{newMinR}.

To derive the form of the new minimal supergravity action, we will use
a duality transform (as discussed in \cite{superspace}) to
transform a chiral compensator to a linear one.
The properly normalized Einstein-Hilbert action is derivable from
\beq
-3 \cdint \Phi_0 \bar\Phi_0
\eeq
after fixing the gauge $\Phi_0=1$.  This action can in turn be
derived from the first-order action
\beq
-3 \cdint \left(X - L \log (X / \Phi_0 \bar\Phi_0) \right)
\eeq
where $L$ is a linear superfield, $X$ is an arbitrary real
superfield of scaling weight 2, and $\Phi_0$ is some 
chiral superfield of scaling weight 1. (Although the theory
seems to depend on $\Phi_0$, this is illusory since the
components of $\Phi_0$ are modified by the redefinition
$\Phi_0 \rightarrow \Phi_0 e^{F/3}$ for chiral $F$ under
which the first-order action is invariant.) Since a linear
superfield $L$ can be written as
$L = \CCD^\alpha \BCCD^2 \Omega_\alpha + \hc$ for
$\Omega_\alpha$ with $\Delta=1/2$ and $w=-1$, an action of
the form $L Z$ has an $L$ equation of motion
which sets $Z = S + \bar S$ for chiral field $S$ of vanishing
conformal weight. Thus varying $L$ gives
$X = \Phi_0 \bar \Phi_0$, up to chiral and antichiral
fields which can be absorbed into a redefinition of
$\Phi_0$. This in turn restores the original action.
On the other hand, we may vary $X$ to conclude $X=L$,
which gives the action
\beq
-3 \cdint \left(L - L \log (L / \Phi_0 \bar\Phi_0) \right)
	= \cdint L V_R
\eeq
where we have defined $V_R \equiv 3\log(L / \Phi_0 \bar \Phi_0)$
and dropped the term linear in $L$ since a linear superfield
has vanishing D-term. $V_R$ is a scalar field with
vanishing conformal and chiral weights, although it does possess
a symmetry $V_R \rightarrow V_R - F - \bar F$ with chiral field $F$.

The prior gauge choice $\Phi_0=\bar\Phi_0=1$ which gave
a properly normalized Einstein-Hilbert term here corresponds to
$L=1$. Choosing this gauge gives the simple action $\dint V_R$
where $V_R = -3 \log(\Phi_0 \bar\Phi_0)$. It is fairly simple to see
now what sort of object $V_R$ is. Since we have gauge-fixed
the scale symmetry in addition to fixing $B=0$, the structure
group of our space differs only from Poincar\'e supergravity by
the presence of a $U(1)$ R-symmetry. These fields $\Phi_0$ and
$\bar\Phi_0$ are covariantly chiral with respect to a derivative
containing the corresponding $U(1)$ connection. Any $U(1)$ theory
of covariantly chiral superfields $\Phi$ ($\CD_\dalpha \Phi=0$)
may be related to a theory with Einstein chiral superfields $\phi$
($D_\dalpha \phi = {E_\dalpha}^M \partial_M \phi$) and a $U(1)$
prepotential $V$,
\[
\bar\Phi \Phi \rightarrow \bar\phi e^{-V/3} \phi
\]
By choosing $F$ appropriately, one may eliminate $\phi$, arriving
at $V_R = V$.

While this is the simplest explanation for what $V_R$ is, it is
somewhat unsatisfying since throughout this paper we have avoided
discussing prepotentials. To arrive at the some point by a rather
more circuitous route, one begins by partially fixing the $U(1)$ gauge
which at the moment is still a full superfield symmetry. We choose
$\Phi_0 = \bar\Phi_0$; that is, set their relative
phase to zero. The symmetry $\Phi_0 \rightarrow \Phi_0 e^{F/3}$
must be compensated with a chiral rotation with parameter
$\Omega = \frac{i}{4} (F - \bar F)$. We have now fixed the
unconstrained $U(1)$ parameter to the imaginary part of
a chiral parameter, and we see immediately that
$V_R$ transforms suspiciously as if it were the prepotential
of such a chiral version of R-symmetry.
If we evaluate the spinorial derivatives of $V_R$, we find
this is exactly so. Begin with
\begin{align*}
\CD_\alpha V_R = -3 \frac{1}{\Phi_0} \CD_\alpha \Phi_0
	= -3\frac{D_\alpha \Phi_0}{\Phi_0} + 2i A_\alpha
\end{align*}
and then note that since as functions $\Phi_0 = \bar\Phi_0$,
\[
D_\alpha \Phi_0 = D_\alpha \bar\Phi_0 = -\frac{2i}{3} A_\alpha \bar \Phi_0
	= -\frac{2i}{3} A_\alpha \Phi_0
\]
where we have used the chirality condition of $\bar\Phi_0$. It follows that
\begin{align}
\CD_\alpha V_R = 4 i A_\alpha, \;\;\;
\CD_\dalpha V_R = -4i A_\dalpha.
\end{align}
$V_R$ plays here the role of the $U(1)$ R-symmetry
prepotential, and so the term $\dint V_R$ is nothing more than the
$U(1)$ Fayet-Iliopolous term.

From our point of view, evaluating the D-term of $V_R$ is
particularly easy. One considers $V_R$ in its original form involving
$\Phi$. One can evaluate the D-term component Lagrangian directly.
After integrating a number of terms by parts, one arrives at\footnotemark
\begin{align}
e^{-1} \dint V_R =& \frac{1}{2} \CD^\alpha X_\alpha - \frac{i}{2} (\psi_m \sigma^m)_\dalpha X^\dalpha
	- \frac{i}{2} (\bar\psi_m \bsigma^m)^\alpha X_\alpha \eol
	& + (A_p + \frac{3}{2} e_p{}^c G_c)
		\times \left(-4 G^b e_b{}^p + i\eps^{pnm\ell} (\psi_n \sigma_m \bar\psi_\ell)\right)
\end{align}
The combination $A_p + \frac{3}{2} e_p{}^c G_c$ can be thought of as the $U(1)$ connection
if one chooses to define the bosonic derivative so that $F_{\alpha \dalpha}$ vanishes.
(Recall that $F_{\alpha \dalpha} = -3i G_{\alpha \dalpha}$ in our convention.)
\footnotetext{The calculation of this total expression can be simplified by noting that any terms
which shift under the chiral transformation of $\Phi$, such as $\CD_\alpha \log\Phi$,
must have vanishing coefficients.}

Using the definition for the lowest component of $G_b$, one finds
\begin{align}
e^{-1} \dint V_R =& \frac{1}{2} \CD^\alpha X_\alpha - \frac{i}{2} (\psi_m \sigma^m)_\dalpha X^\dalpha
	- \frac{i}{2} (\bar\psi_m \bsigma^m)^\alpha X_\alpha \eol
	& - \eps^{pnm\ell} (A_p + \frac{3}{2} e_p{}^c G_c) \partial_n b_{m\ell}
\end{align}
The Einstein-Hilbert action will be contained within $\CD^\alpha X_\alpha$ and the
Rarita-Schwinger action within the terms involving $X_\alpha$ and $X^\dalpha$.
The remaining term, while involving the gauge potential $A_p$ directly, is gauge invariant
when integrated by parts.

Recall that $\CD^\alpha X_\alpha$ is as defined in $U(1)$ superspace \cite{bgg}
and obeys the equality
\[
\CD^2 R + \BCD^2 \bar R = -\frac{2}{3} {R_{ba}}^{ba} - \frac{2}{3} \CD^\alpha X_\alpha
	+ 4 G^a G_a + 32 R \bar R
\]
Since $R=0$, this equation serves to define
\begin{align*}
\frac{1}{2} \CD^\alpha X_\alpha &\equiv - \frac{1}{2} R_{ba}{}^{ba} + 3 G^a G_a \\
	&= -\frac{1}{2} \mathcal R - i (\psi_b \sigma_a T^{ab}) - i (\bar\psi_b \bsigma_a T^{ab})
	- \frac{i}{2} \eps^{k \ell m n} G_k \psi_\ell \sigma_m \bar \psi_n + 3 G^a G_a
\end{align*}
Using $(\psi_m \sigma^m \bar X) = -2 (\psi_m \sigma^m \bsigma^{cb} T_{cb})$ and its conjugate,
it is straightforward to derive
\begin{align*}
e^{-1} \dint V_R =& -\frac{1}{2} \mathcal R
	+ \frac{1}{2} \eps^{k\ell m n} (\bar\psi_k \bsigma_\ell \CD_m' \psi_n)
	- \frac{1}{2} \eps^{k\ell m n} (\psi_k \sigma_\ell \CD_m' \bar\psi_n)
	- \eps^{p n m \ell} A_p' \partial_n b_{m \ell}
\end{align*}
where
\[
A_m' \equiv A_m + \frac{3}{4} e_m{}^a G_m
\]
and $\CD'$ is defined with $A'$ as its $U(1)$ connection. (This latter definition
corresponds to choosing $F_{\alpha \dalpha} = -\frac{3i}{2} G_{\alpha \dalpha}$ in
defining the bosonic derivative.)

In pure new minimal supergravity, the equation of motion of the two-form enforces
the $A'$ connection to (at least locally) be pure gauge, $A' = d \lambda$.
The $A'$ equation of motion on the other hand gives
\[
0 = \eps^{k \ell m n} \left(\partial_\ell b_{mn} + i \psi_\ell \sigma_m \bar\psi_n \right)
\]
Aside from the coupling of the gravitino to the field $A'$, the auxiliary sector
is that of a simple abelian BF model with topological action $\int b \wedge d A'$
and no propagating degrees of freedom.

\subsubsection{New minimal supergravity coupled to matter}
For reference, we include here the simplest couplings of new minimal
supergravity to chiral matter of vanishing $U(1)_R$ charge. (This last
condition forbids a superpotential, so these models are quite simple ones.)
One can derive these by performing a duality transformation from the K\"ahler multiplet,
where $\Psi_0$ is covariantly chiral with respect to a $U(1)_K$. The
modification consists simply of exchanging $\Phi_0$ with $\Psi_0$ in the definition
of $V_R$, which essentially shifts $V_R$ to $V_R + K$. The kinetic matter coupling
of new minimal supergravity is then
\begin{align}
\dint K
\end{align}
as in global supersymmetry. Evaluating this is straightforward. One simply
replaces $X_\alpha$ and $A_m$ associated with $V_R$ with $X_\alpha^K$ and
$A_m^K$. Provided we make the definitions
\begin{align}
X_\alpha^{K} = -\frac{1}{8} \BCD^2 \CD_\alpha K, \;\;\;
X_\dalpha^{K} = -\frac{1}{8} \CD^2 \BCD_\dalpha K
\end{align}
and
\begin{align}
A_m^K = -\frac{1}{2} e_m{}^a \Delta_a K
	+ \frac{i}{4} \psi_m{}^\alpha \CD_\alpha K
	- \frac{i}{4} \psi_m{}_\dalpha \BCD^\dalpha K
\end{align}
one finds
\begin{align}
\dint K = -\frac{1}{2} \CD^\alpha X_\alpha^K + \frac{i}{2} (\psi \sigma \bar X^K)
	+ \frac{i}{2} (\bar\psi \bsigma X^K)
	+ \frac{1}{2} \eps^{k \ell m n} A_k^K \partial_\ell b_{mn}
\end{align}
Unlike in old minimal supergravity, the presence of a K\"ahler potential
does not lead to extra additions to the Einstein-Hilbert term. This is known
to be altered when the chiral matter carries a $U(1)_R$ charge (see for example
\cite{Ferrara:1988qxa}).


\newpage
\section{Conclusion}\label{con}
We have constructed the fully conformal superspace and found its
nonvanishing curvatures to be uniquely described in terms of a single
superfield $W_{\alpha \beta\gamma}$. This is an unsurprising result
since it was long known at the linearized level. Similarly we have
demonstrated how $U(1)$ superspace is related to a certain
gauge of the full conformal superspace; this too is unsurprising,
as it was anticipated in \cite{Muller:1985vg}. Finally, the
various Poincar\'e formulations of superspace have been
explicitly constructed at the superspace level within the
conformal framework in a way more clearly, we believe,
than it had been done in the past. For example, the construction
of \cite{bgg} whereby the K\"ahler potential is absorbed into the
supergravity multiplet seems especially simple in this approach.

Beyond notational and theoretical elegance, is there anything
genuinely new this approach can offer? Perhaps. Noting that the
old minimal to K\"ahler frame conversion was reproduced here
quite easily, one may inquire whether 
in new minimal superspace there exists an analogous
absorbing of the K\"ahler potential into the frame
of superspace. For the simplest chiral models this is not
necessary since it turns out the coupling of chiral matter
does not affect the normalization of the Einstein-Hilbert term.
However, when non-chiral matter is considered, the story becomes more complicated.
In particular, one can show that the condition of $R=0$ in the new minimal model
is relaxed in the presence of certain types of matter, just
as $X_\alpha=0$ is broken in the minimal model when conversion to
the K\"ahler frame is undertaken. We hope to analyze this situation
further in the future.


\section*{Acknowledgments}
I am grateful to Jay Deep Sau, Ben Kain and Mary K. Gaillard for helpful
comments, discussions, and corrections. This work was supported in part by the Director, Office of Science, Office of High Energy and Nuclear Physics, Division of High Energy Physics of the U.S. Department of Energy under Contract DE-AC02-05CH11231, in part by the National Science Foundation under grant PHY-0457315.

\appendix
\section{Geometric preliminaries}

\subsection{Global spacetime symmetries}

The global structure of the conformal symmetry groups of arbitrary
manifolds (with or without torsion and Grassmann coordinates) benefits
from first discussing a simple example: the conformal group on
four dimensional Minkowski (or Euclidean) space. 

\subsubsection{The conformal group}\label{conf_gp}
The flat metric, $ds^2 = dx^m dx^n \eta_{nm}$, is preserved up to
a conformal factor by the differential generators\footnote{The convention used here
for the generators eliminates factors of $i$ in group elements while
making most of the generators anti-Hermitian.}
\begin{align}\label{calg1}
p_a &= \partial_a, \;\;\; &\left(1 + \xi \cdot p\right) x^m &= x^m + \xi^m \eol
m_{ab} &= -x_a \partial_b + x_b \partial_a, \;\;\;
	&\left(1 + \frac{1}{2} \omega^{ba} m_{ab}\right) x^m &= x^m - \omega^{mn} x_n \eol
d &= x \cdot \partial, \;\;\; &\left(1+\lambda d \right) x^m &= x^m + \lambda x^m \eol
k_a &= 2 x_a\, x\cdot \partial - x^2 \partial_a, \;\;\;
	&\left(1 + \epsilon\cdot k \right) x^m &= x^m + 2 (\epsilon \cdot x) x^m - x^2 \epsilon^m
\end{align}
The special conformal generator $k_a$ can also be thought of as
a spatial inversion, followed by a translation and then another
spatial inversion.

These generators are represented on fields by the operators
$P_a$, $M_{ab}$, $D$, and $K_a$ with the following algebra:
\begin{gather}
[M_{ab}, P_c] = P_a \eta_{bc} - P_b \eta_{ac}, \;\;\; 
[M_{ab}, K_c] = K_a \eta_{bc} - K_b \eta_{ac} \\{}
[M_{ab}, M_{cd}] = \eta_{bc} M_{ad} - \eta_{ac} M_{bd} - \eta_{bd} M_{ac} + \eta_{ad} M_{bc}\\{}
[D,P_a] = P_a, \;\;\;
[D,K_a] = -K_a \\{}
[K_a, P_b] = 2 \eta_{ab} D - 2 M_{ab}
\end{gather}
where all other commutators vanish.
The action of such generators on fields is defined by their
action at the origin. One usually takes for conformally primary
fields $\Phi$,
\begin{equation}
P_a \Phi(0) = \partial_a \Phi(0), \;\;\;
M_{ab} \Phi(0) = \mathcal S_{ab} \Phi(0), \;\;\;
D \Phi(0) = \Delta \Phi(0), \;\;\;
K_a \Phi(0) = 0
\end{equation}
Here $\mathcal S_{ab}$ is a differential rotation matrix
appropriate for whatever representation of the rotation group
$\Phi$ belongs to, $\Delta$ is the conformal scaling
dimension, and the vanishing of $K_a$ is called the
\emph{primary} condition. In order to discern the
transformation rules at points beyond the origin, one must
make use of the translation operator $e^{x\cdot P}$
to translate from the origin. This is formally a Taylor expansion:
\begin{align*}
\Phi(x) &= e^{x \cdot P} \Phi(0)
	= \Phi(0) + x^a P_a \Phi(0) + \frac{1}{2} x^a x^b P_a P_b \Phi(0) + \ldots \\
	&= \Phi(0) + x^a \partial_a \Phi(0) + \frac{1}{2} x^a x^b \partial_a \partial_b \Phi(0) + \ldots
\end{align*}
The operator $P_a$ acts only on the field $\Phi$, returning its derivative,
and has no action on the coordinate $x$, which is here just a parameter.
The same is true for the other operators.

If $g$ is any generator of the conformal algebra, the action of $g$
on $\Phi(x)$ can be calculated easily by making use of the
translation operator:
\begin{align}
g \Phi(x) = e^{x\cdot P} e^{-x \cdot P} g e^{x \cdot P} \Phi(0)
	\equiv e^{x \cdot P} \tilde g(x) \Phi(0) 
\end{align}
where $\tilde g(x) \equiv e^{-x\cdot P} g e^{x \cdot P}$ is an
abbreviated notation for the translated $g$. It follows that
\begin{gather}
\tilde P_a(x)  = P_a, \;\;\; \tilde D(x) = D + x^a P_a, \;\;\; \tilde M_{ab}(x) = M_{ab} - x_{[a} P_{b]} \eol
\tilde K_a(x) = K_a + 2 x_a D - 2 x_b M_{ab} + 2 x_a x_b P_b - x^2 P_a
\end{gather}
If these operators are taken to act on a pure function, they
reproduce the derivative representations \eqref{calg1}. It
should be noted that the algebra of the derivative representations
differs by a sign from the algebra of the field representations;
the former can be thought of as a left action on the group manifold
with the latter corresponding to a right action which yields an
opposite sign in the commutator.

On a more general field these expansions involve extra terms appropriate
for $\Phi$'s representation. For a primary field,
\begin{gather}
D \Phi(x) = \Delta \Phi + x^a \partial_a \Phi, \;\;\; 
M_{ab} \Phi(x) = \mathcal S_{ab} \Phi(x) - x_{[a} \partial_{b]}\Phi(x) \eol
K_a \Phi(x) = \left(2 x_a \Delta - 2 x_b \mathcal S_{ab} 
	+ 2 x_a x_b \partial_b - x^2 \partial_a \right) \Phi(x)
\end{gather}
The algebraic relations are simply applied. For example,
\[
D P_a \Phi(x) = [D, P_a] \Phi(x) + P_a \left(\Delta + x^b P_b \right)\Phi(x)
	= (\Delta + 1) P_a \Phi(x) + x^b P_b P_a \Phi(x)
\]
from which one can define the intrinsic scaling dimension of $\partial_a \Phi(x)$
as $\Delta+1$. Similarly can one determine the behavior of the Lorentz rotation 
and special conformal generators:
\begin{align}
M_{bc} P_a \Phi(x) =& \left(\mathcal S_{bc} \delta_a^d + \eta_{a[c} \delta_{b]}^d \right) \partial_d \Phi(x)
 		-  x_{[b}\partial_{c]} \partial_a \Phi(x) \eol
	= & \mathcal S'_{bc} \partial_a \Phi(x) -  x_{[b}\partial_{c]} \partial_a \Phi(x) \\
K_b P_a \Phi(x)
	= & \left(2 \eta_{ba} \Delta - 2 \mathcal S_{ba} \right) \Phi(x)
		+ 2 x_b (\Delta+1) \partial_a \Phi(x) \eol
		&- 2 x_c \left(\mathcal S_{bc} \delta_a^d + \eta_{a[c} \delta_{b]}^d\right) \partial_d \Phi(x)
		+ \left(2 x_b x_c \partial_c - x^2 \partial_b\right) \partial_a \Phi(x) \eol
	= & \kappa_{ba} \Phi(x) + \left(2 x_b \Delta' - 2 x_c \mathcal {S'}_{bc} 
	+ 2 x_b x_c \partial_c - x^2 \partial_b \right) \partial_a \Phi(x)
\end{align}
Both have precisely the forms expected, where $\Delta'$ and ${S'}_{bc}$
are the conformal dimension and rotation matrix appropriate for
$\partial_a \Phi(x)$. The only interesting
feature is that the special conformal generator \emph{removes}
the derivative; at the origin,
$K_b P_a \Phi(0) = \kappa_{ba} \Phi(0) = \left(2 \eta_{ba} \Delta - 2 \mathcal S_{ba} \right) \Phi(0)$.
This same feature is found in the local theory.

The conformal group action we've discussed above involves
transformations only on the fields, leaving the coordinate invariant.
That is, the action of a differential generator $g$ is
\beq
x \rightarrow x, \;\;\; \Phi \rightarrow \Phi'(x) = \Phi(x) + g\Phi(x)
\eeq
If we begin with the action $S = \int d^4x \; \Lag$ (with the Lagrangian
a function of fields and perhaps also the coordinate),
the action of $g$ is only on the fields:
\begin{equation}
\delta_g S = \int d^4x \; \left(\frac{\delta \Lag}{\delta \Phi} g \Phi
	+ \frac{\delta \Lag}{\delta \partial_a \Phi} g \partial_a \Phi \right)
\end{equation}
For the case where $g = \xi \cdot P$, one finds $g \Phi = \xi\cdot\partial\Phi$
and $g \partial_a \Phi = \xi \cdot \partial \partial_a \Phi$. The term in
parentheses is then equivalent to $\frac{d\Lag}{dx} - \frac{\partial \Lag}{\partial x}$.
The first term vanishes as a total derivative; the second must also vanish,
which tells that the Lagrangian cannot contain an explicit dependence on the
coordinate. For the other choices of $g$, the obvious results are recovered:
the Lagrangian must have $\Delta=4$, it must be a Lorentz scalar, and
it must be conformally primary. The simplest conformal action
involving a single primary scalar field of dimension one is
$\Lag = \phi \partial^2 \phi/2 - a \phi^4$. (The only non-trivial
check is to ensure the kinetic term vanishes at the origin
under the action of the special conformal generator.)

The approach outlined above has the feature that it places all the
transformation into the fields themselves. One often finds reference
to a formalism where both the coordinates and the fields transform:
\beq
x \rightarrow x', \;\;\; \Phi(x) \rightarrow \Phi'(x')
\eeq
For example, under translations and finite scalings, one would have
\begin{align}
&x \rightarrow x'=x-a, \;\;\; \Phi(x) \rightarrow \Phi'(x') = \Phi(x) \\
&x \rightarrow x'=e^{-\lambda} x, \;\;\; \Phi(x) \rightarrow \Phi'(x') = e^{\Delta\lambda}\Phi(x)
\end{align}
The part of $g$ which acts as a coordinate
shift has been moved off the fields and onto the coordinate
explicitly; the remaining action of $g$ can be thought of as
a generalized rotation operation, which vanishes if the
field $\Phi$ is a pure function.
The main reason this approach is employed is that it allows
conformal transformations on scalar fields (but \emph{only} scalar
fields) to be compactly written
\beq
x\rightarrow x', \;\;\; \phi(x) \rightarrow \phi'(x') = \left|\frac{\partial x'}{\partial x}\right|^{-\Delta/4} \phi(x).
\eeq
where $\Delta$ is the conformal scaling dimension of $\phi$.
Invariance of the action can then be checked in
one step for all the elements of the conformal group. The 
$\phi^4$ term, for example, transforms as
$
\int d^4x \; \phi(x)^4 \rightarrow \int d^4x' \phi'(x')^4 = \int d^4x J J^{-\Delta} \phi(x)^4
$
where $J = \left|\partial x' / \partial x \right|$. Invariance
is found for $\Delta=1$.

\subsubsection{Constant torsion}
We will ultimately be concerned with a theory containing torsion,
so it is useful to review the effects torsion induces.
Assume the manifold possesses translation
generators $P_a$ with nontrivial (but constant) torsion:
$[P_a,P_b] = -{C_{ab}}^c P_c$.
All other points $x$ relative to the priveleged origin are defined by
the condition $f(x) = e^{x\cdot P} f(0)$
for pure functions $f$.\footnote{The index contraction $x \cdot P$
should be understood as $x^m {\delta_m}^a P_a$. We will
shortly discover a nontrivial vierbein arising from the torsion,
but it does not appear in the translation group element.}
By Taylor's theorem, the $P_a$ in the exponent is playing
the same role as $\partial_a$ and so they are equivalent when
evaluated on the function at the origin. However, since the
$P_a$ do not commute, the operator $e^{x \cdot P}$ acting
on a function $f(y)$ does \emph{not} return $f(x+y)$ since
$e^{x \cdot P} e^{y \cdot P} \neq e^{(x+y) \cdot P}$.

Now let $\Phi$ be a \emph{field} valued on the manifold.
All covariant fields $\Phi$ are simple representations of the
translational isometries, obeying
$\Phi(x) = e^{x\cdot P} \Phi(0)$.
There are three reasonable but inequivalent notions of differentiation,
which we denote the normal, left, and right differentiation:
\begin{align}
\partial_a \Phi(x) &\equiv \frac{\partial}{\partial x^a} \left[e^{x\cdot P} \Phi(0)\right] \\
D_a^{(L)} \Phi(x) &\equiv P_a e^{x\cdot P} \Phi(0) \\
D_a^{(R)} \Phi(x) &\equiv e^{x\cdot P} P_a \Phi(0)
\end{align}
In each of these definitions, the operation on the left is some
sort of derivative on the group translation element $e^{x\cdot P}$
of the general form
\beq
D_a^{(L)} = {{e^{(L)}}_a}^m(x) \partial_m,\;\;\;
D_a^{(R)} = {{e^{(R)}}_a}^m(x) \partial_m,\;\;\;
\eeq
where $\partial_m$ is to be understood as a derivative on the group
parameters $x^m$ and ${{e^{(L)}}_a}^m(x)$ and ${{e^{(R)}}_a}^m(x)$ are
functions of $x$ chosen so that the definitions are satisfied. They
are found most easily by differentiating with respect to $x$
and moving all the $P$'s to the left or to the right:
\[
\partial_m e^{x \cdot P} = {{e^{(L)}}_m}^a(x) P_a e^{x \cdot P}, \;\;\;
\partial_m e^{x \cdot P} = e^{x \cdot P} {{e^{(R)}}_m}^a(x) P_a
\]

It is interesting to note the group commutation rules of these
various derivative operations, which follow directly from their
definitions. The normal differentiation has trivial commutator,
$[\partial_a, \partial_b]=0$, since these operations are simply
derivatives of their parameter. Left differentiation is not so
straightforward. First consider the product of two such operations:
\beq
D_a^{(L)} D_b^{(L)} \Phi(x)
= D_a^{(L)} P_b e^{xP} \Phi
=  P_b D_a^{(L)} e^{xP} \Phi
=  P_b P_a e^{xP} \Phi
\eeq
Since $D_a^{(L)}$ acts only on the translation generator
as a series of derivatives on its parameters, it passes through the
group generators. Here the order of operations has
reversed, which reverses the sign of the commutator:
\beq
[D_a^{(L)}, D_b^{(L)}] \Phi(x) = [P_b,P_a] e^{xP} \Phi =
+{C_{ab}}^c D_c^{(L)} \Phi(x)  
\eeq
A similar calculation with the right differentiation operators
shows that they preserve the order, and we find
\beq
[D_a^{(R)}, D_b^{(R)}] \Phi(x) = -{C_{ab}}^c D_c^{(R)} \Phi(x) 
\eeq

The left and right derivatives formally commute with each other
since they naturally place their corresponding $P_a$ generators on
opposite sides of the translation group element:
\beq
D_a^{(L)} D_b^{(R)} e^{x\cdot P} \Phi = D_a^{(L)} e^{x\cdot P} P_b \Phi
	= P_a e^{x\cdot P} P_b \Phi = D_b^{(R)} D_a^{(L)} e^{x\cdot P} \Phi 
\eeq

While each of these is interesting, only the right derivative
is translationally covariant:
\beq
e^{x\cdot P} D_a^{(R)} \Phi(x_0) = e^{x\cdot P} e^{x_0 \cdot P} P_a \Phi = D_a^{(R)} \Phi(e^{x\cdot P} x_0).
\eeq
(It is a straightforward exercise to show that the other derivative
operations do not obey this rule unless torsion vanishes.)
Therefore we may identify $D_a^{(R)} \equiv D_a$ as the covariant
derivative, and ${{e^{(R)}}_a}^m \equiv {e_a}^m$ as the physical vierbein.
It can be easily calculated by noting
\[
{e_m}^a P_a \equiv e^{-x\cdot P} \partial_m e^{x\cdot P}
\]
The result is\footnote{This result can be generalized in the presence
of local curvatures; see Appendix \ref{normal_gauge}.}
\beq
{e_m}^a = {\delta_m}^a - \frac{1}{2} x^b {C_{mb}}^a + \frac{1}{3!} x^b x^c {C_{mb}}^d {C_{dc}}^a + \ldots
\label{Vbn}
\eeq
where the $C$'s are understood to all possess Lorentz indices.
(That is, the only vierbein in the expression is on the left hand side,
and so this is an explicit, if unclosed, expression for the vierbein.)
The above expansion can be written in a matrix form.
Define the function $f(u) = (e^u-1)/u$; then
$e = f(x C)$ where ${(x C)_a}^b \equiv x^c {C_{ca}}^b$. It
follows that the inverse vierbein can be expanded using the reciprocal:
\beq
{e_a}^m = {(1/f(xC))_a}^m = {\delta_a}^m
	+ \frac{1}{2} x^b {C_{ab}}^m + \frac{1}{12} x^b x^c {C_{ab}}^d {C_{dc}}^m
	+ \ldots
\eeq
This relation for the vierbein can be shown to obey
$\partial_{[n} {e_{m]}}^a = {e_n}^c {e_m}^b {C_{cb}}^a$
which shows that the torsion ${T_{nm}}^a$, in this flat case,
is given in the Lorentz frame by the coefficients ${C_{cb}}^a$.

The above formalism is necessary in order to describe global supersymmetry
in superspace. Begin with a Grassmann manifold with four bosonic dimensions $x^a$
and four fermionic dimensions $\theta^\alpha$ and $\bar\theta_\dalpha$.
The translation isometries consist of the bosonic translations $P_a$ and the
fermionic ones $Q_\alpha$ and $\bar Q^\dalpha$, with a torsion term
$\{Q_\alpha, Q_\dalpha\} = -2i {\sigma_{\alpha \dalpha}}^a P_a$.
The torsion term here is found in the \emph{anticommutator} of the
fermionic $Q$'s. It is useful to think of this anticommutator as just
a normal commutator but with fermionic objects; whenever fermionic objects pass
through each other, a relative sign is introduced, creating the
anticommutator from a commutator.

A superfield $\Phi(x,\theta,\bar\theta)$ is defined by the action at the origin:
$\Phi(x,\theta,\bar\theta) = e^{x\cdot P + \theta Q+ \bar\theta \bar Q} \Phi$.
Since $P$ commutes with $Q$ and $\bar Q$, this can be written
as $\Phi(x,\theta,\bar\theta) = e^{\theta Q + \bar\theta \bar Q} \Phi(x)$.
If we apply a theta derivative to this superfield, there are two avenues for simplification.
One is to move the $Q$ that is brought down all the way to the left, and the other is to move
it all the way to the right. These two calculations are straightforward and yield
\begin{align*}
\partial_\alpha \Phi(x,\theta,\bar\theta) &= \partial_\alpha e^{\theta Q + \bar \theta \bar Q} \Phi(x)
	=\left(Q_\alpha + i \sigma^a_{\alpha \dalpha} \bar \theta^\dalpha P_a \right) e^{\theta Q + \bar \theta \bar Q} \Phi(x) \\
	&= \left(D_\alpha^{(L)} + i \sigma^a_{\alpha \dalpha} \bar \theta^\dalpha P_a \right) \Phi(x,\theta,\bar\theta)
\end{align*}
and
\begin{align*}
\partial_\alpha \Phi(x,\theta,\bar\theta) &= \partial_\alpha e^{\theta Q + \bar \theta \bar Q} \Phi(x) =
	e^{\theta Q + \bar \theta \bar Q}  \left(Q_\alpha - i \sigma^a_{\alpha \dalpha} \bar \theta^\dalpha P_a \right) \phi \\
	&= \left(D_\alpha^{(R)} - i \sigma^a_{\alpha \dalpha} \bar \theta^\dalpha P_a \right) \Phi(x,\theta,\bar\theta)
\end{align*}
From these we see immediately that the various derivatives have the form
\begin{align}
\partial_\alpha \equiv \frac{\partial}{\partial \theta^\alpha}, \;\;\;
D_\alpha^{(L)} \equiv \partial_\alpha - i \sigma_{\alpha \dalpha}^m \bar \theta^\dalpha \partial_m, \;\;\;
D_\alpha^{(R)} \equiv \partial_\alpha + i \sigma_{\alpha \dalpha}^m \bar \theta^\dalpha \partial_m
\end{align}
Note that in the literature \cite{wb}, it is the right derivaive which is $D_\alpha$,
the supersymmetry-covariant derivative. The left derivative is often denoted
$Q_\alpha$ and represents the supersymmetry isometry (it preserves the form of
the vierbein), which is different from the supersymmetry-covariant derivative.
We will discuss this further in the general context \ref{glob_loc}.

\subsubsection{General case}
Let $\mathcal G$ consist of the full set of symmetry transformations
acting on fields on the manifold and $\mathcal H$ denote the subgroup spanned by all the elements
aside from translations.\footnote{When the operators are defined
by their action on the coordinates, one often finds $\mathcal H$
defined as the subgroup which leaves the origin invariant. The manifold
$M$ can therefore be viewed as the coset space $\mathcal G / \mathcal H$,
which is the starting point of the group manifold approach to this
same topic.}  In practice, these normally consist of
rotational, conformal, and any Yang-Mills transformations.

The instrinsic action of $G = \exp g$ on $\Phi$ is defined by $G \Phi(0)$,
its action at the origin. The action of $G$ elsewhere can always be reconstructed
using the translations:
\[
G \Phi(x) \equiv G e^{x\cdot P} \Phi(0) = e^{x\cdot P} \tilde G(x) \Phi(0)
\]
where $\tilde G(x) \equiv e^{-x\cdot P} G e^{x\cdot P}$.
The product group element $e^{x \cdot P} \tilde G$ can be rearranged
into a part depending on $P$ and an element of $\mathcal H$:
\beq\label{temp4}
G \Phi(x) = G e^{x\cdot P} \Phi(0) = e^{\tilde x \cdot P} H_G(x) \Phi(0)
\eeq
where $H_G(x) \in \mathcal H$. All of the translations
have been absorbed in a redefinition of $x \rightarrow \tilde x$.
On a pure function $f(x)$ this would give
$
G f(x) = f(\tilde x),
$
and so $\tilde x$ can be thought of as the action of $G$ induced
on $x$.

The differential version of \eqref{temp4} can be compactly written
\[
g \Phi(x) = e^{x\cdot P} \tilde g(x) \Phi(0)
	= e^{x \cdot P} \left(\xi_g^a(x) P_a + h_g(x)\right) \Phi(0)
\]
where we have separated $\tilde g(x)$ into a part $\xi_g$
consisting only of translation generators and a part $h_g(x)$
consisting only of generators from $\mathcal H$. This formula
can be further simplified by noting the first term involves
the covariant derivative:
\[
g \Phi(x) = \xi_g^a D_a \Phi(x) + e^{x\cdot P} h_g \Phi(0)
	= \xi_g^a {e_a}^m \partial_m \Phi(x) + e^{x\cdot P} h_g \Phi(0)
\]
The action of $g$ thus induces a shift in the coordinate from
$x^m$ to $\tilde x^m = x^m + \xi_g^a(x) {e_a}^m(x)$.

\subsection{Local (gauged) spacetime symmetries}\label{local_g}
In the preceding sections we have discussed the construction of
representations of spacetime symmetry groups which act on fields.
There were several unsatisfying elements to this treatment:
we had to choose a preferred point, the origin; there
existed two alternative methods of describing the transformations,
either as just transforming the fields or transforming the fields
and the coordinates; and there was no clear way to generalize
to local transformations.

Each of these objections can be answered by proceeding to a local
formulation for the manifold. Again let $\Phi(x)$ denote the field
$\Phi$ at the point $x$ on the manifold. Let the symmetry
group $\mathcal G$ consist of generators $X_A$.
The action of such symmetry transformations on a field $\Phi$
is local; they transform the field into other fields at the same spacetime point.
That is, $\delta_g \Phi(x) = g^A(x) X_A \Phi(x)$, where $g^A(x)$ is the
position-dependent transformation. Here we view $X_A$ as an operator and
the product $X_A \Phi$ as a single object. If instead we view $\Phi$
as a column vector in its appropriate representation, then $X_A \Phi$
can be identified as $t_A \Phi$ where $t_A$ is a matrix appropriate
to that representation. The latter objects $t_A$ are what are
normally considered in treatments of Yang-Mills.
It should be noted that their multiplication
rule is backwards from that of the operators. That is,
$X_A X_B \Phi = X_A (t_B \Phi) = t_B X_A \Phi = t_B t_A \Phi$
since the operator $X_A$ passes through the matrix $t_B$. It follows
that if the algebra of the operators is
\[
[X_A, X_B] = -{f_{AB}}^C X_C
\]
then the algebra of the matrices is $[t_A, t_B] = +{f_{AB}}^C t_C$.


The generators can be decomposed into the translation generators
$P_a$ (more precisely, the generators of parallel transport) and
the others $X_{\ul a}$. The existence of purely scalar, non-constant
fields annihilated by $X_{\ul a}$ implies that the commutator of
two such generators cannot give a $P$. In other words,
${f_{\ul c \ul b}}^a = 0$ by assumption. (Supersymmetry
in normal space violates this assumption since two internal
symmetries $Q$ anticommute to give a translation $P$. This is one
advantage of using superspace instead.)

Associated with each generator is a gauge connection ${W_m}^A$,
which can be similarly decomposed into the vierbein ${e_m}^a$ and
the others ${h_m}^{\ul a}$. This decomposition can be written
\beq
{W_m}^A X_A = {e_m}^a P_a + {h_m}^{\ul a} X_{\ul a}
\eeq
The nature of the connection is defined by its action on fields:
\beq
\Phi(x+dx) = (1 + dx^m {W_m}^A(x) X_A) \Phi(x)
\eeq
where $\Phi$ is a scalar on the manifold but possibly nontrivial
in the tangent space. (That is, it may possess Lorentz indices but no Einstein ones.)
This equation is equivalent to
\beq
\partial_m \Phi(x) = {W_m}^A X_A \Phi(x) = {e_m}^a P_a \Phi(x) + {h_m}^{\ul a} X_{\ul a} \Phi(x)
\label{DiffW}
\eeq
which can be read as defining the action of $P_a$ as that of the
covariant derivative:\footnote{$P_a$ is the operator which was frequently
denoted $\Pi_a$ in older literature, the \emph{kinematic} momentum, as opposed
to the canonical momentum.}
\beq
{e_m}^a P_a \Phi(x) = \CCD_m \Phi(x) = \left(\partial_m - {h_m}^{\ul a} X_{\ul a} \right) \Phi(x)
\eeq
Since the vierbein is generally invertible,
$P_a \Phi(x) = {e_a}^m \CCD_m \Phi(x) = \CCD_a \Phi(x)$.
Since $P_a$ is equivalent to the covariant derivative, the
algebra of the $P_a$'s generally develops additional local
elements corresponding to the various curvatures associated
with the manifold. That is, the statement
\[
[\CCD_c, \CCD_b] \Phi = -{R_{cb}}^A X_A \Phi
\]
becomes a property of the algebra itself,
$[P_c, P_b] = -{R_{cb}}^A X_A$.
This alteration of the algebra is the only formal consequence when
passing from a global to a local theory. In the language
of the algebra, ${f_{cb}}^A = {R_{cb}}^A$ become structure
\emph{functions} in a local theory and depend on the
value of the connections. We will see shortly how this comes
about.

Under a gauge transformation,
$
\partial_m (\delta_g \Phi)
	= (\delta_g {W_m}^A) X_A \Phi + {W_m}^A \delta_g X_A \Phi,
$
where $X_A \Phi$ is considered a single object, leading to
the gauge transformation of the connections,
\begin{align}
\delta_g {W_m}^A = \partial_m g^A + {W_m}^B g^C {f_{CB}}^A.
\end{align}

A finite gauge transformation is found by exponentiating an
element of the algebra. That is, for an element $G = \exp(g)$,
$\Phi(x) \rightarrow \Phi'(x) = G(x) \Phi(x)$. Here $G$ is
understood as a power series expansion in
$g = g^A t_A$ where the matrices $t_A$ act only on the fields $\Phi$.
The relation \eqref{DiffW} can also be straightforwardly integrated
using a path-ordered exponential in the matrix language:
\beq
\label{IntW}
\Phi(x) = \mathcal P \exp\left(\int_{x_0}^x W^A t_A\right) \Phi(x_0).
\eeq
This equation is strongly reminiscent of a Wilson line, but extended
to the full symmetry group of the tangent space. It can be
compactly written $\Phi(x) = U(x,x_0) \Phi(x_0)$ where
$U(x,x_0)$ is the path-ordered exponential.
A derivative yields $\partial_m \Phi(x) = {W_m}^A t_A \Phi(x) = {W_m}^A (X_A \Phi)(x)$.
Under a gauge transformation,
\begin{align}
\Phi(x) \rightarrow G(x) \Phi(x), \;\;\;
U(x,x_0) \rightarrow U'(x,x_0) = G(x) U(x,x_0) G(x_0)^{-1}
\end{align}
The integrated rule for the connections can be found by considering
$x$ vanishingly near to $x_0$:
\begin{align}
W(x) \rightarrow W'(x) = -G d G^{-1} + G W G^{-1}
\end{align}

In order for the relation \eqref{IntW} to be path-independent,
any path beginning and ending on the same point must
vanish, $U(x,x)=0$. This is equivalent to the condition that the formal gauge
curvature $\mathcal F^A = dW^A - W^B W^C {f_{CB}}^A$ vanishes.
It serves not as a restriction but as a definition of the covariant curvatures $R$.
An explicit calculation of $\mathcal F$ using $[P_c, P_b] = -{R_{cb}}^A X_A$ yields
\begin{equation}
R^A = dW^A - e^b h^{\ul c} {f_{\ul c b}}^A
		- \frac{1}{2} h^{\ul b} h^{\ul c} {f_{\ul c\ul b}}^A
\end{equation}
as the relation between the covariant curvature (what we normally mean when
we say the ``curvature'') and the gauge fields.\footnote{This is the reverse of the usual approach,
where one simply defines the covariant derivative
and then calculates the curvatures. The condition $\mathcal F=0$ is then
nothing more profound than the commuting of the \emph{coordinate} derivatives.}

Under a $P$-gauge transformation, the vierbein varies as a
covariant Lie derivative:
\begin{align}
\delta_P(\xi) {e_m}^a &= \partial_m \xi^a + \xi^b {R_{bm}}^a - \xi^b {h_m}^{\ul c} {f_{\ul c b}}^a \eol
		&= \xi^n \CCD_n {e_m}^a + \partial_m \xi^n {e_n}^a
\end{align}
where $\xi^m \equiv \xi^a {e_a}^m$.
One recovers the normal Lie derivative by making corresponding gauge transformations
involving the gauge connections:
\begin{align}
\mathcal L_{\xi} {e_m}^a = \left\{\delta_P(\xi^m {e_m}^a) + \delta_H(\xi^m {h_m}^{\ul a}) \frac{}{} \right\} {e_m}^a
	= \delta_{GC} (\xi) {e_m}^a = \xi^n \partial_n {e_m}^a + \partial_m \xi^n {e_n}^a
\end{align}
This rule can be generalized to any function with Einstein indices.
Thus a gauge transformation with gauge parameter $\xi^m W_m$ is equivalent to
a Lie derivative on the field in question. This is precisely the behavior
expected of a diffeomorphism.

\subsubsection{Jacobi and Bianchi identities}\label{jac_bi}
The generators $X_A$ must obey the Jacobi identity:
\beq
0 = [X_C, [X_B, X_A]] + [X_A, [X_C, X_B]] + [X_B, [X_A, X_C]]
\eeq
Assuming this is obeyed for the global theory, the
consequences for the local theory are simple to derive.
Only terms involving the curvatures will differ,
so only two classes of Jacobi identity must be checked:
those with two $P$'s and a generator of $\mathcal H$ and those with three $P$'s.
Taking
\begin{align}
0 = [X_{\ul d}, [P_c, P_b]] + [P_b, [X_{\ul d}, P_c]] + [P_c, [P_b, X_{\ul d}]]
\end{align}
one finds
\begin{align}\label{eq_Rtrans}
X_{\ul d} {R_{cb}}^A = -{R_{cb}}^F {f_{F \ul d}}^A
	- {f_{\ul d [c}}^f {R_{f b]}}^A
	- {f_{\ul d [c}}^{\ul f} {f_{\ul f b]}}^A
\end{align}
The term involving two $f$'s can be eliminated using the
global Jacobi identity, giving\footnotemark
\begin{align}\label{eq_dRtrans}
X_{\ul d} {R_{cb}}^A = -{\Delta R_{cb}}^F {f_{F \ul d}}^A
	- {f_{\ul d [c}}^f {\Delta R_{f b]}}^A
\end{align}
where $\Delta R^A$ represents the difference between the curvature
in the local theory and in the global theory; in the cases we've
discussed, the only curvature in the global theory is the constant
torsion tensor $C$, so $\Delta {R_{cb}}^{\ul a} = {R_{cb}}^{\ul a}$,
but $\Delta {R_{cb}}^a = {T_{cb}}^a - {C_{cb}}^a$.
\footnotetext{This transformation rule can also be derived from the definition
of the $R$'s in terms of the gauge connections, but the above is the
more straightforward path.}

The case of the three $P$'s is also interesting. The rules found there
correspond to the Bianchi identities for the covariant derivative. They
read
\begin{align}
0 &= \sum_{[dcb]} \left(\CCD_d {T_{cb}}^a + {T_{dc}}^f {T_{fb}}^a + {R_{dc}}^{\ul f} {f_{\ul f b}}^a\right) \\
0 &= \sum_{[dcb]} \left(\CCD_d {R_{cb}}^{\ul a} + {T_{dc}}^f {R_{fb}}^{\ul a} + {R_{dc}}^{\ul f} {f_{\ul f b}}^{\ul a}\right)
\end{align}

\subsubsection{Gauge invariant actions over the manifold}\label{actions_m}
An action $S$ in four dimensions is the integral of a Lagrangian
density $\Lag(x)$ over the manifold using the general coordinate
invariant measure $d^4x \,e$. The invariance of the action under
a non-translational symmetry $g^{\ul b}$ relates the transformation
rule of $\Lag$ to that of $e$:
\beq
\delta_g S = \int d^4x \, e\, (\delta_g \Lag + \delta_g {e_m}^a {e_a}^m \Lag)
	= \int d^4x \, e\, (g^{\ul b} X_{\ul b} \Lag + g^{\ul b} {f_{\ul b a}}^a \Lag)
\eeq
One concludes $X_{\ul b} \Lag = -{f_{\ul b a}}^a \Lag$ as a condition
for invariance. One can now check invariance under a translational
symmetry $g^a = \xi^a$, using $\xi^a \CCD_a = \xi^m \CCD_m$:
\beq
\delta_P S = \int d^4x \; e \left({e_b}^n \xi^m \CCD_m {e_n}^b \Lag
	+ \partial_m \xi^m \Lag + \xi^m \CCD_m \Lag \right)
	= \int d^4x \partial_m (\xi^m e \Lag) = 0
\eeq
This is nothing more than the statement that $\delta_P$ is equivalent
to a general coordinate transformation followed by gauge transformations,
under which the action is inert.

A good example of the local approach is again offered by the
conformal group in four dimensions. The non-vanishing part of the
conformal algebra is
\begin{gather}
[M_{ab}, P_c] = P_a \eta_{bc} - P_b \eta_{ac}, \;\;\; 
[M_{ab}, K_c] = K_a \eta_{bc} - K_b \eta_{ac} \eol{}
[M_{ab}, M_{cd}] = \eta_{bc} M_{ad} - \eta_{ac} M_{bd} - \eta_{bd} M_{ac} + \eta_{ad} M_{bc}\eol{}
[D,P_a] = P_a, \;\;\;
[D,K_a] = -K_a \eol{}
[K_a, P_b] = 2 \eta_{ab} D - 2 M_{ab}
\end{gather}
Coupled to each of these generators is a gauge field,
\beq
W_m = {e_m}^a P_a + \frac{1}{2} {\omega_m}^{ba} M_{ab}
	+ b_m D + {f_m}^a K_a
\eeq
such that the action of $P_a$ on physical fields is the
covariant derivative; the other generators are defined by their
intrinsic behavior:
\beq
P_a \Phi = \CCD_a \Phi, \;\;\;
M_{ab} \Phi = \mathcal S_{ab} \Phi, \;\;\;
D \Phi = \Delta \Phi, \;\;\;
K_a \Phi = 0
\eeq
(If $\Phi$ possesses any Einstein indices, we separate them
out with the vierbein and treat only the Lorentz-indexed field as the
actual $\Phi$.) The difference between this and the approach
discussed in the global theory is that these are the behaviors
of the generators at \emph{all} points on the manifold. The
algebra of the generators allows one to calculate the
transformation behavior of any covariant derivative of $\Phi$ by
using the algebra. For example,
\begin{align}
D \CCD_a \Phi &= D P_a \Phi = (\Delta + 1) \CCD_a \Phi \\
K_b \CCD_a \Phi &= K_b P_a \Phi = \left(2 \eta_{ba} \Delta - 2\mathcal S_{ba}\right) \Phi \\
M_{bc} \CCD_a \Phi &= M_{bc} P_a \Phi = \left(\mathcal S_{bc} \delta_a^d + \eta_{a[c} \delta_{b]}^d \right) \CCD_d \Phi
\end{align}
Each of these generators acts locally with no derivative of its
parameter.

The above relations can also be checked using the explicit definition
of the covariant derivative. For that calculation, one would need the
transformation of the gauge connections. For completeness, consider the
arbitrary gauge parameter
\beq
\Lambda^A X_A = \xi^a P_a + \frac{1}{2} \theta^{ba} M_{ab} + \lambda D + \epsilon^a K_a
\eeq
Under a gauge transformation with such a parameter, the gauge connections
transform as
\begin{align}
\delta_G(\Lambda) {e_m}^a &=
	\partial_m \xi^a + \xi^b {\omega_{m b}}^a + \xi^a b_m
	+ \theta^{ab} e_{mb}
	- \lambda {e_m}^a \\
\delta_G(\Lambda) {\omega_m}^{ba} &=
	\partial_m \theta^{ba} + \theta^{[bc} {\omega_{mc}}^{a]}
	-2 \xi^{[b} {f_m}^{a]}
	-2 \epsilon^{[b} {e_m}^{a]}\\
\delta_G(\Lambda) b_m &=
	\partial_m \lambda
	+ 2 \xi^a f_{m a}
	- 2 \epsilon^a e_{ma}\\
\delta_G(\Lambda) {f_m}^a &=
	\partial_m \epsilon^a + \epsilon^b {\omega_{mb}}^a - \epsilon^a b_m
	+ \theta^{ab} f_{mb}
	+ \lambda {f_m}^a
\end{align}
Using these definitions, one can check, for example, that
$\delta_K(\epsilon) \CCD_a \Phi = \left(2 \epsilon_a \Delta - 2 \epsilon^b \mathcal S_{ba}\right) \Phi$
which agrees with the result from the algebra.

If an action $S$ in conformally invariant, the Lagrangian
must obey (using $X_{\ul b} \Lag = -{f_{\ul b a}}^a \Lag$)
\beq
D \Lag = 4 \Lag, \;\;\;
M_{ab} \Lag = 0, \;\;\;
K_a \Lag = 0
\eeq
just as in the global case.
Take as an example the standard $\phi^4$ theory. It is interesting
to note that the conventional way of writing the kinetic term,
$\CCD_a \phi \CCD_a \phi$, is not actually inert under the special
conformal transformations. Rather, one needs to use the covariant
d'Alembertian ($\CCD^a \CCD_a$) to give a gauge-invariant action:
\begin{equation}
S = \int d^4x \, e \left(\frac{1}{2} \phi \CCD^a \CCD_a \phi - a\phi^4 \right)
\end{equation}
It is straightforward to check that this action is inert under all
the gauge transformations. A more interesting question is to ask how the
kinetic action differs from the conventional form. A convenient
starting point is the identity
\beq
\partial_m (e {e_a}^m \phi \CCD^a \phi) 
	= \CCD_m \left(e {e_a}^m \phi \CCD^a \phi \right) + {f_m}^b  K_b \left(e {e_a}^m \phi \CCD^a \phi\right)
\eeq
which follows since the expression in the parentheses is invariant under
every gauge transformation except the special conformal one.
The above expression can be easily evaluated to give
\beq
\partial_m (e {e_a}^m \phi \CCD^a \phi) = 
	e \left(\CCD_a (\phi \CCD^a \phi) + {T_{ba}}^a \phi \CCD^b \phi + 2 {f_a}^a \phi^2\right)
\eeq
This allows one to integrate the action by parts:
\begin{equation}
S = \int d^4x \;e \; \left(\frac{1}{2} \phi \CCD^a \CCD_a \phi - a \phi^4 \right)
	= \int d^4x \; e\; \left(
		-\frac{1}{2} \CCD^a \phi \CCD_a \phi
		- \frac{1}{2} {T_{ba}}^a \phi \CCD^b \phi - {f_a}^a \phi^2 - a\phi^4\right)
\end{equation}
The trace of the torsion tensor usually vanishes in physically interesting
theories, but the term involving the $K$-gauge field ${f_m}^a$ is physically
of interest. In common theories of conformal gravity, it is related to
the Ricci tensor and its trace is proportional to the Ricci scalar.
In such theories, the Lagrangian above can be gauge fixed to yield the
Einstein-Hilbert Lagrangian. (The quartic, if present, would give a
cosmological constant.)

\subsubsection{Global representations from local ones}\label{glob_loc}
We have discussed two ways of implementing the spacetime symmetry group
on the fields. The first involved a selection of a privileged point, the
origin, at which we defined the intrinsic behavior of the fields; the
behavior elsewhere was then calculated by composing the group
element with the translation element. The action of group elements was
taken not only on the fields but also on the translation element, leading
to non-trivial transformation rules for the fields away from the origin.
The second way involved defining gauge connection 1-forms everywhere;
no privileged point was needed, nor
was there any discussion of moving points on the manifold. The advantage of
this latter formulation was that it was trivial to implement \emph{local}
group transformations.  The global structure should be represented by the
local one when restricted to global gauge transformations.

Begin with a vanishing $\mathcal H$-connection and a $P$-connection as defined
in \eqref{Vbn} relative to some origin point 0. Construct a
gauge transformation $\tilde g(x)$ which takes the value $g$ at the origin
but elsewhere is such as to keep the connections invariant.
That is, $\tilde g(x)$ obeys
\begin{equation}
0 = \delta_{\tilde g} {W_m}^A = \partial_m {\tilde g}^A + {e_m}^b {\tilde g}^C {f_{C b}}^A
\end{equation}
This equation can be integrated to give
$\tilde g(x) = e^{-x\cdot P} g e^{+x\cdot P}$
where $x\cdot P \equiv x^m {\delta_m}^a P_a$. To prove this is
correct, recall that to first order in $\xi$,
$1+\xi^m {e_m}^a P_a = e^{-x\cdot P} e^{(x+\xi)\cdot P}$.
It follows then that
\begin{align*}
-\xi^m {e_m}^b {\tilde g}^C {f_{C b}}^A = [\tilde g, \xi^m {e_m}^b P_b] 
	&= e^{-x\cdot P} g e^{(x+\xi)\cdot P} - e^{-x\cdot P} e^{(x+\xi)\cdot P} e^{-x\cdot P} g e^{x\cdot P} \\
	&= e^{-x\cdot P} g e^{(x+\xi)\cdot P} + e^{-(x+\xi)\cdot P} g e^{x\cdot P} - 2 \\
	&= \xi^m \partial_m \tilde g^A
\end{align*}
where the last two equalities hold only to first order in $\xi$.
This gauge transformation, $\tilde g(x)$, is the transformation discussed
in the global approach.

The general form of the locally invariant action
$S = \int d^4x \, e\, \Lag$ obeying
$X_{\ul b} \Lag = - {f_{\ul b a}}^a \Lag$
implies that the globally invariant form must also have that form.
In particular the global measure must be $d^4x\, e$ where
$e$ is nontrivial in the case of a sufficiently complicated (but constant) torsion.
(This is not normally an issue since even global supersymmetry has $E=1$.)
To prove this requirement, consider the global action
$S = \int d^4x \, e \, \Lag$. Under a global gauge transformation $\tilde g$,
the measure is invariant and the Lagrangian changes as
$\delta \Lag = \tilde g^{\ul b} X_{\ul b} \Lag + \tilde g^b P_b \Lag$.
We can first replace
$\tilde g^{\ul b} X_{\ul b}\rightarrow -\tilde g^{\ul b} {f_{\ul b a}}^a$
and then equate that quantity to ${e_a}^m \partial_m \tilde g^a + \tilde g^b {f_{ba}}^a$
using the differential equation for $\tilde g$. Finally note that
$\tilde g^b P_b \Lag = \tilde g^b {e_b}^m \partial_m \Lag$ and we find
\beq
\delta \Lag = {e_a}^m (\partial_m \tilde g^a) \Lag + \tilde g^b {f_{ba}}^a \Lag + 
	\tilde g^b {e_b}^m \partial_m \Lag
	= {e_b}^m \partial_m (\tilde g^b \Lag) + \tilde g^b {f_{ba}}^a \Lag
\eeq
Here by ${f_{ba}}^a$ we mean the trace of the torsion tensor, equivalently
written ${C_{ba}}^a$ or ${T_{ba}}^a$ (these are identical in the global
theory).
The first term can be integrated by parts (if the measure is $e$) to cancel the second,
rendering the action invariant.

The $\tilde g$'s discussed here represent the \emph{isometries} of the flat
space -- the transformations which leave invariant the form of the connections.
Of particular interest is the case where $g = g^a P_a$. There
we find that $\tilde g = \tilde g^a P_a$ (no $\mathcal H$ bits are generated
since the commutator of two $P$'s is another $P$ in the flat, ungauged space),
with the interesting property that $\tilde g^a$ preserves the form of the
vierbein. These are precisely the translation isometries of the space; that
is, they are the diffeomorphisms which preserve the vierbein. We may
write them as a coordinate transformation:
\beq
x^m \rightarrow x^m + \tilde g^a {e_a}^m, \;\;\;
e^a \rightarrow e^a, \;\;\; D_a \rightarrow D_a
\eeq
Recall that the vierbein used here was the one associated with
right differentiation. The action of left differentiation was an
isometry which preserved the form of the vierbein 1-form $e^a$
and the right derivative operator $D_a$. We have recovered this
isometry above; it represents the general form of the translation
isometry of a flat space with torsion.

\subsubsection{Normal gauge}\label{normal_gauge}
In general relativity, there exists a preferred gauge
for the metric, the choice of Riemann normal coordinates, which
expands the metric in terms of the curvature and derivatives thereof.
Similarly in Yang-Mills theories, there exists a preferred gauge,
the Fock-Schwinger gauge, which gives the gauge connection in
terms of the gauge curvature and derivatives thereof.
It is possible to generalize both of these conditions to the sort
of theory discussed here.

Recall that a field at a point $x$ is related to the field at
a fixed point $x_0$ by a Taylor expansion:
\begin{align}
\phi(x) &= \exp\left((x-x_0) \cdot \partial\right) \phi(x_0) \eol
	&= \phi(x_0) + (x-x_0)^m \partial_m \phi(x_0) 
		+ \frac{1}{2} (x-x_0)^m (x-x_0)^n \partial_n \partial_m \phi(x_0) + \ldots
\end{align}
On the other hand, the parallel transport of the field from $x_0$
with parameter $y$ is
\begin{align}
\phi(x_0; y) &= \exp\left(y^a P_a\right) \phi(x_0) \eol
	&= \phi(x_0) + y^a \CCD_a \phi(x_0) 
		+ \frac{1}{2} y^a y^b \CCD_b \CCD_a \phi(x_0) + \ldots
\end{align}
One can choose a gauge such that these coincide for $x=y+x_0$
for scalar fields; this generalizes Riemann normal coordinates
for non-Riemannian geometries (for example, those with torsion).
The further choice that these should coincide for
\emph{all} fields leads to a generalization of Fock-Schwinger
gauge.

In principle, one can equate these series term-by-term to
determine the gauge fields. A slightly simpler method is to
note that ${e_a}^m \partial_m \phi - {h_a}^{\ul b} X_{\ul b} \phi$
is the covariant derivative; therefore one may equate
\begin{align}
{e_a}^m(y) \frac{\partial}{\partial y^m} \exp\left(y^a P_a\right) \phi(x_0)
	- {h_a}^{\ul b}(y) \exp\left(y^a P_a\right) X_{\ul b} \phi(x_0)
	= \exp\left(y^a P_a\right) P_a \phi(x_0)
\end{align}
This can be rearranged to
\begin{align}
\frac{\partial}{\partial y^m} e^{y \cdot P} \phi(x_0)
	&= {e_m}^a e^{y \cdot P} P_a \phi(x_0)
	 + {h_m}^{\ul b}(y) e^{y \cdot P} X_{\ul b} \phi(x_0) \eol
	&= e^{y \cdot P} {\tilde e_m}{}^a P_a \phi(x_0)
	 + e^{y \cdot P} {\tilde h_m}{}^{\ul b}(y) X_{\ul b} \phi(x_0)
\end{align}
where we have defined ${\tilde e_m}{}^a$ and ${\tilde h_m}{}^{\ul b}$
by conjugation with $e^{y \cdot P}$. Multiplying by an overall
factor gives
\begin{align}
e^{-y \cdot P}\frac{\partial}{\partial y^m} e^{y \cdot P} \phi(x_0)
	&= {\tilde e_m}{}^a P_a \phi(x_0)
	 + {\tilde h_m}{}^{\ul b}(y) X_{\ul b} \phi(x_0)
\end{align}
The term on the left can be straightforwardly evaluated term by term:
\begin{align}
e^{-y \cdot P} \partial_m e^{y \cdot P}
	&= \partial_m + P_m + \frac{1}{2} [P_m, y^a P_a]
		+ \frac{1}{3!} L_{y\cdot P}^2 P_m 
		- \frac{1}{4!} L_{y\cdot P}^3 P_m + \ldots \eol
	&= \partial_m + P_m + \sum_{j=1}^\infty \frac{(-1)^j}{(j+1)!} Q_m(j)
\end{align}
where $L_{y\cdot P} f \equiv [y^a P_a,f] = y^a [P_a, f]$
and $Q_m(j) \equiv L_{y\cdot P}^j P_m$. In this expansion
the $y^a$ are to be treated as group parameters, inert under the action of
the generators, and the explicit derivative $\partial_m$ is with respect
to the $y$ only. One may formally solve for the gauge fields by defining
\begin{align}
{\tilde e_m}{}^a &= \delta_m^a + \sum_{j=1}^\infty \frac{(-1)^j}{(j+1)!} {Q_m}^a(j) \eol
{\tilde h_m}{}^{\ul b} &= \sum_{j=1}^\infty \frac{(-1)^j}{(j+1)!} {Q_m}^{\ul b}(j)
\end{align}
where we have expanded $Q_m = {Q_m}^a P_a + {Q_m}^{\ul b} X_{\ul b}$.
Then conjugating by the group element $\exp(y\cdot P)$ generates the
actual gauge fields:\footnotemark
\begin{align}
{e_m}^a &= \delta_m^a + \sum_{j=1}^\infty \frac{(-1)^j}{(j+1)!} \sum_{k=0}^\infty \frac{1}{k!} L_{y \cdot P}^k {Q_m}^a(j) \eol
{h_m}^{\ul b} &= \sum_{j=1}^\infty \frac{(-1)^j}{(j+1)!} \sum_{k=0}^\infty \frac{1}{k!} L_{y \cdot P}^k {Q_m}^{\ul b}(j)
\end{align}
Note that the conjugation generates covariant derivatives of the
listed terms; for example,
$L_{y \cdot P} {Q_m}^a(j) = y^b \nabla_b {Q_m}^a(j)$.

\footnotetext{In the case where there are no curvatures except for
constant torsion, the above reduce to ${h_m}^{\ul b}=0$ and
${e_m}^a$ given by \eqref{Vbn}. Normal gauge is the appropriate
generalization.}

\subsubsection{Gauge invariant actions over submanifolds}\label{actions_sm}
In the case of global supersymmetry, we know that it is natural to consider
not only integrals over the entire superspace of coordinates $(x,\theta,\btheta)$
but also integrals over a chiral superspace of coordinates $(y, \theta)$
where $y = x + i \theta \sigma \btheta$. It is natural to think of the chiral
superspace as lying on a submanifold characterized by a constant value of $\btheta$.
Then change in coordinates from $x$ to $y$ is naturally understood, since in
those coordinates $D^\dalpha = \partial^\dalpha$ and so chiral superfields
(those annihilated by $D^\dalpha$) naturally live on such a submanifold.

Let us take this point of view seriously and derive some useful results
about actions on submanifolds. We will assume that the space under
consideration is purely bosonic so that our geometric intuition can be trusted.
Let the full manifold $M$ be $D$-dimensional  on which we may define
the parallel transport operators $P_A$, where $A=1,\ldots, D$.
Let $P$ be decomposed as $P_A = (P_\chA, P_{\dalpha})$
where $\chA = 1, \ldots, \chD$ and $\dalpha = \chD+1, \ldots, D$.
We will use Gothic indices $\chA$ to denote the submanifold tangent
space indices. Our object of interest is a submanifold $\chM$ of
dimension $\chD$ defined so that $P_{\dalpha}$ annihilates the functions
naturally integrated over $\chM$.

This can be made more concrete by choosing coordinates
$z^M = (\chz^\chm, \btheta^{\dot \mu})$
so that $\chM$ is parametrized by $\chz^\chm$ with constant $\btheta^{\dot \mu}$;
we will assume $\btheta^{\dot \mu} = 0$ for definiteness, but any constant
will do. In this way the coordinates on $M$ can be related nicely to the coordinates on
$\chM$.\footnote{Of course $\btheta$ here is to be understood as a bosonic coordinate
at the moment.} Then the condition that $P_{\dalpha}$ annihilates the
natural integrands on $\chM$ means $P_{\dalpha} = \partial / \partial \btheta^\dalpha$
when acting on pure functions, or, equivalently, that $\chM$ lies at a constant slice of $\btheta^{\dot \mu}$.
This choice of coordinates has the benefit of simplifying calculations while unfortunately
forcing a breakdown in manifest general coordinate invariance on $M$;
equivalently, this forces one to choose a certain $P$-gauge.
We will therefore avoid making this explicit assumption until it is
absolutely necessary.


Recall that an invariant integral on the whole manifold $M$ is
\begin{align}
S = \int_M E^1 \wedge E^2 \wedge \ldots \wedge E^D \; V = \int d^Dz \,E \,V
\end{align}
where $E = \det({E_M}^A)$ and $V$ is an appropriate integrand to make the action
gauge invariant. We have already shown that invariance under the non-translation
symmetries $\mathcal H$ requires
$\delta_g V = -g^{\ul b} {f_{\ul b A}}^A$, while invariance under
$P$ follows from general coordinate invariance.
An invariant integral over $\chM$ can be very similarly defined:
\begin{align}
\chS = \int_{\chM} E^1 \wedge E^2 \wedge \ldots \wedge E^\chD \; W
	= \int_{\chM} \chE^1 \wedge \chE^2 \wedge \ldots \wedge \chE^\chD \; W
	= \int d^\chD\chz \, \chE \, W,
\end{align}
where $\chE = \det({\chE_\chm}^\cha)$ is the volume measure
and $W$ is an appropriate integrand.
The subvierbein form $\chE^\cha$ is taken to be identical to $E^\cha$
when restricted to the manifold $\chM$.\footnote{
In the special coordinates where $\chM$ corresponds to $\btheta=0$, the vierbein
obeys ${E_{\dmu}}^\cha \vert_{\chM} = 0$. This condition is equivalent to
the conditions $\chE^\cha = E^\cha\vert_{\chM} = d\chz^\chm {\chE_\chm}^\cha$.}
Invariance of this integral under the action of $\mathcal H$
requires $\delta_g W = -g^{\ul \chb} {f_{\ul b \cha}}^\cha W$.
(Note the trace of the structure constant is over the submanifold's
Lorentz indices.)
However, since the integral is over a submanifold, it is not obviously
taken into itself under $P$-gauge transformations.

We check first the requirement of $P_{\dalpha}$ invariance, which means
essentially that such actions should not depend on the constant value of
$\btheta$ used to define $\chM$. The action varies as
\beq
0 = \delta_\xi \chS = \int d^\chD \chz \,
	\chE \left(\xi^{\dalpha} {T_{\dalpha \chm}}^\chb {\chE_\chb}^\chm W +
	\xi^{\dalpha} \CCD_{\dalpha} W \right).
\eeq
(The term ${\chE_\chb}^\chm$ represents the inverse of the \emph{sub}vierbein.
It does not necessarily correspond to ${E_\chb}^\chm$, since the inverse of
a submatrix is not necessarily the submatrix of the inverse unless certain
requirements are placed on the coordinates $\chz$ being used for the submanifold, or
equivalently, the gauge choice for the vierbein.)
Each term should vanish separately. Requiring the second term to
vanish enforces the covariant constancy of $W$ in the direction of $P_{\dalpha}$.
Requiring consistency of $\CCD_{\dalpha} W = 0$ with the algebra gives
several additional constraints:
\begin{align}
0 &= [\CCD_{\dalpha}, \CCD_{\dbeta}] W =
	- {T_{\dalpha \dbeta}}^\chc \CCD_\chc W
	+ {R_{\dalpha \dbeta}}^{\ul c} {f_{\ul c \chd}}^\chd W \\
0 &= [X_{\ul a}, \CCD_{\dbeta}] W = 
	- {f_{\ul a \dbeta}}^\chc \CCD_\chc W
	+ {f_{\ul a \dbeta}}^{\ul c} {f_{\ul c \chd}}^\chd W \label{eq_Wcons}
\end{align}
(The second commutator vanishes since
$\CCD_\dbeta X_{\ul a} W = -\CCD_\dbeta {f_{\ul a \chb}}^\chb W = 0$.)
From this simple result we learn
${T_{\dalpha \dbeta}}^\chc = {f_{\ul a \dbeta}}^\chc = 0$
as well as 
${R_{\dalpha \dbeta}}^{\ul c} {f_{\ul c \chd}}^\chd = 
{f_{\ul a \dbeta}}^{\ul c} {f_{\ul c \chd}}^\chd = 0$.
The other term in the variation of the subaction gives
two new terms which must vanish:
\[
{T_{\dalpha \chm}}^\chb {\chE_\chb}^\chm = {T_{\dalpha \dgamma}}^\chb {E_\chm}^{\dgamma} {\chE_\chb}^\chm
	+ {T_{\dalpha \chb}}^\chb
\]
The first of these,
${T_{\dalpha \dgamma}}^\chb {E_\chm}^{\dgamma} {\chE_\chb}^\chm=0$,
is already a condition for the
existence of a covariantly constant $W$. The second, ${T_{\dalpha \chb}}^\chb=0$,
amounts to an additional constraint on the space.\footnote{
These constraints are stricter than necessary.
One could choose that
$\nabla_\dalpha W = -{T_{\dalpha \chm}}^\chb {\chE_\chb}^\chm W$,
as opposed to requiring each term to separately vanish. We have chosen
to separate them in the way we have since it makes sense that the
conditions we want should be simple conditions on $W$, like chirality,
and simple conditions on the geometry, like vanishing of certain torsions,
as opposed to something more complicated relating the two.}

Next we check $P_\cha$ invariance of the subaction. One finds
\beq
0 = \delta_\xi \chS = \int d^\chD\chz\, \chE \left(
	\CCD_\chm \xi^\cha {\chE_\cha}^\chm W +
	\xi^{\cha} {T_{\cha \chm}}^\chb {\chE_\chb}^\chm W + \xi^{\cha} \CCD_{\cha} W \right).
\eeq
Integrating the first term by parts gives
\beq
0 = \delta_g \chS = \int d^\chD\chz\, \chE \left(
	-\xi^\cha {\chE_\cha}^\chm \CCD_\chm W + \xi^{\cha} \CCD_{\cha} W 
	- \xi^\cha {\chE_\cha}^\chn {T_{\chn \chm}}^\chb {\chE_\chb}^\chm W
	+ \xi^{\cha} {T_{\cha \chm}}^\chb {\chE_\chb}^\chm W
	\right).
\eeq
Invariance holds under the same set of conditions. For example,
${\chE_\cha}^\chm \CCD_\chm W = {\chE_\cha}^\chm {E_\chm}^B \CCD_B W
	= {\chE_\cha}^\chm {E_\chm}^\chb \CCD_\chb W = \CCD_\cha W$
since $W$ is covariantly constant with respect to $P_{\dalpha}$
and ${E_\chm}^\chb$ is equivalent to ${\chE_\chm}^\chb$. A similar
argument demontrates the cancellation of the torsion terms.

The constraints we have found are:
\begin{gather*}
{T_{\dalpha \dbeta}}^\chc =0, \,\,\, {f_{\ul a \dbeta}}^\chc = 0 \\
{R_{\dalpha \dbeta}}^{\ul c} {f_{\ul c \chd}}^\chd = 0, \,\,\,
{f_{\ul a \dbeta}}^{\ul c} {f_{\ul c \chd}}^\chd = 0 \\
{T_{\dalpha \chb}}^\chb=0
\end{gather*}

The next question to consider is whether integrals over a manifold
$M$ can be related to integrals over the submanifold $\chM$, and
vice-versa. We will deal with $M\rightarrow \chM$ first and then
consider the reverse.

\begin{description}
\item[Case 1: $M\rightarrow \chM$] $\;$\newline
Consider the integration of a function $V$ over the whole manifold:
$
\int_M d^Dz \, E \, V.
$
We would like to decompose it into an integral of some other
function $W$ over the submanifold $\chM$. The most straightforward
way to do this is to adopt the coordinates (equivalently, choose the
$P$-gauge) so that $z^M = (\chz^\chm, \btheta^{\dmu})$ and $\chM$
corresponds to $\btheta=0$. Note that it is rather trivial to
choose ${E_{\dmu}}^\cha\vert_{\chM}=0$; it can be shown
that the conditions we derived for the invariance of the subactions
over $\chM$ allow us to extend this condition over all of $M$.\footnotemark
We then can assume a gauge choice where ${E_{\dmu}}^\cha = 0$
everywhere, as well as the additional requirements
${h_{\dmu}}^{\ul b} {f_{\ul b \cha}}^\cha = 0$. These two
conditions mean that $\CCD_{\dalpha} W = 0$ is equivalent to
$\partial_{\dmu} W = 0$.\footnotetext{The construction will be
given when needed for the explicit case of $\mathcal N=1$ superspace.}
Given these, one may easily show that $\chE$ is itself independent of $\btheta$:
\beq
\partial_{\dmu} \chE = \partial_{\dmu} {E_\chn}^\cha {\chE_\cha}^\chn
	= \CCD_{\dmu} {E_\chn}^\cha {\chE_\cha}^\chn = {T_{\dmu \chn}}^\cha {\chE_\cha}^\chn = 0
\eeq
This is important since the gauge choice for the vierbein implies
$E = \chE \bar\Sigma$, where $\bar\Sigma \equiv \det({E_{\dmu}}^{\dalpha})$.
Then $E$ separates into a part ($\chE$) independent of $\btheta$ and
another ($\bar\Sigma$) which is an appropriate density in $\btheta$.

Under these assumptions, we find
\begin{align}\label{DtoF}
\int_M d^Dz \, E \, V = \int_{\chM} d^\chD\chz \, \chE \, \CP[V]
\end{align}
where
\begin{align}\label{cp}
\CP[V] \equiv \int d^{\dot d} \btheta \, \bar\Sigma \, V.
\end{align}
Note that $\CP[V]$ is covariantly constant with respect to $P_{\dalpha}$
for a quite trivial reason: by construction, $\CP[V]$ is independent
of $\btheta$ and so $\partial_{\dmu} \CP[V] = 0$ in a gauge where
$\partial_\dalpha = \CCD_\alpha$. This operation
can be extended to any gauge by first evaluating it in the special gauge
used here and then transforming to the desired gauge using
$\delta_g \CP[V] = -g^{\ul b} {f_{\ul b \cha}}^\cha \CP[V]$.

\item[Case 2: $\chM\rightarrow M$] $\;$\newline
In principle an integral over a submanifold $\chM$ can be defined
by an integral over the whole manifold $M$ using an appropriate
delta function $\chDelta$. Then
\beq
\int_{\chM} d^\chD\chz \, \mathcal E \, W = \int_M d^Dz \, E \, W \chDelta
\eeq
That both sides remain gauge invariant under $\mathcal H$ implies
$\delta_g \chDelta = -g^{\ul b} {f_{\ul b \dalpha}}^{\dalpha} \chDelta$.
The simplest way to describe the constraints is to choose the
coordinates $z$ to decompose as $z^M = (\chz^\chm,\btheta^{\dmu})$
where the submanifold $\chM$ lives at $\btheta^{\dmu} = 0$.
In this special gauge, $\chDelta$ takes the simple form
\beq
\chDelta = \frac{\delta^{\dot d} (\btheta)}{\bar\Sigma}.
\eeq
This is not the only such $\chDelta$ that will work;
an entire family is permissible, of the form
\beq
\chDelta = \frac{X}{\CP[X]}.
\eeq
The choice $X = \delta^{\dot d} (\btheta)$ reproduces the simplest
example. If, however, $\CP[1]$ is a simple enough object, the
choice $X=1$ becomes extremely attractive. \footnote{The above construction
applies very nicely to Poincar\'e supergravity, where if one chooses $X=1$, one finds
$\Delta = 1 / 2 R$.}
\end{description}

That both of these results should hold implies
\beq
\int_M d^Dz \, E \, V 
	= \int_{\chM} d^\chD\chz \, \chE \, \CP[V]
	= \int_M d^Dz \, E \, \CP[V] \chDelta 
\eeq
Since $\chDelta$ can be placed in the form $X / \CP[X]$, the
equivalence of the first and third forms implies $\CP$ is a
self-adjoint operation under the full integration.

While it is self-adjoint, $\CP$ is not actually a projector, as it is
not idempotent (that is, $\CP^2 \neq \CP$). The true projector
(in the special gauge) is $\chPi$, which is defined by
\beq
\chPi[V] \equiv \int d^{\dot d} \btheta \,\bar\Sigma\, V \chDelta.
\eeq
This formula is a very complicated way of saying a simple thing:
$\chPi[V]$ is formally identical (in this gauge) to $V\vert_{\btheta=0}$
provided we use the simplest $\chDelta$.
The advantage of the more cumbersome form $X / \CP[X]$ is that it can
be extended to any other gauge since the gauge transformation properties
of the various objects are well-defined.\footnote{Applying
this to the case of Poincar\'e supergravity, one finds
$\CP = -\frac{1}{4} (\BCD^2 - 8R)$ and
$\chPi = -\frac{1}{8 R} (\BCD^2 - 8 R)$.}

\newpage
\section{Implicit grading} \label{imp_g}
We make use of the convention of \cite{wb} with respect to superspace
indices and their contractions. Furthermore, we adopt an implicit
grading scheme to avoid cumbersome notation. In any formula involving
capital Roman superindices ($A$, $B$, $C$,\ldots), an order is set
by the uncontracted indices of the first term; all other terms, if not in the order given,
must be supplemented with a grading to flip the indices to the appropriate
order. In addition, all index contractions are to be done
high to low between adjacent indices; any other configuration of
indices must be swapped into this configuration.

A few examples help a good deal. First the commutator:
\[
[\nabla_B, \nabla_A] = -R_{BA}
\]
Explanding this out gives
\[
\nabla_B \nabla_A - \nabla_A \nabla_B = -R_{BA}
\]
The first term sets the order to be $B$ then $A$; the second term
has the wrong order and so a grading must be inserted. The final
result is
\[
\nabla_B \nabla_A - (-)^{AB} \nabla_A \nabla_B = -R_{BA}
\]
The commutator is really an anticommutator if both $A$ and $B$
are fermionic.

Next, a more involved example:
\[
{V_C}^B \nabla_B W_A + {V_C}^B \nabla_A W_B = {F_{AB}}^{B D} G_{CD}
\]
The first term sets the order: $C$ then $A$. The $B$ contraction is
properly done, so no grading is necessary for the first term.
The second term has $C$ and $A$
in the correct order, but the $B$ contraction is done through the $A$.
One must swap the $A$ with either $B$ to get an adjacent
contraction, giving a grading $(-)^{AB}$.
The third term on the right side has the $B$ contraction done in the wrong
order. This requires we place a grading of $(-)^B$.
In addition, the $D$ contraction is done through the index $C$, giving
a grading of $(-)^{CD}$.
Finally, the overall order of indices is $A$ then $C$; swapping them to
the correct order gives a grading $(-)^{AC}$. The final result with the
gradings restored is
\[
{V_C}^B \nabla_B W_A + (-)^{AB} {V_C}^B \nabla_A W_B = (-)^{B + CD + AC} {F_{AB}}^{B D} G_{CD}.
\]
Now suppose $G$ were a two-form. Then the form indices $CD$ can be swapped
at the cost of a sign if they are not both fermionic; this gives
\[
{V_C}^B \nabla_B W_A + (-)^{AB} {V_C}^B \nabla_A W_B = -(-)^{B + AC} {F_{AB}}^{B D} G_{DC}.
\]
We would have compactly written this without the gradings as
\[
{V_C}^B \nabla_B W_A + {V_C}^B \nabla_A W_B = -{F_{AB}}^{B D} G_{DC}.
\]
which is equal to the first equation, provided we take
$G_{DC} = -G_{CD}$ which is true modulo the grading.

The advantage of this notational method is that in any calculation
involving superindices, one may naively treat them as if they were
all regular bosonic indices. Then when one wishes to actually
insert the components, the gradings can be added on the fly subject
to the rules we have given.

\newpage
\section{Global superconformal transformations}\label{global_sc}
In the literature on the conformal group, the generators on the
fields in the global approach are given at an arbitrary point $x$.
For example, $D$ is defined as $\Delta + x \cdot \partial$. (See
for example \cite{df}.) For completeness, we present the global
superconformal generators in the same global picture.

The action of a generator $g$ on a field
$\Phi$ may be defined at the origin. One takes the defining relations
for a primary superfield $\Phi$ as
\begin{gather}
P_a \Phi(0) = \partial_a \Phi(0), \;\;\;
Q_\alpha \Phi(0) = D_\alpha \Phi(0), \;\;\;
\bar Q^\dalpha \Phi(0) = \bar D^\dalpha \Phi(0) \eol
M_{ab} \Phi(0) = \mathcal S_{ab} \Phi(0), \;\;\; D \Phi(0) = \Delta \Phi(0), \;\;\; A \Phi(0) = i w \Phi(0) \eol
K_a \Phi(0) = 0, \;\;\;
S_\alpha \Phi(0) = 0, \;\;\;
\bar S^\dalpha \Phi(0) = 0
\end{gather}
The action of the supersymmetry translation generators $Q_\alpha$ at the origin
are formally defined to be the same as $D_\alpha$. This is certainly allowed by
the discussion in Wess and Bagger since both are equivalent to
$\partial_\alpha$ there; however, it will soon be apparent that
the \emph{intrinsic} action of $Q_\alpha$ on a field anywhere is to
be found by the action of $D_\alpha$.

In order to find the action of $g$ elsewhere,
conjugation by the translation operator is necessary. That is,
in order to calculate $g\Phi(z)$, one must commute $g$ past the
translation element, $g \Phi(z) = g e^{zP} \Phi(0) = e^{zP} \tilde g(z) \Phi(0)$
where $\tilde g(z) \equiv e^{-zP} g e^{zP}$, and the elements in the
expansion of $g'$ are to be taken to act on $\Phi$ at the origin.
One may calculate the effect of conjugation by the translation
element on each of the generators:
\begin{align}
\tilde P_a(z)  =& P_a \eol
\tilde Q_\alpha(z) =& Q_\alpha - 2i (\sigma^a \bar \theta)_\alpha P_a \eol
\tilde {\bar Q}^\dalpha(z) =& \bar Q^\dalpha - 2i (\bsigma^a \theta)^\dalpha P_a \eol
\tilde D(z) =& D + x^a P_a + \frac{1}{2} \theta Q + \frac{1}{2} \btheta \bar Q \eol
\tilde A(z) =& A - i \theta Q + i \btheta \bar Q - 2 (\theta \sigma^a \btheta) P_a \eol
\tilde M_{ab}(z) =& M_{ab} - x_{[a} P_{b]} + (\theta \sigma_{ab} Q) + (\btheta \bsigma_{ab} \bar Q)
                  + P_c \epsilon_{abcd} (\theta \sigma_d \btheta) \eol
\tilde K_a(z) =& K_a + 2 x_a D - 2 x_b M_{ab} + i (\theta \sigma_a \bar S) + i (\btheta \bsigma_a S)
          + 2 x_a x_b P_b - x^2 P_a \eol
       & + x_a (\theta Q) -2 x_b (\theta \sigma_{ab} Q)
          + x_a (\btheta \bar Q) - 2 x_b (\btheta \bsigma_{ab} \bar Q)
	  + 3 \sohn_a A + \epsilon_{abcd} \sohn_b M_{cd} \eol
       & - 2 \epsilon_{abcd} \sohn_b x_c P_d
	  - 2i \sohn_a (\theta Q) + 2i \sohn_a (\btheta \bar Q)
	  - 2 \sohn^a P_a \eol
\tilde S_\alpha(z) =& S_\alpha + i x_a (\sigma_a \bar Q)_\alpha - 2 \theta_\alpha D
                      + 3 i \theta_\alpha A + 2 (\sigma^{ba} \theta)_\alpha M_{ab} \eol
		 &   - 2 \theta_\alpha x^a P_a + 4 (\sigma^{ab} \theta)_\alpha x_a P_b
		      - 2 \theta^2 Q_\alpha - 2 \theta_\alpha (\btheta \bar Q)
		      + 2 i \theta^2 (\sigma^a \btheta)_\alpha P_a \eol
\tilde {\bar S}^\dalpha(z) =& \bar S^\dalpha + i x_a (\bsigma_a Q)^\dalpha - 2 \btheta^\dalpha D
                      - 3 i \btheta^\dalpha A + 2 (\bsigma^{ba} \btheta)^\dalpha M_{ab} \eol
		 &   - 2 \btheta^\dalpha x^a P_a + 4 (\bsigma^{ab} \btheta)^\dalpha x_a P_b
		      - 2 \btheta^2 \bar Q^\dalpha - 2 \btheta^\dalpha (\theta Q)
		      + 2 i \btheta^2 (\bsigma^a \theta)^\dalpha P_a
\end{align}
where $\sohn^a \equiv \theta\sigma^a\btheta$.

The first set of definitions imply
\beq
P_a \Phi(z) = \partial_a \Phi(z), \;\;\;
Q_\alpha \Phi(z) = D_\alpha \Phi(z) - 2i (\sigma^a \bar \theta)_\alpha \partial_a \Phi(z), \;\;\;
\bar Q^\dalpha \Phi(z) = D^\dalpha \Phi(z) - 2i (\bsigma^a \theta)^\dalpha \partial_a \Phi(z)
\eeq
which is consistent with the standard definitions in the literature.

\newpage
\section{Solution to the Bianchi identities} \label{app_Bianchi}
\subsection{General solution to gauge constraints}
The constraints chosen for conformal supergravity include a set of
constraints we shall call the ``gauge'' constraints for their
similarity to the constraints imposed on internal gauge theories in superspace:
\begin{gather*}
\{\nabla_\alpha, \nabla_\beta \} = \{\nabla_\dalpha, \nabla_\dbeta \} = 0 \\
\{\nabla_\alpha, \nabla_\dalpha \} = -2i \nabla_{\alpha \dalpha}
\end{gather*}
where $\nabla_A \equiv {E_A}^M \left(\partial_M - {h_M}^{\ul b} X_{\ul b} \right)$
is the covariant derivative. Here $X_{\ul b}$ is any non-translation symmetry generator;
for the conformal group it consists of scalings $D$, chiral rotations $A$,
Lorentz rotations $M_{ab}$, and the special conformal transformations $K_C$.
In principle, it may also include any internal symmetries (eg. Yang-Mills), but
we will not be explicitly concerned with those here. Since they commute with the conformal
group, it is quite easier to add these symmetries later when needed.

The gauge constraints enforce relationships between the various fermionic connections.
One could attempt to solve these constraints in terms of prepotentials and then
give all the connections and curvatures in terms of these prepotentials.
In the case of internal symmetries, this is quite straightforward to do;
one finds the prepotentials take the form of adjoint Hermitian superfields
$V = V^r X_r$ where $X_r$ is the internal symmetry generator. These in turn
possess a gauge invariance of the form $V \rightarrow V + \Lambda + \bar\Lambda$
for chiral superfields $\Lambda$.
When the symmetry group fails to commute with translations, this approach
is more difficult (though not impossible).
Moreover, in practice one is only concerned with
calculating the curvatures themselves. It turns out the simpler procedure is
usually to derive the constraints the curvatures obey and to solve the
curvatures in terms of some unconstrained superfields. In this latter procedure,
one finds chiral gaugino superfields $\W = \W^r X_r$ whose lowest components are
the gauginos and which transform homogeneously under the gauge transformation.
(These, of course, can be written in terms of the gauge prepotentials, but this
is usually not necessary to do.) It is this latter procedure which we will follow
here.

The starting point to deriving constraints on the curvatures is the Bianchi identity
\[
0 = \sum_{[ABC]} [\nabla_A, [\nabla_B, \nabla_C]]
\]
where the sum is over (graded) cyclic permutations of the indices. Both
the permutation and the commutator carry an implicit grading which
gives an extra sign whenever two fermionic indices are pushed past
each other. We shall examine each case in turn, in a treatment
roughly analogous to that of \cite{wb}.

The case of $\alpha \beta \gamma$ is trivial. All terms in the sum vanish.

The second case is $\alpha \beta \dgamma$. The Bianchi identity reads
\begin{align*}
0 & = [\nabla_\alpha, \{\nabla_\beta, \nabla_\dgamma\}]
	+ [\nabla_\dgamma, \{\nabla_\alpha, \nabla_\beta\}]
	+ [\nabla_\beta, \{\nabla_\dgamma, \nabla_\alpha\}] \\
& = -2i [\nabla_\alpha, \nabla_{\beta \dgamma}] + 0 - 2i [\nabla_\beta, \nabla_{\alpha \dgamma}] \\
& = +2i R_{\alpha (\beta \dgamma)} + 2i R_{\beta (\alpha \dgamma)}
\end{align*}
This implies the curvature is antisymmetric in the undotted indices.
We therefore may define the ``gaugino'' superfield $\W$ by
\begin{gather}
R_{\alpha (\beta \dbeta)} = 2i \eps_{\alpha \beta} \W_\dbeta, \;\;\;
R_{\dalpha (\beta \dbeta)} = 2i \eps_{\dalpha \dbeta} \W_\beta
\end{gather}
We have included the analogous formulae for the complex conjugate. Note that
$\W_{\beta}^\dag = -\W^{\dbeta}$ under this definition.

The third case of interest is $\alpha \beta c$. One finds
\begin{align*}
0 & = \{\nabla_\alpha, [\nabla_\beta, \nabla_c]\}
	+ [\nabla_c, \{\nabla_\alpha, \nabla_\beta\}]
	- \{\nabla_\beta, [\nabla_c, \nabla_\alpha]\} \\
& = -\{\nabla_\alpha, R_{\beta c}\} + 0 - \{\nabla_\beta, R_{\alpha c}\}
\end{align*}
Writing $R$ in terms of $\W$ and contracting with $\sigma^c_{\gamma \dgamma}$
gives
\[
0 = -2i \eps_{\beta \gamma} \{\nabla_\alpha,\W_\dgamma\}
	- 2i \eps_{\alpha \gamma} \{\nabla_\beta,\W_\dgamma\}
\]
A further contraction with $\eps^{\gamma \beta}$ gives
\begin{align}
0 = \{\nabla_\alpha, \W_\dalpha\} = \{\nabla_\dalpha, \W_\alpha \}
\end{align}
where we have included the conjugate result as well. This generalizes
the chirality condition of the normal Yang-Mills case, but this is not
quite the conventional chirality. To wit,
\[
0 = \{\nabla_\alpha, {\W_\dalpha}^B X_B\}
	= (\nabla_\alpha \W_\dalpha^B) X_B - {\W_\dalpha}^C {f_{C \alpha}}^B X_B
\]
$\W_\dalpha$ is antichiral in the conventional sense only when the
second term vanishes, which is the case when the symmetry group under
consideration is internal (ie. one that commutes with
translations). Nevertheless, it is useful to retain the term ``chiral'' to
describe $\W_\alpha$ and ``antichiral'' for $\W_\dalpha$.

The fourth case of interest is $\alpha \dbeta c$. We find
\begin{align*}
0 & = \{\nabla_\alpha, [\nabla_\dbeta, \nabla_c]\}
	+ [\nabla_c, \{\nabla_\alpha, \nabla_\dbeta\}]
	- \{\nabla_\dbeta, [\nabla_c, \nabla_\alpha]\} \\
& = -\{\nabla_\alpha, R_{\dbeta c}\} -2i [\nabla_c, \nabla_{\alpha \dbeta}] - \{\nabla_\dbeta, R_{\alpha c}\}
	= - \{\nabla_\alpha, R_{\dbeta c}\} + 2i R_{c (\alpha \dbeta)} - \{\nabla_\dbeta, R_{\alpha c}\}
\end{align*}
which serves to define the bosonic curvature:
\begin{align*}
2i R_{b (\alpha \dalpha)} = \{\nabla_\alpha,R_{\dalpha b}\} + \{\nabla_\dalpha,R_{\alpha b}\}
\end{align*}
Rewriting the right-hand side in terms of $\W$ gives
\begin{align*}
R_{(\beta \dbeta) (\alpha \dalpha)} = +\eps_{\dalpha\dbeta} \{\nabla_\alpha,\W_\beta\}
	+\eps_{\alpha\beta} \{\nabla_\dalpha,\W_\dbeta\}
\end{align*}
The left-hand side is antisymmetric under interchange of the pairs $(\beta \dbeta)$ 
and $(\alpha \dalpha)$ and so the right-hand side must be as well. It is
easy to check that this requires the additional condition
\begin{align}
\{\nabla^\alpha, \W_\alpha\} = \{\nabla_\dalpha, \W^\dalpha\}
\end{align}
This generalizes the analogous property for the Yang-Mills case much as the chirality
condition has been generalized.
Using this constraint one may rewrite the curvature in the manifestly antisymmetric form
\begin{align}
R_{(\beta \dbeta) (\alpha \dalpha)} = -\frac{1}{2} \eps_{\dbeta\dalpha} \{\nabla_{\{\beta},\W_{\alpha\}}\}
	- \frac{1}{2} \eps_{\beta\alpha} \{\nabla_{\{\dbeta},\W_{\dalpha\}}\}
\end{align}

The remaining cases to check are $\alpha b c$ and $a b c$.
These turn out to follow from the previous conditions on $\W$ (just as in
the Yang-Mills case) and so we do not include them here. All other
cases are conjugates of those given above, and so the constraints have
been solved.

It is useful to derive how the symmetry generator $X_{\ul d}$ acts on $\W_\beta$.
In order to do this, it is helpful to have a set of constraints on the structure
constants consistent with the Jacobi identities. The easiest way to proceed is
from the general formula \eqref{eq_Rtrans}, specializing to the cases of $CB$ equal to $\gamma \beta$
and $\gamma \dbeta$. For the first case, one finds
\begin{align}\label{eq_Jac1}
0 = \sum_{(\gamma \beta)}
	\left(-{f_{\ul d \gamma}}^F {R_{F \beta}} - {f_{\ul d \gamma}}^{\ul f} {f_{\ul f \beta}}^A X_A\right)
\end{align}
where $R_{F \beta} = {R_{F \beta}}^A X_A$ where $X_A$ in this and the above
formula consists of both the translations $P_A$ and the non-translation
symmetries $X_{\ul a}$. For the second case, one finds
\begin{align}\label{eq_Jac2}
0 = 2i {f_{\ul c (\beta \dbeta)}}^A X_A
	- {f_{\ul c \beta}}^D R_{D \dbeta} - {f_{\ul c \dbeta}}^D R_{D \beta}
	- {f_{\ul c \beta}}^{\ul d} {f_{\ul d \dbeta}}^A X_A
		- {f_{\ul c \dbeta}}^{\ul d} {f_{\ul d \beta}}^A X_A
\end{align}
(We have relabelled $\ul d$ to $\ul c$ and $\gamma$ to $\beta$ since $\beta$
and $\dbeta$ naturally go together to form a vector index.)

One set of additional constraints is also useful. For any theory in
superspace, we would like to be able to write down chiral integrals;
the existence of these implies the structure constant constraints \eqref{eq_Wcons}
\[
{f_{\ul a \beta}}^c = f_{\ul a \beta \dgamma} = 0, \;\;\;
{f_{\ul a \beta}}^{\ul c} \left({f_{\ul c d}}^d + {f_{\ul c \ddelta}}^\ddelta \right) = 0
\]
as well as their complex conjugates
\[
{f_{\ul a \dbeta}}^c = {f_{\ul a \dbeta}}^\gamma = 0, \;\;\;
{f_{\ul a \dbeta}}^{\ul c} \left({f_{\ul c d}}^d - {f_{\ul c \delta}}^\delta \right) = 0
\]

Applying these constraints to \eqref{eq_Jac1} gives
\begin{align}
0 = \sum_{(\gamma \beta)}
	\left(-{f_{\ul d \gamma}}^\nu {R_{\nu \beta}} - {f_{\ul d \gamma}}^{\ul f} {f_{\ul f \beta}}^A X_A\right)
	= - \sum_{(\gamma \beta)}{f_{\ul d \gamma}}^{\ul f} {f_{\ul f \beta}}^A X_A
\end{align}
which is a further constraint on the structure constants. Note that this constraint
is equivalent to
\begin{align}
{f_{\ul d \gamma}}^{\ul f} {f_{\ul f \beta}}^A X_A =
	\frac{1}{2} \eps_{\gamma \beta} {f_{\ul d}}^{\phi \ul f} {f_{\ul f \phi}}^A X_A
\end{align}

Applying the constraints to \eqref{eq_Jac2} gives ${f_{\ul c b}}^A$ in terms
of ${f_{\ul c \beta}}^A$ and ${f_{\ul c \dbeta}}^A$:
\begin{gather} \label{eq_Jac3}
f_{\ul c (\beta \dbeta)(\alpha \dalpha)} = 2 \eps_{\dbeta \dalpha} f_{\ul c \beta \alpha}
	- 2 \eps_{\beta \alpha} f_{\ul c \dbeta \dalpha} \\
{f_{\ul c (\beta \dbeta)}}^\alpha = -\frac{i}{2} {f_{\ul c \dbeta}}^{\ul d} {f_{\ul d \beta}}^\alpha \\
{f_{\ul c (\beta \dbeta)}}^\dalpha = -\frac{i}{2} {f_{\ul c \beta}}^{\ul d} {f_{\ul d \dbeta}}^\dalpha \\
{f_{\ul c (\beta \dbeta)}}^{\ul a} = -\frac{i}{2} {f_{\ul c \beta}}^{\ul d} {f_{\ul d \dbeta}}^{\ul a}
	-\frac{i}{2} {f_{\ul c \dbeta}}^{\ul d} {f_{\ul d \beta}}^{\ul a}
\end{gather}

We are now in a position to derive the general gauge transformation
property of $\W_\dbeta$. To proceed, first note that in principle
$R_{\gamma (\beta \dbeta)} = {f_{\gamma (\beta \dbeta)}}^A X_A + \Delta R_{\gamma (\beta \dbeta)}$
where the first term on the right is a structure constant in the global theory
and the second term is the local correction. (In practice, the first
term usually vanishes.) It follows that a similar decomposition of
$\W$ takes place, giving ${\W_{\dbeta}}^A = {f_{\dbeta}}^A + \Delta{\W_{\dbeta}}^A$.
Since the first term is a structure constant, it necessarily is
gauge invariant; we therefore need only calculate the
gauge transformation of the local correction. Using equation \eqref{eq_dRtrans},
for the case of $CB = \gamma b$ gives
\begin{align}
2i \eps_{\gamma \beta} X_{\ul d} {\W_\dbeta}^A = -2i \eps_{\gamma \beta} {\Delta \W_{\dbeta}}^F {f_{F \ul d}}^A
	+ 2i f_{\ul d \gamma \beta} \Delta {\W_\dbeta}^A
	- i f_{\ul d (\beta \dbeta)(\gamma \dgamma)} \Delta \W^{\dgamma A}
\end{align}
Using \eqref{eq_Jac3} allows one to show the right-hand size is proportional to
$\eps_{\gamma \beta}$. The final result is
\begin{align*}
X_{\ul d} {\W_\dbeta}^A = -{\Delta \W_{\dbeta}}^F {f_{F \ul d}}^A
	- {f_{\ul d \phi}}^\phi \Delta {\W_\dbeta}^A
	- {f_{\ul d \dbeta \dgamma}} \Delta \W^{\dgamma A}
\end{align*}
The first term on the right hand size can be combined with the left-hand side
to yield the compact formula
\begin{align}
[X_{\ul d}, {\Delta \W_\dbeta}] =
	- {f_{\ul d \phi}}^\phi \Delta {\W_\dbeta}
	- {f_{\ul d \dbeta \dgamma}} \Delta \W^{\dgamma}
\end{align}
The complex conjugate is
\begin{align}
[X_{\ul d}, {\Delta \W_\beta}] =
	+ {f_{\ul d \dphi}}^\dphi \Delta {\W_\beta}
	- {f_{\ul d \beta}}^\gamma \Delta \W_{\gamma}
\end{align}

We include the precise definition of the covariant derivative of the local
gaugino superfields for completeness:
\begin{align}
\nabla_C {\Delta\W_\beta}^A = {E_C}^M \partial_M {\Delta\W_\beta}^A
	+ {h_C}^{\ul d} \left({\Delta\W_\beta}^F {f_{F \ul d}}^A
		- {f_{\ul d \dphi}}^\dphi \Delta{\W_\beta}^A
		+ {f_{\ul d \beta}}^\gamma \Delta {\W_\gamma}^A \right)\\
\nabla_C {\Delta\W_\dbeta}^A = {E_C}^M \partial_M {\Delta\W_\dbeta}^A
	+ {h_C}^{\ul d} \left({\Delta\W_\dbeta}^F {f_{F \ul d}}^A
		+ {f_{\ul d \phi}}^\phi \Delta{\W_\dbeta}^A
		+ {f_{\ul d \dbeta \dgamma}} \Delta \W^{\dgamma A} \right)
\end{align}
(The covariant derivative of the constant part of $\W$ vanishes.)

\subsection{Conformal supergravity solution}
From the result of the previous section, we may define \emph{maximal} conformal
supergravity as the theory with the Yang-Mills constraints
\begin{gather*}
\{\nabla_\alpha, \nabla_\beta \} = \{\nabla_\dalpha, \nabla_\dbeta \} = 0 \\
\{\nabla_\alpha, \nabla_\dalpha \} = -2i \nabla_{\alpha \dalpha}.
\end{gather*}
It follows that the remaining curvatures are of the form
\begin{align*}
R_{\alpha (\beta \dbeta)} &= 2i \eps_{\alpha \beta} \W_\dbeta \\
R_{\dalpha (\beta \dbeta)} &= 2i \eps_{\dalpha \dbeta} \W_\beta \\
R_{(\beta \dbeta) (\alpha \dalpha)} &= -\frac{1}{2} \eps_{\dbeta\dalpha} \{\nabla_{\{\beta},\W_{\alpha\}}\}
	- \frac{1}{2} \eps_{\beta\alpha} \{\nabla_{\{\dbeta},\W_{\dalpha\}}\}
\end{align*}
where the superfields $\W$ obey the constraints
\begin{gather*}
\{\nabla_\dalpha,\W_\alpha\} = \{\nabla_\alpha, \W_\dalpha\} = 0\\
\{\nabla^\alpha,\W_\alpha\} = \{\nabla_\dalpha, \W^\dalpha\}
\end{gather*}
The $\W$ here is understood as
\[
\W_\alpha = {\W(P)_\alpha}^B P_B + \frac{1}{2} {\W(M)_\alpha}^{c b} M_{bc}
	+ {\W(D)_\alpha} D + {\W(A)_\alpha} A + {\W(K)_\alpha}^B K_B
\]
That is, there is a $\W$ associated with each symmetry in the conformal group.
These $\W$ are not conformally primary but are rotated into each other
by the action of the conformal group. In this case, the global
theory is characterized by $\W=0$ and so no decomposition of $\W$ into global
and local parts is necessary.

The chirality condition $\{\nabla_\dalpha, \W_\alpha\} = 0$ reads
\begin{align}
0 &= \nabla_\dalpha {\W(P)_\alpha}^B \nabla_B
	- {\W(P)_\alpha}^C {T_{C \dalpha}}^B \nabla_B + \W(M)_{\alpha \dalpha \dbeta} \nabla^\dbeta
	+ \frac{1}{2} \W(D)_{\alpha} \nabla_\dalpha + i \W(A)_\alpha \nabla_\dalpha \eol
0 &= \frac{1}{2} \nabla_\dalpha {\W(M)_\alpha}^{cb} M_{bc}
	- \frac{1}{2} {\W(P)_\alpha}^D {R_{D \dalpha}}^{cb} M_{bc}
	- 2 \W(K)_{\alpha \dbeta} {M^\dbeta}_\dalpha \eol
0 &= \nabla_\dalpha {\W(K)_\alpha}^B K_B
	- {\W(P)_\alpha}^C {R(K)_{C \dalpha}}^B K_B
	+ i {\W(K)_{\alpha (\dalpha}}^{\beta)} S_\beta \eol
0 &= \nabla_\dalpha \W(D)_\alpha - {\W(P)_\alpha}^B R(D)_{B \dalpha} - 2 \W(K)_{\alpha \dalpha} \eol
0 &= \nabla_\dalpha \W(A)_\alpha - {\W(P)_\alpha}^B R(A)_{B \dalpha} - 3i \W(K)_{\alpha \dalpha} \label{eq_chirality}
\end{align}
For the last two equations we have omitted the generators $D$ and $A$ respectively.
The curvatures in these expressions are defined in terms of $\W$;
therefore, these formulae possess both linear and quadratic terms in $\W$.

The condition $\{\nabla^\alpha, \W_\alpha \} = \{\nabla_\dalpha, \W^\dalpha \}$ reads
\begin{gather}
\nabla^\alpha {\W(P)_{\alpha}}^B \nabla_B
	+ \W(P)^{\alpha C} {T_{C \alpha}}^{B} \nabla_B - {{\W(M)^\alpha}_\alpha}^\beta \nabla_\beta
	- \frac{1}{2} \W(D)^{\alpha} \nabla_\alpha + i \W(A)^\alpha \nabla_\alpha \eol
	 = \nabla_\dalpha \W(P)^{\alpha B} \nabla_B
	+ {\W(P)_\dalpha}^C {T_C}^{\dalpha B} \nabla_B - {{\W(M)_{\dalpha}}^\dalpha}_\dbeta \nabla^\dbeta
	- \frac{1}{2} \W(D)_{\dalpha} \nabla^\dalpha - i \W(A)_\dalpha \nabla^\dalpha \label{eq_herm1}
\end{gather}
\begin{gather}
\frac{1}{2} \nabla^\alpha {\W(M)_\alpha}^{cb} M_{bc}
	+ \frac{1}{2} \W(P)^{\alpha D} {R_{D \alpha}}^{cb} M_{bc}
	+ 2 \W(K)^{\alpha \beta} M_{\beta \alpha} \eol
	 = \frac{1}{2} \nabla_\dalpha \W(M)^{\dalpha cb} M_{bc}
	+ \frac{1}{2} {\W(P)_\dalpha}^D {R_{D}}^{\dalpha cb} M_{bc}
	+ 2 \W(K)_{\dalpha \dbeta} M^{\dbeta \dalpha}\label{eq_herm2}
\end{gather}
\begin{gather}
\nabla^\alpha {\W(K)_\alpha}^{B} K_B
	+ \W(P)^{\alpha C} {R(K)_{C\alpha}}^{B} K_B
	- i {{\W(K)^{\dalpha}}_\dalpha}^{\beta} S_\beta \eol
	= \nabla_\dalpha \W(K)^{\dalpha B} K_B
	+ {\W(P)_\dalpha}^C {R(K)_{C}}^{\dalpha B} K_B
	- i {\W(K)_{\dalpha}}^{(\dalpha \beta)} S_\beta \label{eq_herm3}
\end{gather}
\begin{gather}\label{eq_herm4}
\nabla^\alpha \W(D)_\alpha + {\W(P)}^{\alpha B} R(D)_{B\alpha} + {2 \W(K)^\alpha}_\alpha
	= \nabla_\dalpha \W(D)^\dalpha + {\W(P)_\dalpha}^B {R(D)_B}^\dalpha + {2 \W(K)_\dalpha}^\dalpha
\end{gather}
\begin{gather}\label{eq_herm5}
\nabla^\alpha \W(A)_\alpha + {\W(P)}^{\alpha B} R(A)_{B\alpha} - 3i {\W(K)^\alpha}_{\alpha}
	 = \nabla_\dalpha \W(A)^\dalpha + {\W(P)_\dalpha}^B {R(A)_B}^\dalpha + 3i {\W(K)_\dalpha}^{\dalpha}
\end{gather}

This is a very complicated structure that simplifies a great deal
when we apply the further constraints of conformal supergravity. These
are $F_{\alpha b} = 0$, $H_{\alpha b} = 0$, 
and ${T_{\gamma b}}^A = 0$ along with their complex conjugates.
(In addition, we want ${T_{cb}}^a = 0$ but this turns out to be a
consequence of the other constraints.) These constraints clearly force
$\W(A)_\alpha$, $\W(D)_\alpha$, and ${\W(P)_\alpha}^B$ to zero.
Since this set of constraints is conformally invariant (ie.
$S_\gamma \W(D)_\beta = +2 \W(P)_{\beta \gamma} = 0$), it follows
that the covariant derivative of any of these also vanishes.

The only non-vanishing gaugino superfields are then $\W(M)$ and $\W(K)$.
In terms of these, the chirality constraints \eqref{eq_chirality} read
\begin{align*}
0 &= \W(M)_{\alpha \dalpha \dbeta} \nabla^\dbeta \\
0 &= \frac{1}{2} \nabla_\dalpha {\W(M)_\alpha}^{cb} M_{bc}
	- 2 \W(K)_{\alpha \dbeta} {M^\dbeta}_\dalpha \\
0 &= \nabla_\dalpha {\W(K)_\alpha}^B K_B
	+ i {\W(K)_{\alpha (\dalpha}}^{\beta)} S_\beta \\
0 &= - 2 \W(K)_{\alpha \dalpha} \\
0 &= - 3i \W(K)_{\alpha \dalpha}
\end{align*}
It follows that $\W(M)_{\alpha \dbeta \dgamma}$ and
$\W(K)_{\alpha \dalpha}$ vanish. Furthermore, $\W(M)_{\alpha \beta \gamma}$
is chiral and $\nabla_\dalpha {\W(K)_\alpha}^\beta = -i {\W(K)_{\alpha \dalpha}}^\beta$.

Considering the remaining constraints, we have \eqref{eq_herm1}
\begin{gather*}
- {{\W(M)^\alpha}_\alpha}^\beta \nabla_\beta = - {{\W(M)_{\dalpha}}^\dalpha}_\dbeta \nabla^\dbeta
\end{gather*}
This implies that ${\W(M)^\alpha}_{\alpha \gamma} = 0$. Therefore,
$\W(M)_{\alpha \beta \gamma}$ is totally symmetric in its indices.
Similarly for the conjugate.

Next is \eqref{eq_herm2}
\begin{gather*}
\frac{1}{2} \nabla^\alpha {\W(M)_\alpha}^{cb} M_{bc}
	+ 2 \W(K)^{\alpha \beta} M_{\beta \alpha}
	 = \frac{1}{2} \nabla_\dalpha \W(M)^{\dalpha cb} M_{bc}
	+ 2 \W(K)_{\dalpha \dbeta} M^{\dbeta \dalpha}
\end{gather*}
which implies that $\W(K)_{\beta \gamma} = -\frac{1}{4} \nabla^\alpha \W(M)_{\alpha\beta \gamma}$
(as well as its conjugate). Since we already know that
$\W(K)_{\beta (\alpha \dalpha)} = i \nabla_\dalpha \W(K)_{\beta \alpha}$, it follows
that $W(K)_{\beta (\alpha \dalpha)} = -\frac{1}{2} {\nabla^\phi}_\dalpha \W(M)_{\phi \beta \alpha}$.

Equation \eqref{eq_herm3} implies
\begin{gather*}
\nabla^\alpha {\W(K)_\alpha}^{B} K_B
	- i {{\W(K)^{\dalpha}}_\dalpha}^{\beta} S_\beta
	= \nabla_\dalpha \W(K)^{\dalpha B} K_B
	- i {\W(K)_{\dalpha}}^{(\dalpha \beta)} S_\beta
\end{gather*}
which, when we insert our existing formulae, gives a new identity
\[
\nabla^\beta {\nabla^\phi}_\dalpha \W(M)_{\phi \beta \alpha}
	= \nabla^\dbeta {\nabla^\dphi}_\alpha \W(M)_{\dphi \dbeta \dalpha}
\]

Finally, we note that the final two constraints \eqref{eq_herm4} and \eqref{eq_herm5}
give
\[
+ {2 \W(K)^\alpha}_\alpha = {2 \W(K)_\dalpha}^\dalpha
\]
and
\[
- 3i {\W(K)^\alpha}_{\alpha} = + 3i {\W(K)_\dalpha}^{\dalpha},
\]
which are satisfied trivially. (Both sides vanish.)

All of the curvatures are then specified in terms of a single
totally symmetric chiral superfield $\W(M)_{\alpha \beta \gamma}$ as well as its conjugate,
which together obey a Bianchi identity. Furthermore, from the transformation
rules of the $\W$ found in the previous section, $\W(M)_{\alpha \beta \gamma}$ is
conformally primary of scale dimension $3/2$ and $U(1)$ weight $+1$.
To make contact with the conventional normalizations and reality
conditions, we define a new superfield $W_{\alpha \beta \gamma}$ via
$\W(M)_{\alpha \beta \gamma} = -2 W_{\alpha \beta \gamma}$ and
$\W(M)_{\dalpha \dbeta \dgamma} = +2 W_{\alpha \beta \gamma}$
and summarize our results in terms of it:
\begin{gather*}
{\W(P)_\alpha}^B = {\W(P)_\dalpha}^B = 0 \\
{\W(D)_\alpha} = {\W(D)_\dalpha} = 0 \\
{\W(A)_\alpha} = {\W(A)_\dalpha} = 0 \\
\W(M)_{\alpha \dbeta \dgamma} = \W(M)_{\dalpha \beta \gamma} = 0 \\
\W(M)_{\alpha \beta \gamma} = -2 W_{\alpha \beta \gamma}, \;\;\;
\W(M)_{\dalpha \dbeta \dgamma} = +2 W_{\dalpha \dbeta \dgamma} \\
\W(K)_{\alpha \beta} = \frac{1}{2} \nabla^\phi W_{\phi \alpha \beta}, \;\;\;
\W(K)_{\dalpha \dbeta} = \frac{1}{2} \nabla^\dphi W_{\dphi \dalpha \dbeta} \\
\W(K)_{\alpha \dbeta} = \W(K)_{\dalpha \beta} = 0 \\
\W(K)_{\alpha (\beta \dbeta)} = {\nabla^\phi}_\dbeta W_{\phi \alpha \beta}, \;\;\;
\W(K)_{\dalpha (\beta \dbeta)} = {\nabla^\dphi}_\beta W_{\dphi \dalpha \dbeta}
\end{gather*}

$W_{\alpha \beta \gamma}$ is a totally symmetric chiral primary superfield
obeying a Bianchi identity
\begin{align*}
\nabla^\beta {\nabla^\phi}_\dalpha W_{\phi \beta \alpha}
	= - \nabla^\dbeta {\nabla^\dphi}_\alpha W_{\dphi \dbeta \dalpha}
\end{align*}

From the above definitions of $\W$ and of the curvatures $R$ in terms
of $\W$, one can quite easily derive the curvatures in terms of $W$.
These are given fully in Section \ref{curv_solns}.


\end{document}